\documentclass[pre,aps,onecolumn,showpacs,10pt]{revtex4-1}
\usepackage{amsmath,amssymb,graphicx}
\usepackage{epsfig}
\usepackage{textcomp}
\usepackage{color}
\usepackage{cases}
\usepackage{epstopdf}
\usepackage{multirow}
\usepackage{array}
\usepackage{ulem}
\usepackage[left=1.3in,right=1.8in,top=1.2in,bottom=1.8in]{geometry}
\usepackage{hyperref}
\hypersetup{pdfpagemode=UseNone,colorlinks=true,linktoc=page,pdfborder={0 0 0},linkcolor=blue,citecolor=blue}

\begin{document}
\title{WKB theory of large deviations in stochastic populations}

\author{Michael Assaf  and Baruch Meerson}

\affiliation{Racah Institute of Physics, Hebrew University
of Jerusalem, Jerusalem 91904, Israel}
\email{michael.assaf@mail.huji.ac.il, meerson@mail.huji.ac.il}

\begin{abstract}
Stochasticity can play an important role in the dynamics of biologically relevant populations. These span a broad range of scales: from intra-cellular populations of molecules to population of cells and then to groups of plants, animals and people. Large deviations in stochastic population dynamics -- such as those determining population extinction, fixation or switching between different states --
are presently in a focus of attention of statistical physicists.  We review recent progress in applying different variants of dissipative WKB approximation (after Wentzel, Kramers and  Brillouin) to this class of problems. The WKB approximation allows one to evaluate the mean time and/or probability of population extinction, fixation and switches resulting from
either intrinsic (demographic) noise, or a combination of the demographic noise and environmental variations,
deterministic or random. We mostly cover well-mixed populations, single and multiple, but also briefly consider populations on heterogeneous networks and spatial populations. The spatial setting also allows one to study large fluctuations of the speed of biological invasions. Finally, we briefly discuss possible directions of future work.

\vspace{0.5cm}

\end{abstract}

\maketitle

\tableofcontents

\section{Introduction}
\label{intro}

It was realized long ago that stochasticity can play an important role in the dynamics of biologically relevant populations. These span a broad range of scales: from intra-cellular populations of molecules to population of cells and then to groups of plants, animals and people  \cite{Delbrueck,Bartlett,Karlin,Bailey,Nisbet,Nasellbook,Gabriel,Andersson,Allen,Nasellbook2,Tsimring,Allen2015}.
The stochasticity, or noise may have different origins. Two main types of noise have been identified: intrinsic, or demographic noise and extrinsic, or environmental noise. The demographic noise
reflects natural ``quantization" of individuals the population is made of, and random character of elemental transitions such as births, deaths, interactions and movement of the individuals. Environmental variations (not necessarily stochastic) are usually described in terms of time-dependent elemental transition rates. If one ignores the  noise, a steady-state population size corresponds to an attractor in the space of population sizes as described by a \textit{deterministic} rate equation, or equations. Taking the stochasticity into account, one observes random excursions of the population sizes around the attractor. For \textit{established} populations the typical excursions around the attractor are small. However, from time to time a rare large excursion occurs, which may lead to extinction of one of more populations, or to a population switch to the vicinity of another attractor. As a result, stochasticity can make deterministically stable attractors \textit{metastable}. In such situations it is interesting, and often biologically important, to determine the mean time to extinction or switch, when starting from the vicinity of an attractor. This review will mostly deal with problems of this type. We will confine ourselves to individual-based Markov population models and not consider continuum models of population dynamics, based on Langevin-type equations, see \textit{e.g.} Ref. \cite{Spagnolo}. The fact that typical fluctuations are small implies the presence of a small parameter, usually coming from a disparity of the elemental transition rates. The WKB approximations -- the focus of this review -- utilize this small parameter in a smart way in order to approximately solve the master equation which is considered to be the exact model of the population dynamics. There are three
recent reviews
on closely related subjects. The
review \cite{OM2010} on the WKB approximation in stochastic population models was primarily intended for population biologists and ecologists, and it dealt only with extinction of well-mixed populations.  The review \cite{WeberFrey} provided a comprehensive and pedagogic introduction to the path integral representation of master equations and to approximate methods of solution. Finally, the review \cite{Bressloff2016} is a survey of different mathematical methods of analysis of biological switching processes in a variety of systems, at both the genotypic and phenotypic levels.

In a well-mixed single population -- the simplest paradigm of stochastic population dynamics -- the only generic type of attractor of the deterministic rate equation is a stable fixed point. As we will see, even in this simple situation there are two different scenarios of population extinction caused by demographic noise.
Scenario A is observed in the absence of an Allee effect, when the population is monostable.  Scenario B corresponds to bistability caused by an Allee effect. In the context of population switches due to a weak noise one always deals with bistability.

When there are two interacting populations (predator and prey, competition or symbiosis, susceptible and infected populations, etc.), a generic attractor can be either a stable fixed point (a node or a focus), or a stable limit cycle, and we will review these cases separately. As the number of interacting populations increases, additional types of attractors may appear, including chaotic attractors \cite{Ott}. Noise-induced escape from chaotic attractors has been studied theoretically \cite{Grassberger,Graham1991} and experimentally \cite{Khovanov} for continuous noisy systems, as described by Langevin equations, but not for stochastic populations.

The WKB (Wentzel-Kramers-Brillouin) approximation is best known to physicists in the context of quantum mechanics. In the time-dependent WKB theory a WKB ansatz leads to an approximate description of the time-dependent wave function as described by the non-stationary Schr\"{o}dinger equation: for example, for a (quasi-classical) quantum particle in a time-dependent potential, see Ref. \cite{LLQM}, Chapter III, Sec. 17.  In its turn, the stationary WKB theory can be used for approximate calculations of the eigenvalues  and eigenfunctions corresponding to highly excited states
as described by the stationary Schr\"{o}dinger equation, for evaluating the (small) probability of tunneling through a potential barrier, and for other purposes, see Chapter VII of Ref. \cite{LLQM}.  Similarly, there are time-dependent and stationary versions of the WKB approximation in stochastic problems as we will see below.

For stochastic classical systems with a \textit{continuous} space of states a WKB theory was developed by Freidlin and Wentzel \cite{FW}, Dykman \cite{Dykman} and Graham \cite{Graham}. As populations are naturally ``quantized" -- they have an integer number of individuals -- the WKB approximation here has some important differences.  As of present,  there are two main strategies of applying the WKB approximation to stochastic populations. In the more straightforward ``real-space" WKB method (where by ``real space" is actually meant the space of population sizes) one applies a WKB ansatz directly to the master equation, or to the equation describing the stationary or quasi-stationary distribution \cite{Kubo,KMST1985,DMR1994,RFRSM2005,vEjnden,KS2007,DSL2008,MS2008,EK2009,KD2009,AM2010,
MobA2010,AMob2010,KDM2010,B2010,AM2011,LM2011,BMcK2011,ARS2011,GM2012,N2012,BTG2012,KMKS2012,
GMR2013,ERADS2013,ARSG2013,VLACB2013,AMR2013,RBBSA2015,BAM2015,N2015,MACH2015,SM2016}.
In the ``momentum space" WKB method one first derives an exact evolution equation -- a linear partial differential equation (PDE) -- for the probability generating function. For example, for a well-mixed single population the probability generating function is defined as \cite{Gardiner,vanKampen}
\begin{equation}\label{genfun}
G(p,t) = \sum_{n=0}^{\infty} p^n P_n(t),
\end{equation}
where $P_n(t)$ is the probability of observing $n$ individuals at time $t$, and $p$ is an auxiliary variable. Then one applies a WKB ansatz to the evolution equation for $G(p,t)$ or to the stationary equation describing the eigenstates of $G(p,t)$ \cite{EK2004,AM2006,AM2006a,AM2007,EscuderoRodriguez,KM2008,AKM2008,AKM2009,AMS2010,BAA2016,BA2016}.  For established populations a small parameter required for the WKB approximation is $1/N$, where $N\gg 1$ is a typical population size characterizing the (quasi)stationary  distribution. In well-mixed single populations each of the two WKB methods, in combination with additional perturbation techniques, using the same small parameter $1/N$, yields accurate and controllable results,  in the leading and subleading orders of $N$,  for the mean time to extinction/switch. Even the leading-order WKB results, which miss pre-exponential factors, are usually much more accurate than the results obtained with the more traditional ``diffusion approximation" based on the van Kampen system-size expansion \cite{Nisbet}. Indeed, as was shown in many studies \cite{Gaveau,EK2004,Doering,KS2007,AM2006a,AM2007}, the mean time to extinction/switch, obtained with the diffusion approximation, usually involves an error that is exponentially large in $N$.

For multiple populations, even the leading-order results for the mean time to extinction/switch are, in general, unavailable in analytical form. Still, the WKB method proves very useful here, as it yields, albeit in a numerical form, \textit{the optimal path}: a special trajectory of the system
in the phase space of population sizes and the conjugate momenta that gives a dominant contribution to the specified large deviation such as extinction or switch. In its turn, the mean time to extinction/switch is given (again, in a numerical form) by the ``classical action" evaluated along the optimal path. In this class of problems analytical progress is  possible  if, besides $N$, there is an additional  small parameter in the system, coming from an additional disparity in the elemental transition rates \cite{DSL2008,KM2008,KD2009,MS2009,KDM2010,KMKS2012}.

As we already mentioned, environmental/extrinsic variations are usually taken into account by allowing some of the elemental transition rates to vary with time. In some cases these variations are deterministic: describing seasonality of births and deaths, ``catastrophes" and other effects in population biology, and vaccination in epidemiology. In other cases the variations can be viewed as stochastic, reflecting multiple concurrent mechanisms. In the leading order of the WKB approximation these problems are similar to multi-population problems. Here too the WKB theory yields the optimal path of the population to the specified large deviation \cite{EscuderoRodriguez,KMS2008,AKM2008,AKM2009,LM2013,B2015,ARSG2013} and, for a stochastic rate variation, the optimal realization of the environmental noise that gives a dominant contribution to this large deviation \cite{KMS2008,LM2013,ARSG2013}.

When the number of interacting sub-populations increases,  their topological heterogeneity becomes very important
in determining the large deviation properties. This is a largely unexplored subject, and we will review some recent
work in this direction \cite{HS2016}.

Introducing migration of populations in space (traditionally modeled as random walk of the individuals) makes the large deviation problems both richer and more difficult. Here the WKB method, once applicable,  provides a convenient classical-field-theory framework for studying large deviations: from population extinction \cite{EK2004,MS2011} to large velocity fluctuations of biological invasion fronts \cite{MSK2011,MS2011a,MVS2012}.

In the last few years the WKB approximation in stochastic population dynamics has been adopted by a growing number of practitioners. This review (which may have a natural bias toward our own work) does not attempt to cover all of them.

\section{Extinction, Fixation and Switching in Established Single Populations}
In this section we will briefly review the two WKB methods -- the real-space and the momentum-space -- on several examples of escape from a long-lived metastable state corresponding to an established well-mixed single population.
The elemental transitions do not need to be single-step: they can involve transitions between a state with $n$ individuals and a state with $n+r$ individuals, where $r=\pm 1,\pm 2,\dots$. These transitions occur with the rate $W_r(n)$, so the master equation reads
\begin{eqnarray}\label{master0}
\frac{d P_n(t)}{dt} &=& \sum_{r} \left[W_r(n- r)P_{n-r}(t) -W_r(n)P_n(t)\right],
\end{eqnarray}
where $P_{k<0}=0$.

\begin{figure}[ht]
\includegraphics[width=6.5cm,height=2.5cm,clip=]{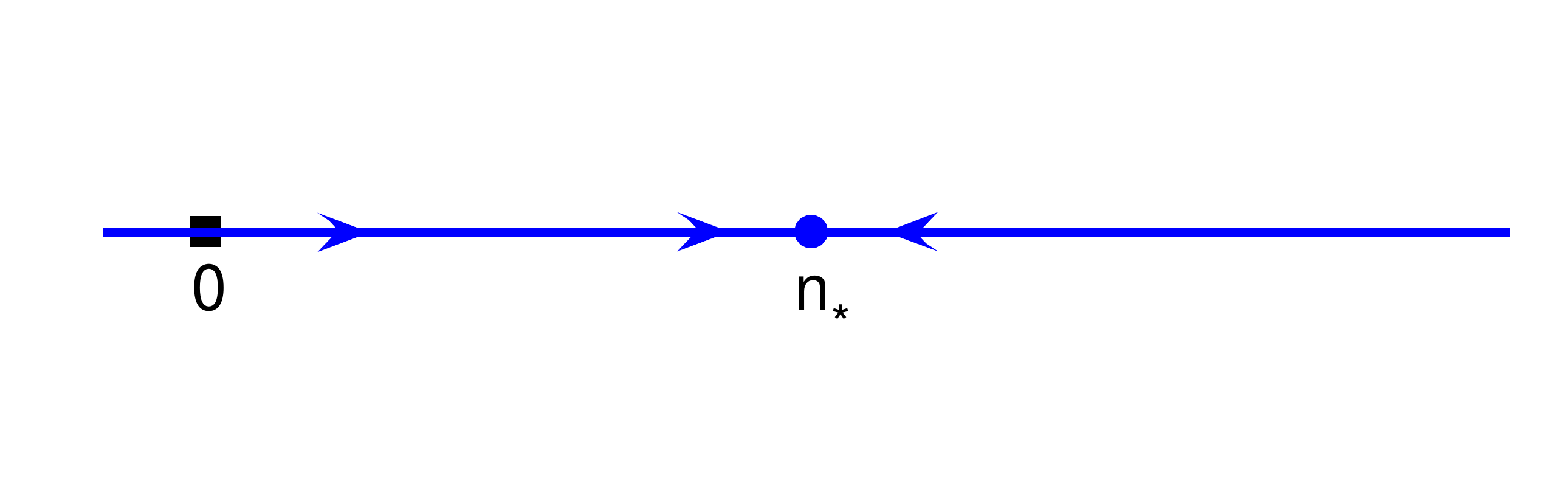}
\includegraphics[width=6.5cm,height=2.5cm,clip=]{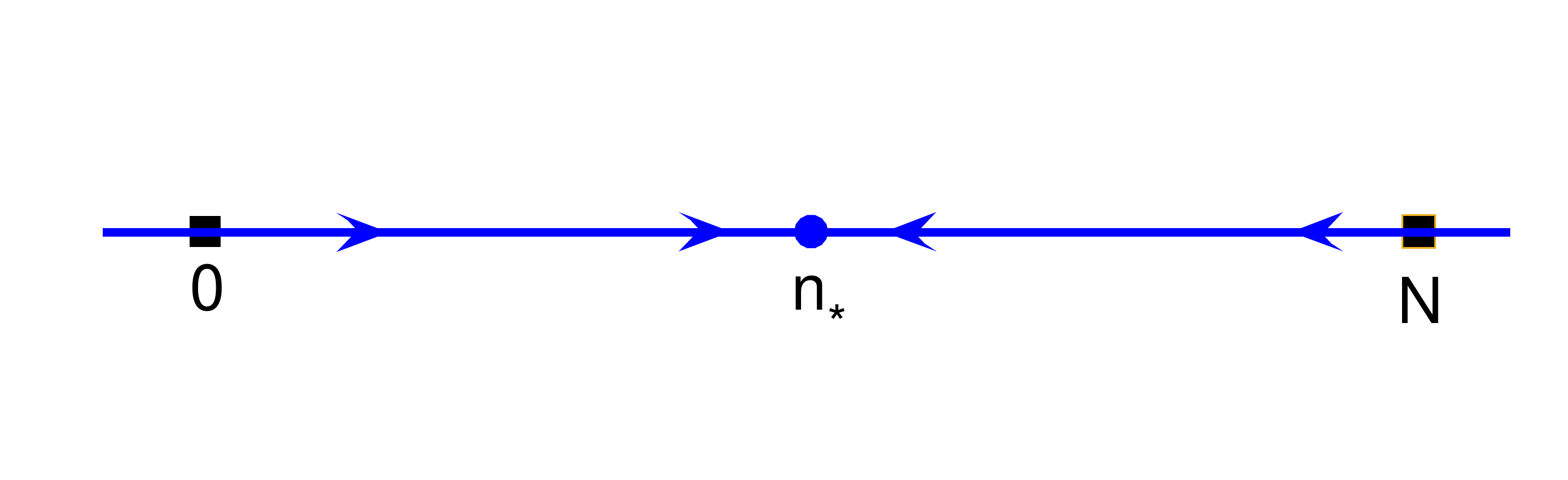}
\caption{Scenario A. The full circle denotes an attracting fixed point of the deterministic rate equation, the full square denotes an absorbing state of the stochastic process. Left panel: extinction. Right panel: fixation.} \label{phasespaceA}
\end{figure}

Typically, there are two escape routes, or escape scenarios, from a metastable state of a single population~\cite{AM2010}. Let us denote the attracting fixed point, in the vicinity of which the population resides, by $n_*$. In escape scenario A there exist one or two adjacent repelling fixed points
of the deterministic rate equation that are also absorbing states of the stochastic dynamics. A single repelling point $n=n_u=0$ corresponds to extinction; two repelling points correspond to fixation, see Fig.~\ref{phasespaceA}.

\begin{figure}[ht]
\includegraphics[width=6.5cm,height=2.5cm,clip=]{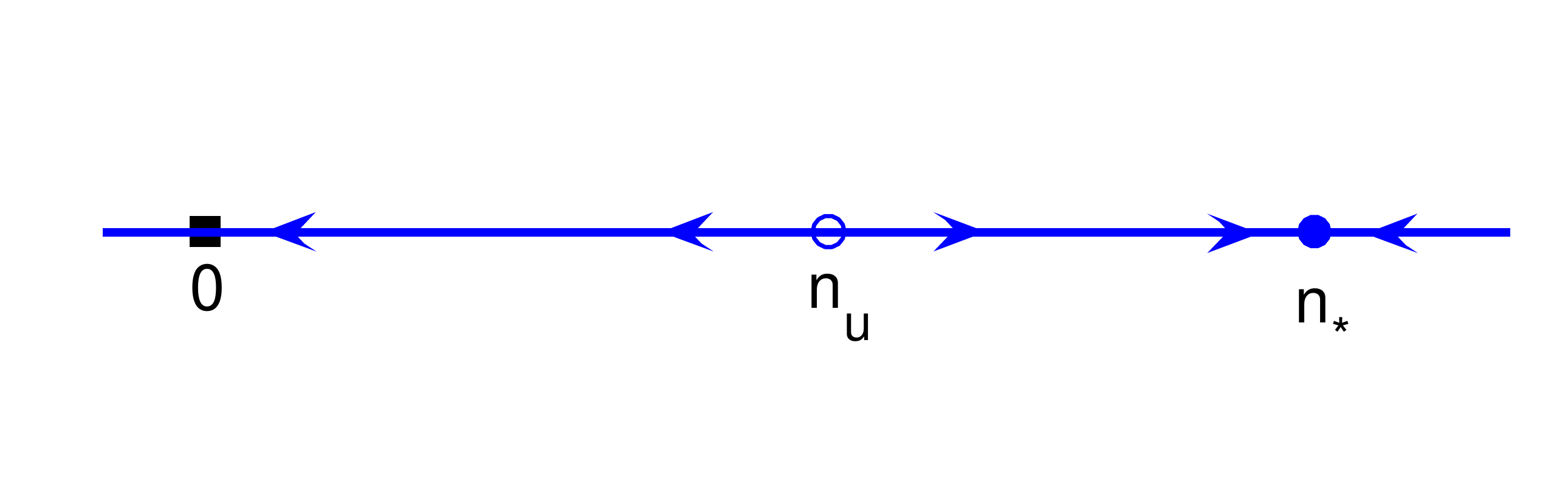}
\includegraphics[width=6.5cm,height=2.5cm,clip=]{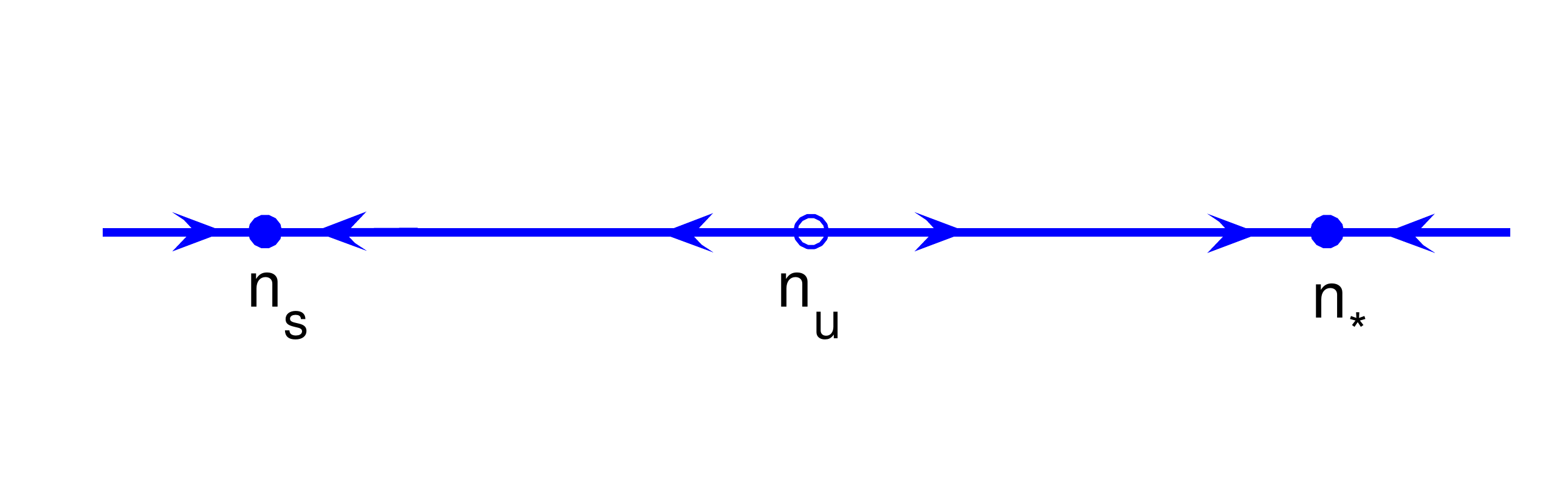}
\caption{Scenario B. The full circle denotes an attracting fixed point, the full square denotes an absorbing state, and the empty circle denotes a repelling fixed point. Left panel: extinction. Right panel: switching. } \label{phasespaceB}
\end{figure}

In escape scenario B, in addition to the repelling point $n_u>0$, which is now \textit{non}-absorbing, there exists a second attracting fixed point of the deterministic rate equation, $n=n_s$, see Fig.~\ref{phasespaceB}. If $n_s=0$ and absorbing, this is extinction. If $n_s>0$ and non-absorbing, this is switching from $n_*$ to $n_s$. Each of the scenarios A and B can be dealt with by the real-space WKB approach, while the momentum-space approach is limited to scenario A \cite{AMS2010}.

Let us derive a general approximate expression for the mean time to escape. After a short relaxation time $t_r$, determined by the deterministic rate equation,  the population gets established as a long-lived metastable distribution peaked near $n=n_*$. The metastable distribution very slowly decays in time because of the probability ``leakage" into the absorbing state(s). The long-time decay is characterized by the lowest positive eigenvalue
$1/\tau$ of the master equation (\ref{master0}), so that $\tau\gg t_r$. That is, at times $t\gg t_r$, one has~\cite{DMR1994,KS2007,MS2008,EK2009,AM2010}
\begin{equation}\label{qsdintro}
P_{n>0}(t\gg t_r)\simeq \pi_n e^{-t/\tau},\;\;P_{0}(t\gg t_r)\simeq
1-e^{-t/\tau}.
\end{equation}
Here we have assumed for concreteness a single absorbing state at $n=0$, corresponding to extinction. The case of fixation will be dealt with shortly.  The function $\pi_n$ ($n=1,2, \dots $) describes the quasi-stationary distribution (QSD) of the population. For metastable populations the decay time $\tau$ is an accurate approximation to the mean
time to extinction (MTE) or fixation.
Using Eq.~(\ref{qsdintro}), one arrives at an eigenvalue problem for the QSD $\pi_n$, $n=1,2\dots$:
\begin{equation}\label{qsdmaster}
\sum_{r} \left[W_r(n- r)\pi_{n-r} -W_r(n)\pi_n\right] = -\pi_n/\tau.
\end{equation}
At large $N$ the eigenvalue  $1/\tau$ is exponentially small, see below. Therefore, the term in the right hand
side of Eq.~(\ref{qsdmaster})
can be neglected,
and we arrive at a quasi-stationary equation
\begin{equation}\label{qsdmaster1}
\sum_{r} \left[W_r(n- r)\pi_{n-r} -W_r(n)\pi_n\right] =0, \;\;\;n=1,2, \dots \,.
\end{equation}
Once the QSD is known, we can plug Eq.~(\ref{qsdintro}) into Eq.~(\ref{master0}) for $n=0$, and obtain the MTE from the relation
\begin{equation}\label{E}
1/\tau = \sum_{r>0} W_r(r) \pi_{r}.
\end{equation}
Equation~(\ref{E}) holds for both scenarios A and B. The QSD itself, however, is scenario-dependent, and this ultimately determines the MTE $\tau$. In the following we will derive the expressions for the QSD for each of the two escape scenarios.

\subsection{Extinction and Fixation: Scenario A}
In Sec.~\ref{extinctionA} we will determine the QSD in the case of extinction via scenario A. Section~\ref{fixationA} will deal with fixation.

\subsubsection{Extinction}
\label{extinctionA}

We will illustrate some general results, that we will present shortly, on a typical example: a variant of the stochastic Verhulst model~\cite{Na2001} with the (rescaled) birth and death rates
\begin{equation}\label{ratesVer}
W_{+1}(n)=B n\;,\;\;\;W_{-1}(n)=n+B\frac{n^2}{N},\quad N\gg 1.
\end{equation}
The quadratic corrections account for competition for resources. For this model, the master equation~(\ref{master0}) is
\begin{eqnarray}\label{master}
\frac{d P_n(t)}{dt} &=& W_{+1}(n-1)P_{n-1}(t)+W_{-1}(n+1)P_{n+1}(t)-[W_{+1}(n)+W_{-1}(n)]P_n(t),\nonumber\\
\frac{d P_0(t)}{dt} &=& W_{-1}(1)P_1(t) \,.
\end{eqnarray}
The deterministic rate equation can be obtained by multiplying both sides of Eq.~(\ref{master}) by $n$ and summing over all $n$'s. Using the mean-field assumption $\langle n^2\rangle=\langle n\rangle^2$, this procedure yields \cite{Gardiner,vanKampen}
\begin{equation}
\dot{\bar{n}}=(B-1)\bar{n}-B\frac{\bar{n}^2}{N},
\label{Verhulstdeter}
\end{equation}
where $\bar{n}$ denotes the mean population size. This equation (that also emerges from the WKB formalism) has two fixed points:  at $\bar{n}=n_u=0$ and at $\bar{n}=n_*=N(B-1)/B$. We will assume that $B>1$, so these fixed points are repelling and attracting, respectively. As a result,  for $B>1$ the population size will flow to the attracting fixed point at $n_*$ when starting from any nonzero value of $n$ at $t=0$. The metastable population distribution is peaked at about $n=n_*$, whereas the repelling fixed point $n=0$ is an absorbing state of the stochastic process.

Let us determine the QSD by using the real-space WKB method.  To this end we express the QSD in terms of the rescaled population size $q=n/N$ and look for the solution of Eq.~(\ref{qsdmaster1}) at  $n\gg 1$ by making the WKB ansatz~\cite{DMR1994,KS2007,MS2008,EK2009,AM2010}
\begin{equation}\label{fastmode}
\pi_n\equiv \pi(q) \simeq A e^{-NS(q) - S_1(q)-\dots},
\end{equation}
where $S(q)$ and $S_1(q)$ are assumed to be $O(1)$, and a constant prefactor $A$ is introduced for convenience. The leading-order calculations, where one neglects $S_1$, are straightforward as we show shortly. The sub-leading order calculations demand some effort. This is because the subleading WKB solution for the QSD has to be complemented by a recursive solution of the master equation in the vicinity of the absorbing state. In the case of extinction the subleading WKB solution has to be matched with a recursive solution valid at $n=O(1)$~\cite{KS2007,AM2010}, see below.

\begin{figure}
\includegraphics[scale=0.2]{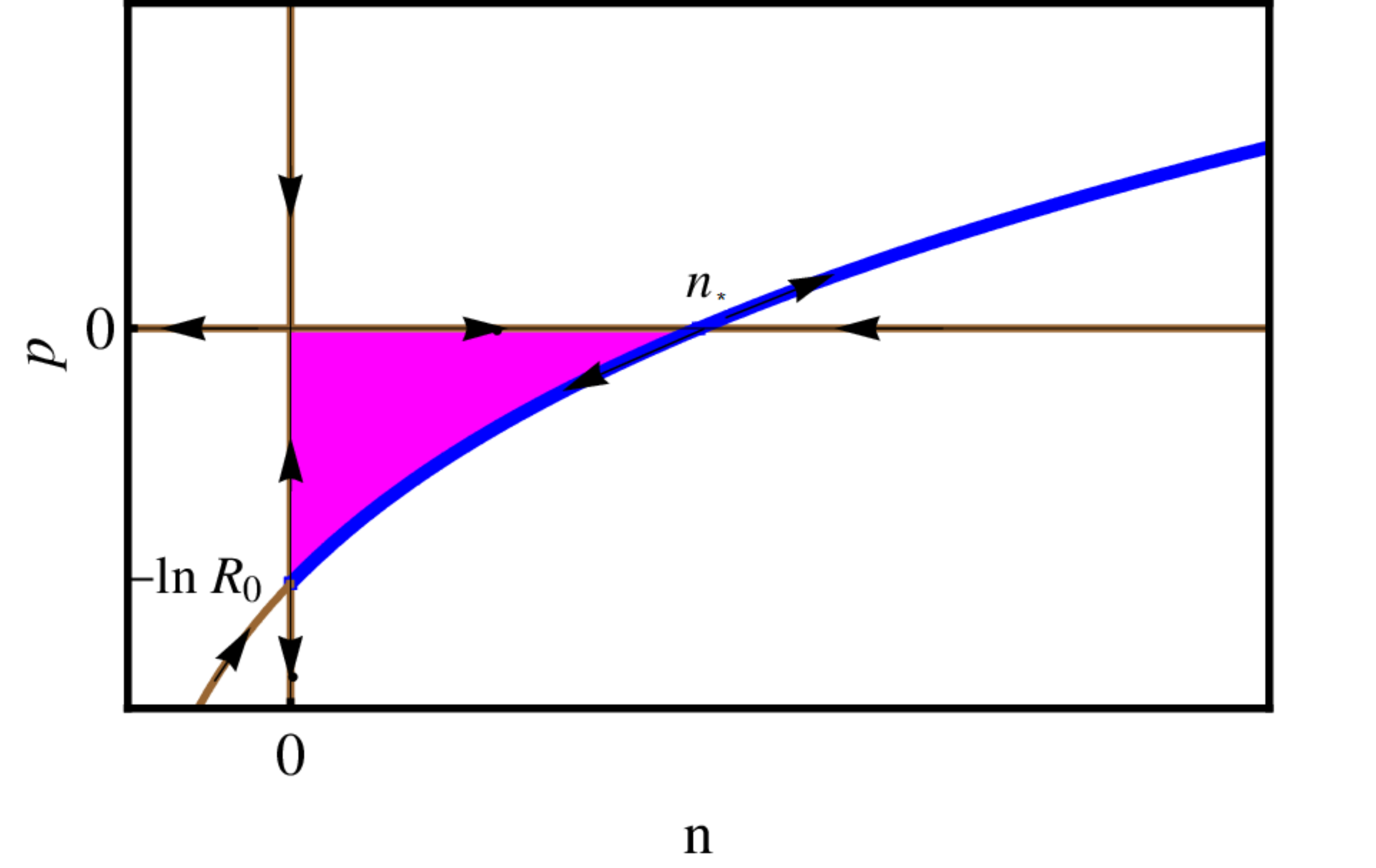}
\includegraphics[scale=0.222]{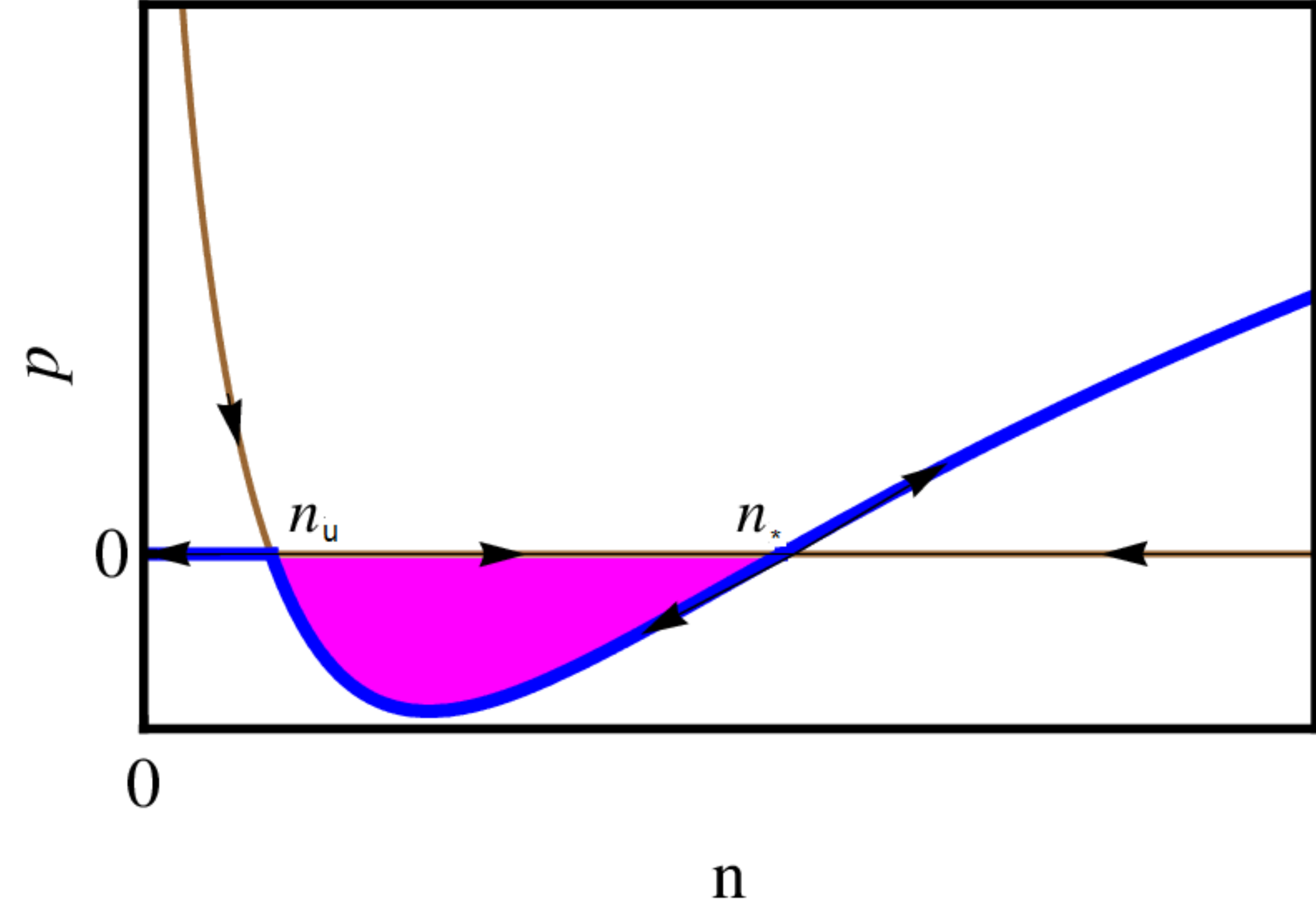}
\caption{Extinction scenarios A and B \cite{AM2010}.
Left panel: Scenario A (no Allee effect): the repelling fixed point $\bar{n}=0$ is an absorbing state
of the stochastic process. The stochastic
extinction occurs via a large fluctuation which brings the established population
from a vicinity of the attracting fixed point $\bar{n}=n_*$ directly to the
absorbing state $n=0$.  Right panel: Scenario B (Allee effect): the absorbing state at $n=0$ is an attracting
fixed point. The stochastic
extinction occurs via a large fluctuation which brings the established population
from a vicinity of the \textit{next} attracting fixed point $\bar{n}=n_*$  to a vicinity of the
repelling fixed point $\bar{n}=n_u$. From there the population size flows ``downhill"
to the absorbing state $n=0$ almost deterministically. The thick curves show the corresponding optimal paths
$p=p(q)$.}
\label{scAB}
\end{figure}

\begin{figure}[ht]
\includegraphics[scale=0.3]{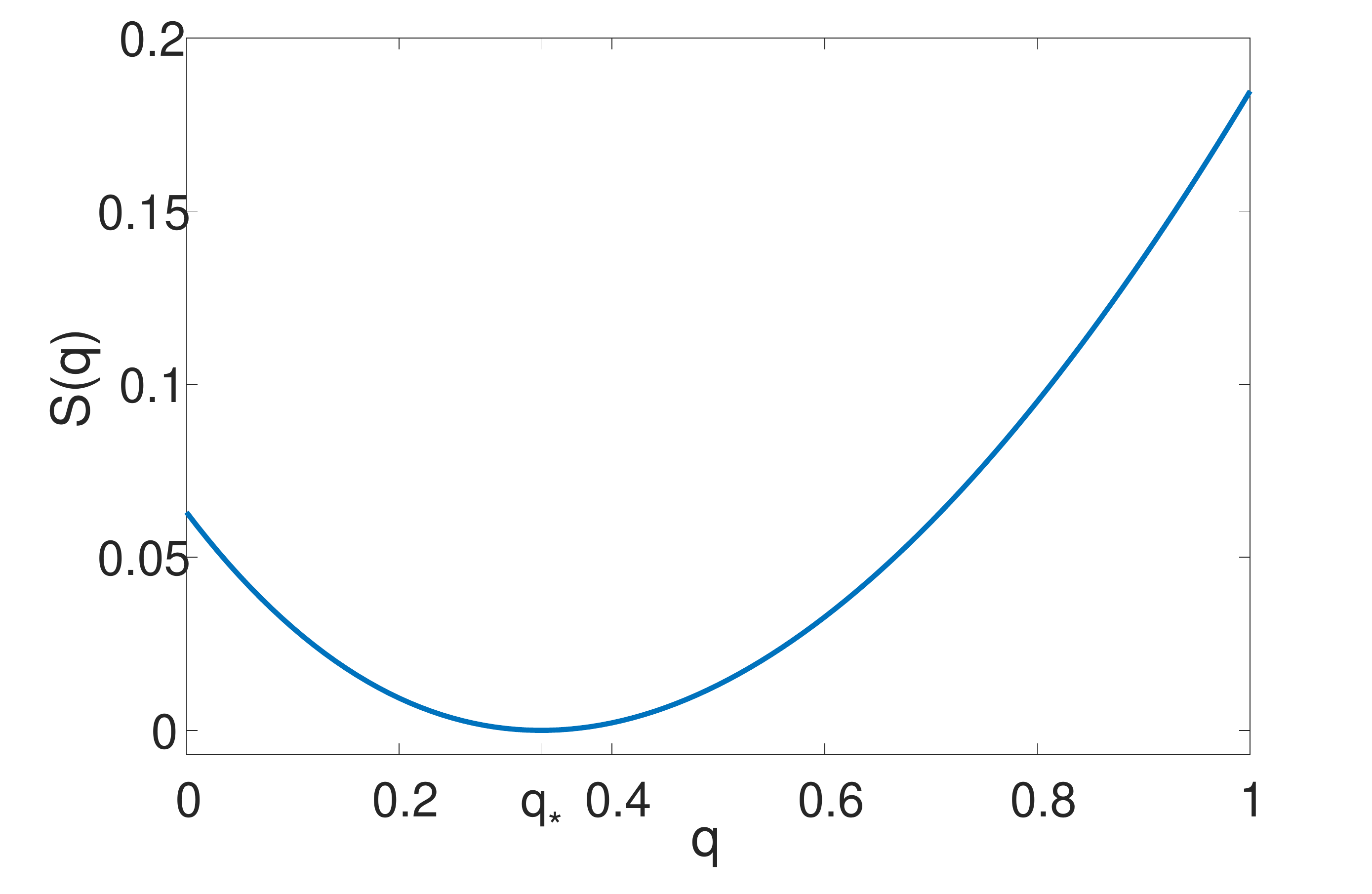}
\caption{The action function $S(q)$ for the Verhulst model, given by Eq.~(\ref{SVerhulst}), for $B=1.5$. Here $q_*=1-1/B=1/3$.} \label{actionScenarioA}
\end{figure}

We plug the WKB ansatz~(\ref{fastmode}) into Eq.~(\ref{qsdmaster1})  and, assuming $n\gg 1$,
 Taylor-expand the functions of $q\pm 1/N$ around $q$. In the leading and subleading orders we obtain
\begin{equation}\label{SS1}
S(q)=\int^q p_a(q)dq,\quad S_1(q)=\frac{1}{2}\ln[w_+(q)w_-(q)].
\end{equation}
Here $w_+(q)=W_+(n)/N=Bq$ and $w_-(q)=W_-(n)/N=q(1+Bq)$ are the rescaled transition rates, see Eq.~(\ref{ratesVer}). Furthermore, $p_a(q)=\ln[w_-(q)/w_+(q)]=\ln[(1+Bq)/B]$ is the so called fast-mode WKB solution \cite{AM2010}, see Fig.~\ref{scAB}. This solution describes the optimal path: the nontrivial zero-energy trajectory of the effective classical Hamiltonian
\begin{equation}\label{hamil}
H(q,p)=w_+(q)\left(e^{p}-1\right)+w_-(q)\left(e^{-p}-1\right) ,
\end{equation}
where $p=dS/dq$ is the momentum.  The constant $A$ in Eq.~(\ref{fastmode}) can be found by normalizing the QSD in the vicinity of the attracting fixed point $q_*=n_*/N=1-1/B$, where the QSD can be approximated by a Gaussian. This yields
\begin{equation}\label{EqA}
A=\sqrt{\frac{S^{\prime\prime}(q_*)}{2\pi N}}e^{NS(q_*)+S_1(q_*)}.
\end{equation}
[Note that there is an additional,  slow-mode WKB solution \cite{AM2010} for which  $S(q)=0$ and $S_1(q)=\ln[w_+(q)-w_-(q)]$.  In scenario A this solution gives a negligible contribution in the entire WKB region, so it should be discarded.  In scenario B, however, this solution plays an important role, see below.] Plugging Eqs.~(\ref{SS1}) and (\ref{EqA}) into Eq.~(\ref{fastmode}), we obtain the WKB approximation for the QSD for the Verhulst model \cite{AM2010}:
\begin{equation}\label{QSDVer}
\pi_n=\pi(q)=\frac{B-1}{\sqrt{2\pi N B q^2(1+Bq)}}\,e^{-N S(q)},
\end{equation}
where
\begin{equation}\label{SVerhulst}
S(q) = 1-\frac{1}{B}-q+\left(q+\frac{1}{B}\right)\ln\left(q+\frac{1}{B}\right).
\end{equation}
The action function $S(q)$ is depicted in Fig.~\ref{actionScenarioA}.

\begin{figure}[ht]
\includegraphics[scale=0.35]{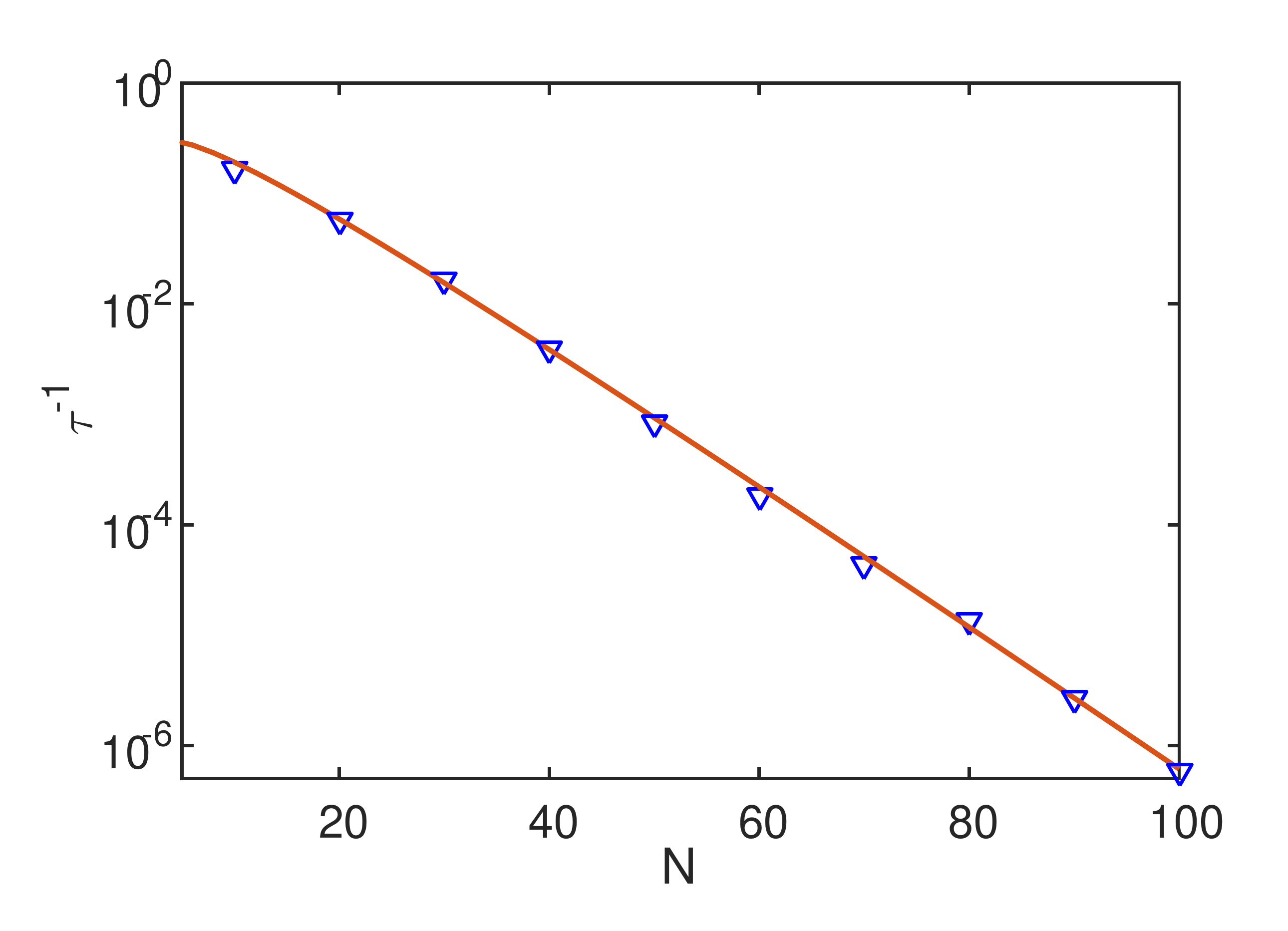}
\caption{The mean extinction rate $\tau^{-1}$ versus $N$ on a semi-log scale for $B=2$. Line: theoretical result~(\ref{tausingleA}), symbols: results of Monte-Carlo simulations.} \label{extscenarioA}
\end{figure}

The approximation (\ref{QSDVer}) and (\ref{SVerhulst}) is invalid at $n=O(1)$, or $q=O(1/N)$. Here one has to solve the quasi-stationary master equation recursively. Fortunately,  one can neglect the nonlinear terms in the transition rates \cite{KS2007,AM2010}. The recursive solution is
\begin{equation}\label{recursive}
\pi_n=\frac{(B^n-1)}{\tau(B-1)n}\,,
\end{equation}
where $\tau$ is the MTE we are after.  Matching this recursive solution with the WKB solution~(\ref{QSDVer}) in their joint region of validity $1\ll n\ll \sqrt{N}$ and using Eqs.~(\ref{E}) and (\ref{SVerhulst}), we find~\cite{AM2010}
\begin{equation}
\tau =\sqrt{\frac{2\pi B}{N}}\frac{1}{(B-1)^2}\exp\left[\frac{N}{B}\left(B-1-\ln B\right)\right] \equiv \sqrt{\frac{2\pi B}{N}}\frac{1}{(B-1)^2}e^{N S(0)}.
\label{tausingleA}
\end{equation}
As expected, $\tau$ is exponentially large in $N$. As one can clearly see now, the leading-order WKB action $S(q)$ plays the role of (and sometimes called) the non-equilibrium potential. It describes an effective exponential barrier to extinction. The pre-exponential factor is important in this case, as it includes the large parameter $N^{-1/2}$. The presence of a large parameter in the pre-exponent is a typical feature of the extinction scenario A. Figure~\ref{extscenarioA} shows a comparison of the mean extinction rate $1/\tau$ from Eq.~(\ref{tausingleA}) with the mean extinction rate measured in Monte-Carlo simulations at different $N\gg 1$.
Very good agreement is observed.

Close to the transcritical bifurcation of the deterministic rate equation, $\delta \equiv B-1\ll 1$, but still at
$N\delta^2 \gg 1$, Eq.~(\ref{tausingleA}) simplifies to
\begin{equation}
\tau\simeq \sqrt{\frac{2\pi}{N}}\, \frac{e^{N \delta^2/2}}{\delta^2} .
\label{tausingleAclose}
\end{equation}
This close-to-the-bifurcation result is universal: it applies to a whole class of population models
that do not include an Allee effect \cite{AM2010}.

Now, continuing using the Verhulst model as an example, we will show that the often used van-Kampen system size expansion
(also called the diffusion approximation) generally yields an error in the MTE that is exponentially large in $N$ \cite{Gaveau,EK2004,Doering,KS2007,AM2006a,AM2007}. The van-Kampen system size expansion approximates the exact master equation (\ref{master}) by a Fokker-Planck equation. The Fokker-Planck  equation can be obtained from Eq.~(\ref{master}) in the following way. One defines $j_{\pm}(n)=W_{\pm 1}(n)P_n(t)$ and Taylor expands $j_{\pm}(n\mp 1)$ around $n$ up to the second order. Then, introducing the rescaled population size $q=n/N$, one arrives at the Fokker-Planck equation
\begin{equation}
\frac{\partial P(q,t)}{\partial t}=-\frac{\partial}{\partial q}\left[v(q)P(q,t)\right]+\frac{1}{N}\frac{\partial^2}{\partial q^2}\left[d(q)P(q,t)\right].
\end{equation}
Here
$$
v(q)=w_{+}(q)-w_{-}(q)=q(B-1-Bq)
$$
is the deterministic drift velocity, and
$$
d(q)=w_{+}(q)+w_{-}(q)=q(B+1+Bq)
$$
is the diffusivity that describes an effective multiplicative noise. Now we
continue exploiting the large parameter $N\gg 1$. We  make the ansatz  $P(q,t)=\pi(q)e^{-t/\tau}$
and use the WKB approximation, see Eq.~(\ref{fastmode}), to analyze the quasi-stationary equation for $\pi(q)$. This procedure yields, in the leading order in $N$, a Hamilton-Jacobi equation $H(q,dS/dq)=0$ with the Hamiltonian
\begin{equation}
H(q,p)=p\left[v(q)+\frac{d(q)p}{2}\right].
\end{equation}
The resulting action function is
\begin{equation}
S^{FP}(q)=-\int_{q_*}^q \frac{2v(q')}{d(q')}dq'=2\left[q-1+\frac{1}{B}-2\ln\frac{(1+B+Bq)}{2B}\right].
\end{equation}
This expression (which is then multiplied by a large $N$ inside the exponent of the MTE) differs from the asymptotically exact expression~(\ref{SVerhulst}) thus invalidating the Fokker-Planck equation as a controlled approximation for the purpose
of dealing with large deviations.

The van-Kampen system size expansion may become accurate close to bifurcation points of the deterministic model. In our Verhulst example, the bifurcation is at $B=1$. At $B-1\ll 1$, the stable fixed point is $q_*\simeq B-1\ll 1$. Therefore we can expand $v(q)$ and $d(q)$ in the vicinity of $q=0$, which yields
\begin{equation}\label{HFPbif}
H\simeq pq\left[v'(0)+\frac{v''(0)}{2}q+\frac{d'(0)}{2}p\right]\simeq pq(B-1-q+p).
\end{equation}
As one can check, this Hamiltonian coincides with the asymptotically exact WKB Hamiltonian (\ref{hamil}) close to the bifurcation at $B=1$.  Naturally, the zero-energy activation trajectory,
\begin{equation}\label{activFPbif}
p(q)=-\frac{2v'(0)+qv''(0)}{d'(0)}=q+1-B,
\end{equation}
is also the same as the one following from the Hamiltonian (\ref{hamil}), and $S^{FP}(0)$ is given by $(1/2)(B-1)^2$, which coincides with the exponential term in Eq.~(\ref{tausingleAclose}). Performing the subleading-order calculations, one arrives at a pre-exponent which also coincides with that of Eq.~(\ref{tausingleAclose}) (see Ref.~\cite{AM2010} for details).

We now present, using the same example of the Verhulst model,  the momentum-space WKB method.  Multiplying both sides of the master equation~(\ref{master}) by $p^n$ and summing over all $p$ one arrives at an exact evolution equation for the probability generating function $G(p,t)$ from Eq.~(\ref{genfun}):
\begin{equation}\label{genPDE}
\frac{\partial G}{\partial t}=(p-1)\left\{\left[Bp-\left(1+\frac{B}{N}\right)\right]\frac{\partial G}{\partial p}-\frac{Bp}{N}\frac{\partial^2 G}{\partial p^2}\right\}.
\end{equation}
One way of using the WKB approximation would be to look for the time-dependent solution as $G(p,t)=\exp[-S(p,t)]$ and neglecting, in the leading  WKB order, the second derivative $\partial_p^2 S$~\cite{EK2004}.
This would suffice if we limit ourselves to the leading WKB order. For more accurate calculations, that yield pre-exponents, it is convenient to first exploit the known spectral properties of the metastability problem~\cite{AM2006,AM2006a,AM2007,AMS2010}. Indeed, the metastable solution, corresponding to the established population whose distribution is sharply peaked around $n_*=N(1-1/B)$, is described by the lowest excited eigenmode $\phi(p)$ of the differential operator in the right hand side of Eq.~(\ref{genPDE}): the mode that decays with the rate $1/\tau$,
\begin{equation}
G(p,t)=1-\phi(p)e^{-t/\tau}.
\label{WKBp}
\end{equation}
At $t\to\infty$ one obtains $G(p,t)=1$, corresponding to population extinction, $P_n(t\to \infty)=\delta_{n,0}$.  We plug the ansatz (\ref{WKBp}) into Eq.~(\ref{genPDE}) and arrive at an ordinary differential equation (ODE) for $\phi(p)$:
\begin{equation}\label{phiODE}
(p-1)\left\{\left[Bp-\left(1+\frac{B}{N}\right)\right]\phi'(p)-\frac{Bp}{N}\phi''(p)\right\}=\frac{\phi}{\tau},
\end{equation}
where the prime denotes the derivative with respect to the argument.
The probability conservation yields $G(1,t)=1$ [see Eq.~(\ref{genfun})], therefore $\phi(1)=0$. Analyticity of $\phi(p)$ in the vicinity of $p=0$ brings about an additional condition:
$$
\left(1+\frac{B}{N}\right)\phi'(0)-\frac{\phi(0)}{\tau} =0
$$
which, in view of the expected exponential smallness of $1/\tau$, can be replaced by $\phi'(0)=0$. The ``self-generated" boundary conditions $\phi'(0)=\phi(1)=0$ define the natural interval  $0\le p\leq 1$ on which the boundary value problem for Eq.~(\ref{phiODE}) should be considered~\cite{AM2006,AM2006a,AM2007,AMS2010}.  In order to find the MTE, we can consider two separate regions and match the asymptotic solutions in their joint region of validity. In the bulk -- that is, not too close to $p=1$ -- we can assume that $\phi(p)$ is almost constant, and look for a perturbative solution
\begin{equation}
\phi(p)=1+\delta\phi (p),\quad \delta \phi(p)\ll 1.
\label{linearize}
\end{equation}
Plugging this ansatz into Eq.~(\ref{phiODE}), and defining $u(p)=\phi'(p)$, the perturbative solution in the bulk can be written as
\begin{equation}\label{bulkU}
u^{b}(p)=\frac{N}{B\tau p}e^{-NS(p)}\int_{0}^p \frac{e^{NS(p)}}{p-1}dp,
\end{equation}
where $S(p)=(1/B)\ln p -p+1$, and the boundary condition $u(0)=0$ is satisfied.

\begin{figure}[ht]
\includegraphics[scale=0.35]{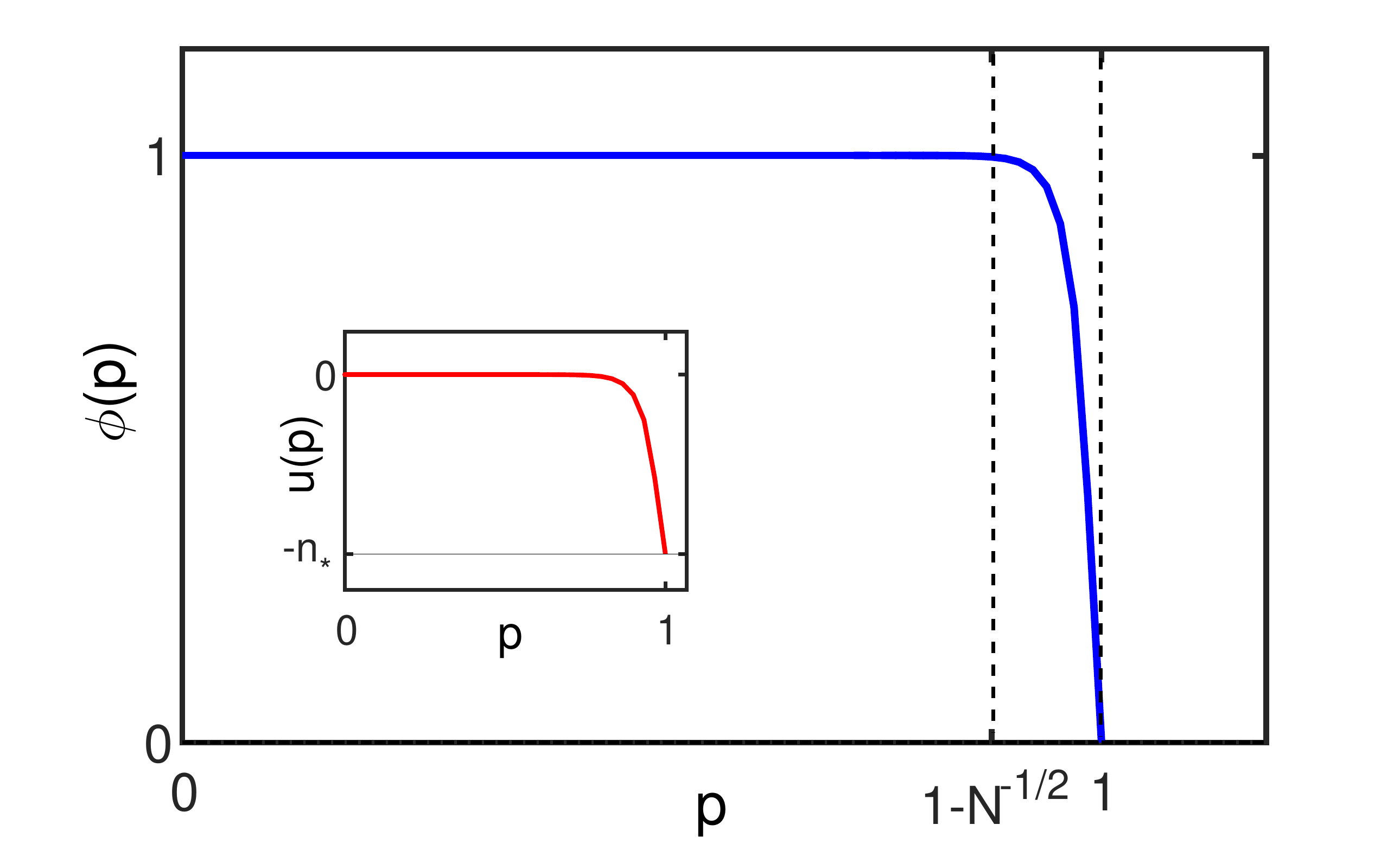}
\caption{An illustration of the function $\phi(p)$ in the region of interest $p\in[0,1]$. The boundary layer is schematically denoted by the region bounded by the vertical dashed lines. The inset shows $u(p)=\phi'(p)$.} \label{momentum-phi}
\end{figure}

In the boundary-layer region $1-p\ll 1$ we notice that $\phi(p)$ rapidly falls from a value close to unity to zero over a narrow region of size $O(N^{-1/2})\ll 1$. In the boundary layer, the approximate ODE reads
\begin{equation}
\left[Bp-\left(1+\frac{B}{N}\right)\right]u(p)-\frac{Bp}{N}u'(p)=0.
\label{BLeq}
\end{equation}
Here we look for a WKB-type solution $u(p)=a(p)e^{-NS(p)}$ where $S(p)$ is the action in the momentum space, and $a(p)$ is an amplitude. Plugging this ansatz in Eq.~(\ref{BLeq}), we demand that the terms cancel each other separately in the zeroth and first order in $1/N$.  What is the boundary condition for $u(p)$ at $p=1$? At times $t\ll \tau$ we have, on the one hand, $\partial_p G(p=1,t)=n_*=N(B-1)/B$ and, on the other hand, $\partial_p G(p=1,t)=-u(1)$, so $u(1)=-n_*$. As a result, the boundary-layer solution is
\begin{equation}\label{blU}
u^{bl}(p)=-\frac{n_*}{p}e^{-NS(p)}.
\end{equation}
The solutions to $\phi(p)$ and $u(p)$ are illustrated in Fig.~\ref{momentum-phi}.

In order to find the MTE $\tau$, one can match the bulk solution (\ref{bulkU}) and the boundary-layer  solution (\ref{blU}) in their joint region of validity $N^{-1/2}\ll 1-p\ll 1$~\cite{AM2006,AM2006a,AM2007,AMS2010}. For $1-p\gg N^{-1/2}$, the integral in the bulk solution~(\ref{bulkU}) can be evaluated by a saddle-point approximation, where the saddle point is found at $p=1/B$. This gives
\begin{equation}\label{bulkUint}
\int_{0}^p \frac{e^{NS(p)}}{p-1}dp\simeq -\sqrt{\frac{2\pi B}{N}}\frac{e^{-NS(1/B)}}{B-1}.
\end{equation}
Plugging this into Eq.~(\ref{bulkU}) and matching with Eq.~(\ref{blU}) we again arrive at Eq.~(\ref{tausingleA}).

Notice that, although the boundary-layer solution (\ref{blU}) has a recognizable WKB form, the actual use of the WKB approximation here was unnecessary. The only approximation  used was to
linearize  Eq.~(\ref{phiODE}), that is to replace $\phi(p)$ by $1$ in the right hand side and neglect the exponentially small term $\tau^{-1}$. This is a typical situation for the elemental transitions leading to a second-order ODE for $\phi(p)$, like Eq.~(\ref{phiODE}). Yet, already for third-order ODEs (such as the one that appears, for example, when one  of the transitions is triple annihilation  $3A\to \emptyset$), the WKB approximation becomes indispensable \cite{AMS2010}.

\subsubsection{Fixation}
\label{fixationA}

Consider a population consisting of $n$ mutants of type $A$ and $N-n$ wild-type individuals of type $B$, which can be genes, cells or even animals. What is the probability that the mutants take over (fixate) the entire population, causing the extinction of the wild-type individuals? This question can be addressed in the framework of evolutionary game theory (EGT) which allows to describe how successful strategies spread by imitation or reproduction~\cite{No2006}.

In its simplest form, the EGT framework includes two reactions $A+B\to A+A$ at a rate $f_A$, and $A+B\to B+B$, at a rate $f_B$. Here, an individual chosen proportionally to its fitness (reproduction potential) produces an identical offspring which replaces a randomly chosen individual, and $f_A$ and $f_B$ denote the fitness of types $A$ and $B$, respectively. If, in addition, the fitness of $A$ and $B$ depends on the current number of mutants, the deterministic rate equation -- the replicator dynamics~\cite{No2006} -- describing the mean number of mutants, can be written as
\begin{equation}\label{RE}
\dot{n}=n(N-n)[f_A(n)-f_B(n)].
\end{equation}
The simplest form of $f_A$ and $f_B$, which guarantees an intermediate coexistence state, is when $f_A(n)-f_B(n)\sim n_*-n$, corresponding to the linear Moran model~\cite{Mo1964}. This yields an attracting point at $n=n_*$ and two repelling fixed points: at $n=0$ and $n=N$.

The stochastic version of this model is described by the master equation
\begin{equation}
\frac{d P_n(t)}{dt} = W_{+1}(n-1)P_{n-1}(t)+W_{-1}(n+1)P_{n+1}(t)-[W_{+1}(n)+W_{-1}(n)]P_n(t),
\end{equation}
where $W_{+1}(n)=n(N-n)f_A(n)$ and $W_{-1}(n)=n(N-n)f_B(n)$~\cite{No2006,MobA2010,AMob2010,BTG2012,BTG2012}. The stochastic dynamics of this problem resemble those of the Verhulst model. After a short relaxation time $t_r$, the system enters a long-lived metastable coexistence state centered about $n_*$. Here, however, the system can escape to either the absorbing state $n=0$ corresponding to \textit{extinction} of the mutants, or to $n=N$ corresponding to \textit{fixation} of the mutants. As a result, the dynamics of the probability distribution function at $t\gg t_r$ satisfy the ansatz~\cite{MobA2010,AMob2010,AM2011,BTG2012}
\begin{equation}
P_{1\leq n\leq N-1}=\pi_n e^{-t/\tau}, \;\;\;P_0(t)=\varphi(1-e^{-t/\tau}),\;\;\;P_N(t)=(1-\varphi)(1-e^{-t/\tau}),
\end{equation}
where the mean time to fixation (of either the wild type or the mutants), $\tau$, and the probability of mutant extinction, $\varphi$, satisfy
\begin{equation}
\varphi=W_{-1}(1)\pi_1\tau,\;\;\;\tau=[W_{-1}\pi_1+W_{+1}(N-1)\pi_{N-1}]^{-1}.
\end{equation}

The solution of this problem can be found using the real-space WKB approach, and follows the same lines as the solution presented above for extinction. As before, the WKB solution breaks down close to the boundaries $n=0$ and $n=N$, where one has to use recursive solutions by linearizing the reaction rates close to $n=0$, and $n=N$, respectively; see Refs.~\cite{MobA2010,AMob2010,AM2011}.

The scenario we have presented above contains wild-type individuals and those having a single mutation. However, there exist more complicated scenarios of fixation with multiple mutations, when \textit{e.g.}, a beneficial mutation (with a higher fitness than the wild type's) is obtained from a wild-type species by going through a detrimental mutation (with a lower fitness than the wild type's). Such problems, which include in their simplest form a two-locus genotype space, and in which the negative effects of
two single mutations are overcompensated by a higher-fitness double mutant, can also be dealt with by using the real-space WKB approach, see \textit{e.g.}, Ref.~\cite{AFKS2011}.

\subsection{Extinction and Switching: Scenario B}\label{Alleeeffect}
We will now consider escape scenario B, see Fig.~\ref{phasespaceB}. In Sec. \ref{extinctionB} we will study extinction ($n=n_s=0$ is absorbing), whereas Sec. \ref{switchingB} will deal with switching between two metastable states ($n=n_s$ is non-absorbing).

\subsubsection{Extinction}
\label{extinctionB}
Following Ref. \cite{AM2010}, we use the following set of
reactions: binary reproduction $2A\stackrel{\lambda}{\rightarrow} 3A$, the reverse
transition $3A\stackrel{\sigma}{\rightarrow} 2A$, and  linear decay $A\stackrel{\mu}{\rightarrow} \emptyset$. Here
\begin{eqnarray}
\hspace{-4mm}W_1=\frac{\lambda n(n-1)}{2}\,,\;W_{-1}=\mu n +\frac{\sigma n(n-1)(n-2)}{6}.\label{ratesAllee}
\end{eqnarray}
The deterministic rate equation is \cite{AM2010}
\begin{equation}
\dot{\bar{n}}=\frac{\lambda}{2} n^2-\mu n-\frac{\sigma}{6}n^3.
\label{det2}
\end{equation}
The fixed point at $n=0$ is now attracting. Furthermore, denoting by $\delta^2=1-8\sigma \mu/(3 \lambda^2)>0$, and $N=3\lambda/(2 \sigma)$, Eq.~(\ref{det2}) has two additional fixed points $n_u=N(1-\delta)$ and $n_*=N(1+\delta)$. An established stochastic population resides in the vicinity of the fixed point $n_*$. This simple model accounts for the Allee effect that refers to a variety of processes that reduce the per-capita growth
rate at small population densities. The Allee effect has been long known in ecology and population biology  \cite{Alleeetal}.
Very recently, its importance has been also appreciated in mathematical cancer biology \cite{Alleetumor}.

\begin{figure}[ht]
\includegraphics[scale=0.35]{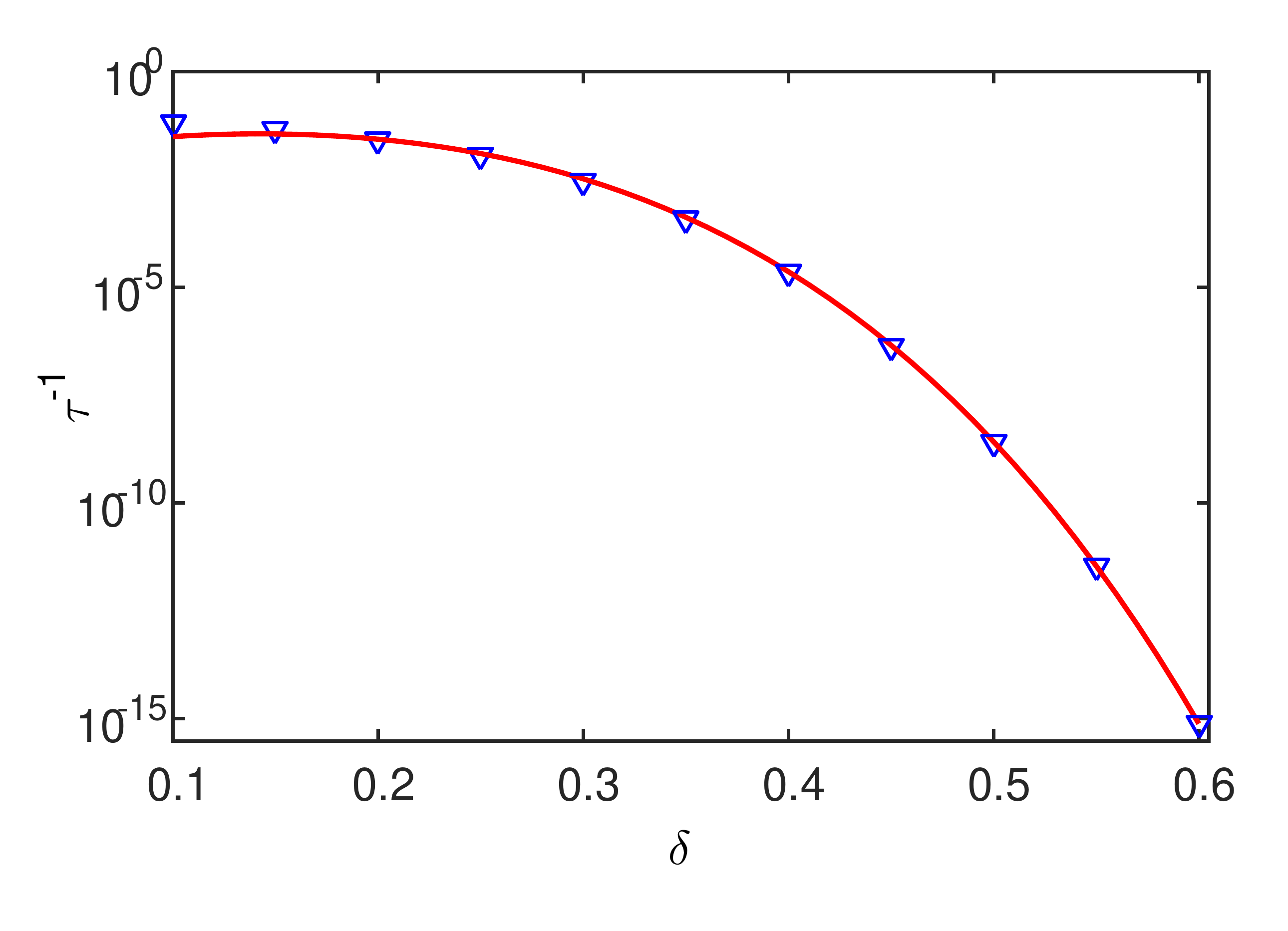}
\caption{The mean extinction rate $\tau^{-1}$ versus $\delta$ on a semi-log scale. Line: theoretical result~(\ref{exttimeB}). Symbols: results of a numerical solution of the master equation set for $N=200$ and $\mu=1$.} \label{extscenarioB}
\end{figure}

In contrast to scenario A, where the WKB solution is determined solely by the non-trivial zero energy trajectory (the fast mode),  the WKB solution in scenario B includes two modes. The fast-mode WKB solution dominates at $n>n_u$ (but not too close to $n_u$), whereas the slow-mode WKB solution (see below) dominates at $0<n<n_u$ (again, not too close to  $n=n_u$)~\cite{MS2008,EK2009,AM2010}, see Fig.~\ref{scAB}. Moreover, because the slow-mode solution diverges at $n=n_u$, one has to go beyond the WKB approximation in a boundary layer $|n-n_u|\ll n_u$ where the fast and slow WKB modes are strongly coupled. Fortunately, in this boundary layer the quasi-stationary master equation can be approximated by a quasi-stationary Fokker-Planck equation \cite{MS2008,EK2009,AM2010}. As a result, the QSD at $n\gg 1$ involves \textit{three} distinct asymptotes which need to be matched with one another. Finally, at $n=O(1)$, where the WKB  approximation breaks down, a recursive solution for the QSD can be found by neglecting the non-linear terms in the master equation~\cite{AM2010}. The recursive solution, however, is only needed if one wants to determine the MTE in a straightforward manner, by using Eq.~(\ref{E}). There is, however, an important shortcut which does not require the small-$n$ recursive solution \cite{EK2009}. This is because the metastable distribution develops a  \textit{constant} probability current toward small $n$.  It is this current that determines the escape rate from the metastable state $n=n_*$.  One can determine this  current in a close vicinity of $n=n_u$, without the need to address the region of $n=O(1)$.  The resulting calculations yield a general expression for the MTE \cite{AM2010}, which turns out to be the same  as the mean time to switch \cite{EK2009}, and even as the mean time to population explosion if the attracting fixed point is at $n=\infty$ \cite{MS2008}.  In our example of the three reactions with an Allee effect the resulting MTE is the following:
\begin{equation}\label{exttimeB}
\tau=\frac{\pi(1-\delta)}{\mu\delta}e^{N\Delta S},\quad
\Delta S=2\left(\delta-\sqrt{1-\delta^2}\arctan\frac{\delta}{\sqrt{1-\delta^2}}\right).
\end{equation}
As in scenario A, $\tau$ is exponentially large in $N$. Now, however, the pre-exponential factor is independent of $N$. In Fig.~\ref{extscenarioB} one can see a comparison between the theoretical formula~(\ref{exttimeB}) and numerical Monte-Carlo simulations, and excellent agreement is observed.

Close to the saddle-node bifurcation of the deterministic rate equation, $\delta \ll 1$, but still at
$N\delta^3 \gg 1$, Eq.~(\ref{exttimeB}) becomes \cite{EK2009,AM2010}
\begin{equation}\label{mte2bif}
\tau=\frac{\pi}{\mu\delta}\exp\left(\frac{2}{3}N \delta^3\right)\,.
\end{equation}
Equation~(\ref{mte2bif}) can be also obtained from a Fokker-Planck equation, and it is universally applicable to a whole class of well-mixed single-population models that have a very strong Allee effect. This result also holds for a whole class of continuous stochastic systems \cite{Dykman}.

\subsubsection{Switching}
\label{switchingB}
The difference between switching and extinction is that, in the case of switching, the target state, corresponding to the second attracting point $n_s$, is non-absorbing. As explained in the previous subsection, the mean time to escape in scenario B does not depend on the exact nature of the target state. Still, for the sake of completeness, we will briefly show how to evaluate the mean switching time using, as an example, a genetic switch.

We will consider a gene regulatory network that displays a deterministically bistable behavior. Gene regulatory networks are responsible for regulating the production of proteins.
During gene expression a DNA segment -- the gene -- is transcribed into an mRNA molecule which is then translated into a protein. This process is often regulated via transcription factors (which are also proteins) that can bind to the DNA promoter site and affect the mRNA transcription rate, and thereby the protein translation rate. Here control of the mRNA transcription is done by either recruiting or blocking RNA polymerase -- an enzyme that performs the transcription of genetic information from DNA to RNA~\cite{P1992}.

In some cases the dynamics of the transcription factor, that controls a specific protein, are strongly affected by the protein itself, so there is feedback. A \textit{positive} feedback can give rise to a bistable behavior, which has been shown to occur \textit{e.g.} in the \textit{lac operon} circuit~\cite{RMOBL2011,Bress2014} and also in the context of competence in \textit{Bacillus subtilis}~\cite{MMW2008}. Bistability can also occur if two different proteins \textit{negatively} regulate each other, see \textit{e.g.}, Refs.~\cite{P1992,GCC2000,AWT2005,LLBB2006,N2012,BiA2015}. Such feedback-based genetic switches are abundant in cell biology. They regulate diverse decision-making processes such as microbial environmental adaptation, developmental pathways and nutrient homeostasis, see Ref.~\cite{RMOBL2011} and references therein. In such systems it has been shown that the lifetime of these different gene-expression states (phenotypes) is determined by stochastic fluctuations of mRNA and proteins during gene expression that can yield spontaneous switching, even in the absence of a deterministic signal~\cite{ShS2008,ARS2011}.

We will consider a simple genetic switch in which the protein of interest positively regulates the transcription of the mRNA molecule that is responsible for its own production, giving rise to a positive feedback loop~\cite{ShS2008,ARS2011,VLACB2013,Bress2014}. In many situations the lifetime of the mRNA is short compared to that of the protein~\cite{PE2000}. Here one can make the simplifying assumption that the mRNA species instantaneously equilibrates, and the switching time depends solely on the protein dynamics. (We will revisit this assumption in Sec.~\ref{Switch2D} where we will explicitly consider the mRNA dynamics as well.) This gives rise to the so-called ``self-regulating gene" which is a protein-only model. In this model the protein's production rate is given by a function $f(n)$  which depends on the current protein number $n$, while degradation, mainly due to cell division, occurs at a rate $n$~\cite{ARSG2013}. Here time is measured in cell cycle units.

When positive feedback is at play, the function $f(n)$ increases with a growing protein number. A widely used positive-feedback function is the Hill function~\cite{biochemistry,RMOBL2011}
\begin{equation}\label{hillFun}
f(n)=a_0+(a_1-a_0)\frac{n^h}{n^h+n_0^h},
\end{equation}
where $h$ is the Hill exponent, $a_0$ is the baseline production rate, $n_0$ is its midpoint, and the production rate saturates at $a_1$ at $n\to\infty$. For $h\geq 2$ the deterministic rate equation for the mean number of proteins,
\begin{equation}
\dot{n}=f(n)-n
\end{equation}
can give rise to bistability. In the bistable case, the rate equation admits (at least) three fixed points: a lower attracting point $n_*$, an intermediate repelling point $n_u$, and a higher attracting point at $n_s$. In cellular biology the attracting fixed points represent  `off' and `on' phenotypes,  in which the cell produces a low/high number of proteins, respectively. When stochasticity is accounted for, one observes noise-driven switching between the two phenotypic states, each of which now becoming metastable.

The stochastic description of the self-regulating gene is given in terms of the master equation
\begin{equation}
\frac{d P_n(t)}{dt} = f(n-1)P_{n-1}(t)+(n+1)P_{n+1}(t)-[f(n)+n]P_n(t).
\end{equation}
What is the mean switching time (MST) to the on state $n=n_s$ when starting from the vicinity of the off state $n=n_*$?
As usual, we first determine the QSD $\pi_n$ via the ansatz $P_{n<n_u}(t)=\pi_n e^{-t/\tau}$, where $\tau$ is the MST, and employ the WKB ansatz~(\ref{fastmode})~\cite{EK2009,ARSG2013}. In the leading order, this gives rise to a Hamilton-Jacobi equation $H(q,dS/dq)=0$, where
\begin{equation}\label{hamilswitch}
H(q,p)=\tilde{f}(q)\left(e^{p}-1\right)+q\left(e^{-p}-1\right)
\end{equation}
is the Hamiltonian, $\tilde{f}(q)=f(n)/N$, $q=n/N$ is the rescaled protein concentration, and $N$ is the typical protein population size. As a result, the MST in the leading order, $\tau\sim e^{N\Delta S}$, is given by
\begin{equation}
\Delta S=\int_{q_*}^{q_u}\ln[q/\tilde{f}(q)]dq.
\end{equation}
As the target state $n_s$ is \textit{not} absorbing, there is actually a backward probability current towards $n_*$. However, when starting from the vicinity of the off state $n=n_*$, and at intermediate times $t_r\ll t\ll \tau$, the backward current is exponentially small compared to the forward current, and therefore, does not affect the MST~\cite{EK2009,AM2010}.

As discussed above, the subleading-order corrections to the MST  can be calculated by matching the fast-mode WKB solution with a boundary-layer solution valid in the vicinity of $n_u$, and subsequently, matching the boundary-layer solution to the slow-mode WKB solution. This allows the calculation of the probability current through the repelling point and subsequently the MST~\cite{EK2009,AM2010}.

\section{Extinction Conditioned on Non-Establishment}
\label{nonest}
In this section we consider a failure of establishment. The question we are asking is the following: What is the probability that a population, with initial population size $n_0\gg 1$, goes extinct \textit{before} it gets established at an attracting fixed point $n_*$? This probability is exponentially small and amenable to a WKB treatment. Here too there are two extinction scenarios: A and B. We will consider scenario A in some detail on the example of the Verhulst model with $W_{\pm 1}(n)$ given by Eq.~(\ref{ratesVer}). The repelling fixed point $n=0$ is an absorbing state of our stochastic process. In order to answer the question we have posed, we declare the attracting fixed point $n=n_*$ absorbing, and find the probability of reaching $n=0$ rather than $n=n_*$.  Let us denote by ${\cal P}_{n}$ this conditional extinction probability starting from $n$ individuals.  We can write the following recursive equation~\cite{Gardiner}:
\begin{equation}\label{extProb}
{\cal P}_{n}=\frac{W_{+1}(n)}{W_{+1}(n)+W_{-1}(n)}{\cal P}_{n+1}+ \frac{W_{-1}(n)}{W_{+1}(n)+W_{-1}(n)}{\cal P}_{n-1}.
\end{equation}
This equation follows from the observation that the probability of extinction starting from $n$ individuals equals the sum of the probability of extinction starting from $n+1$ individuals multiplied by the probability to reach state $n+1$ and the probability of extinction starting from $n-1$ individuals multiplied by the probability to reach state $n-1$ \cite{Gardiner}. The boundary conditions for ${\cal P}_{n}$ are ${\cal P}_{0}=1$ and ${\cal P}_{n_*}=0$. Equation~(\ref{extProb}) can be rewritten as
\begin{equation}\label{extProb1}
W_{+1}(n)R_{n}-W_{-1}(n)R_{n-1}=0,
\end{equation}
where $R_{n}={\cal P}_{n+1}-{\cal P}_{n}$.
The ensuing single-step problem is in fact exactly solvable, see e.g. \cite{Gardiner}. This is not so, however, in more complicated situations,
where multi-step transitions are present. Therefore, we employ a WKB approximation and sketch the leading-order WKB solution.  The WKB ansatz is
$R_{n}\equiv R(q)=A_1e^{-N{\cal S}(q)}$, $q=n/N$, and $A_1$ is a constant. Plugging it into Eq.~(\ref{extProb1})~\cite{MobA2010,AMob2010} we arrive, in the leading order, at the problem of finding the non-trivial zero energy trajectory, generated by the Hamiltonian
\begin{equation}\label{extProbHam}
H(q,p)=w_{+1}(q)-w_{-1}(q)e^{p},
\end{equation}
where $w_{\pm}(q)=W_{\pm}(n)/N$, and $p={\cal S}'(q)$ is the momentum. The solution is $p(q)=\ln[w_{+1}(q)/w_{-1}(q)]$, and therefore $R(q)=A_1e^{-N{\cal S}(q)}$, where
\begin{equation}\label{actNonEstab}
{\cal S}(q)=\int_q^0 p(q)dq=-q+q\ln\left(\frac{1}{B}+q\right)+ \frac{1}{B}\ln\left(1+Bq\right).
\end{equation}
The approximation we have made is valid at $n\gg 1$, or $q\gg N^{-1}$, and $N\gg 1$. As a result, we obtain
\begin{equation}\label{extProb2}
{\cal P}_n=A_2+\sum_{m=0}^{n-1}R_m\simeq A_2+NA_1\int_{0}^{q}R(q)dq,
\end{equation}
where we have replaced the sum by an integral. $A_1$ and $A_2$ are constants which can be found using the boundary conditions ${\cal P}_{0}=1$ and ${\cal P}_{n_*}=0$, and we obtain
\begin{equation}
{\cal P}_n\simeq \frac{\int_{q}^{q_*}e^{-N{\cal S}(q')}dq'}{\int_{0}^{q_*}e^{-N{\cal S}(q')}dq'}.
\end{equation}
As $N\gg 1$, the integrals in the numerator and denominator can be evaluated by Taylor-expanding ${\cal S}(q')$ around the lower boundaries $q$ and $0$, respectively. Then, putting $n=n_0<n_*$, we arrive at the final result~\cite{MobA2010,AMob2010}
\begin{equation}\label{extProbfinal}
{\cal P}_{n_0}\simeq C(q_0) e^{-N{\cal S}(q_0)},
\end{equation}
where $q_0=n_0/N$, $C(q_0)=A_3{\cal S}'(0)/{\cal S}'(q_0)$, and $A_3$ is an additional pre-factor that we will not present here. It comes from two sources. The first is the subleading WKB term  ${\cal S}_1(q)$. The second is the boundary correction term in the Euler-Maclaurin formula of replacing a sum by an integral, that we neglected in Eq.~(\ref{extProb2}).

The optimal path to extinction conditioned on non-establishment can also be found by solving a modified problem where a \textit{reflecting wall} is put at $n=n_0<n_*$ so that the population cannot have more than $n_0$ individuals. In the latter case the population \textit{cannot} reach the metastable state at $n_*$.  The pre-exponential factors in these two problems are, however, different.

In scenario B there are three fixed points of the deterministic theory: the attracting points $n=0$ and $n=n_*$ and a repelling point $n=n_u$. Starting from $n=n_0>n_u$, one is interested in the probability of extinction prior to reaching the established state at $n_*$. By similar arguments, the leading-order WKB result is
\begin{equation}
{\cal P}_{n_0}\simeq C(q_0) e^{-N[{\cal S}(q_u)-{\cal S}(q_0)]},
\end{equation}
where $C(q_0)$ is a preexponential factor which can be calculated.

\section{Bursty Reactions}\label{bursty}
Sometimes production or death of individuals occurs in ``bursts" of random size. This happens, for example, in living cells. When the life-time of mRNA molecules is  short compared to the cell cycle, proteins are synthesized in geometrically-distributed bursts, see \textit{e.g.} Ref.~\cite{PE2000}.  This fact alters protein statistics and may drastically decrease switching times between different phenotypic states~\cite{ARS2011}. Additional examples include bursty viral production from infected cells, see \textit{e.g.} \cite{PKP2011}, and variations in the number of offspring in animals, which has been shown to decrease the extinction risk \cite{GR2013}.

We will illustrate the problem of bursty reactions by considering a modification of the Verhulst model, presented above. (A similar approach can be taken to treat the problem of bursty influx, or arrival in groups, see Ref.~\cite{BAA2016}.) The microscopic dynamics are defined by the following ensemble of reactions and their corresponding rates~\cite{BA2016}
\begin{equation}
	\label{a}
	A\xrightarrow{BnD(k)/\langle k\rangle}A+kA,\quad A\xrightarrow{n+Bn^2/N}\emptyset.
\end{equation}
Here, $N\gg1$ is the typical population size, $B\gtrsim1$ is the average reproduction rate, and $k=0,1,2,\dots$ is the offspring number per birth event, which is sampled from a normalized burst size distribution $D(k)$ with the first and second moments denoted by $\langle k\rangle$, and $\langle k^2\rangle$, respectively.

In this case the master equation reads
\begin{eqnarray} \label{masterburst}
\dot{P}_n &=& \frac{B}{\langle k\rangle} \left[\sum_{k=0}^{n-1}D(k)(n-k)P_{n-k}-\sum_{k=0}^\infty D(k)n P_n\right] \nonumber\\
&+&(n+1)\left[1+\frac{B(n+1)}{N}\right]P_{n+1}-n\left[1+\frac{Bn}{N}\right]P_n,
\end{eqnarray}
whereas the deterministic rate equation coincides with Eq.~(\ref{Verhulstdeter}). The long-lived metastable state is peaked about the attracting fixed point $n_*=N(1-1/B)$, and our aim is to calculate the MTE. The WKB machinery leads to $H=0$, with Hamiltonian~\cite{BA2016}
\begin{equation}\label{Hamburst}
H(q,p)=(1-e^{-p}) B q \left[e^{p}F(p)-1/B-q\right],
\end{equation}
where
\begin{equation}
F(p)=\frac{\sum_{k=0}^\infty e^{kp}D(k)-1}{\langle k\rangle (e^{p}-1)}.
\end{equation}
When $D(k)$ is specified, the activation trajectory $p_a(q)$ can be found explicitly. Having found $p_a(q)$, one can show that the QSD's variance satisfies $\sigma^2=N[1+F'(0)]$, where $F'(0)=(1/2)(\langle k^2\rangle/\langle k\rangle-1)$. The MTE can be evaluated as ~\cite{BA2016}
\begin{equation}\label{tauburst}
\tau  \sim e^{N\Delta S}, \quad \Delta S=\frac{p_a(0)}{B}-\int_0^{p_a(0)}e^{p}F(p)dp.
\end{equation}

\begin{figure}[ht]
\includegraphics[scale=0.9]{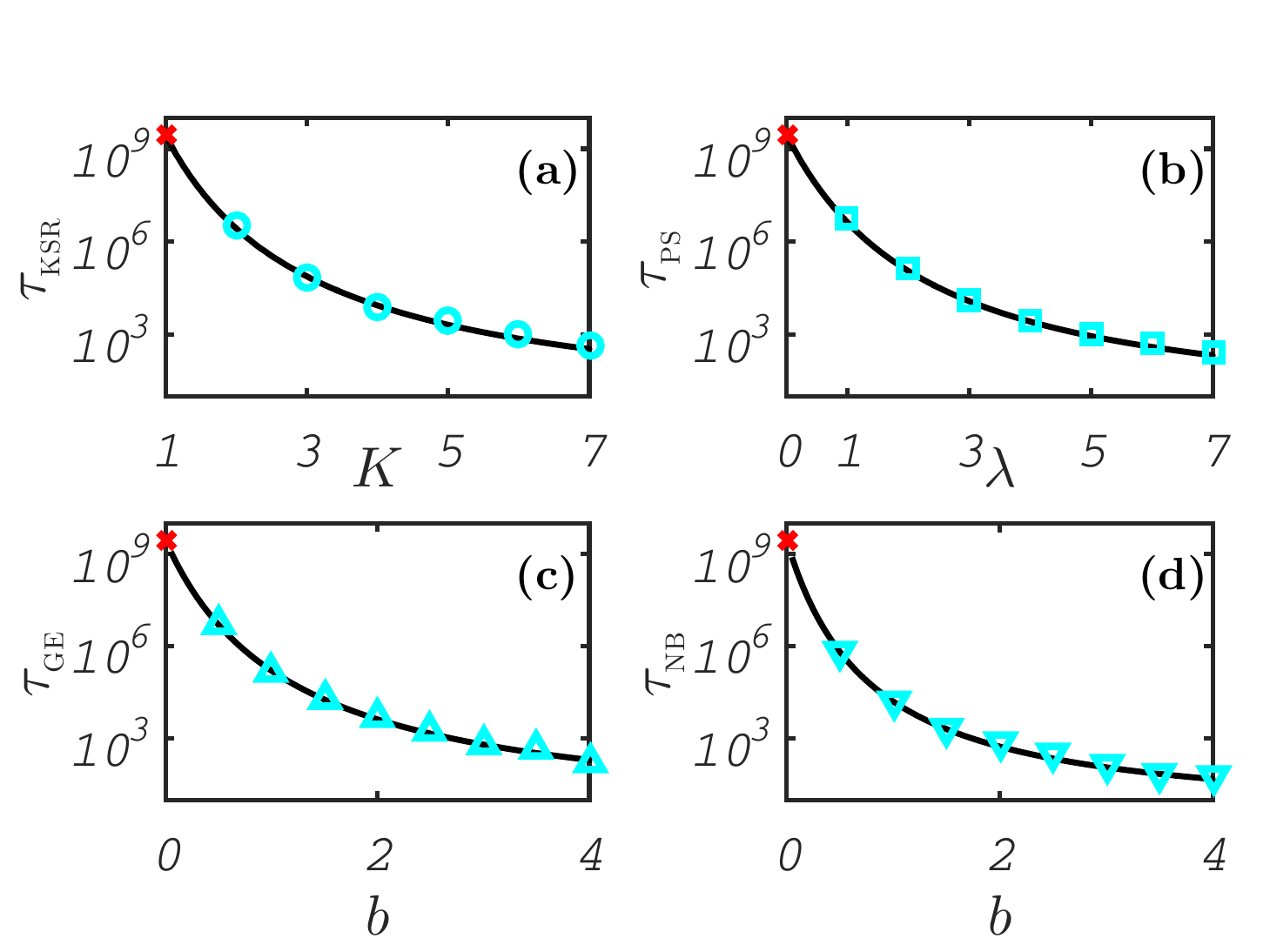}
\caption{MTE for different burst size distributions as a function of their characteristic parameter: (a) $K$-step reproduction (KSR) $D(k)=\delta_{k,K}$; (b) Poisson (PS) $D(k)=\lambda^k e^{-\lambda}/k!$; (c) geometric (GE) $D(k)=\left(1/(1+b)\right)\left(b/(1+b)\right)^k$; (d) negative-binomial (NB) $D(k)=(k+a-1)!/(k!(a-1)!)\left(1/(1+b)\right)^a\left(b/(1+b)\right)^k$.  The solid lines denote theoretical results given by Eq.~(\ref{tauburstfull}),  the ($\circ$), ($\Box$), ($\bigtriangleup$) and ($\bigtriangledown$)  markers denote results of Monte-Carlo simulations, while the (x) marker denotes the theoretical value for the single-step case with $D(k)=\delta_{k,1}$. Parameters are $a=2$, $B=2$, and $N=150$~\cite{BA2016}.}
\label{burstytau}
\end{figure}

Finding the subleading correction by the real-space WKB method would require matching of the subleading-order WKB solution with
a recursive solution at $n={\cal O}(1)$. It turns out, however, that the method of calculating the recursive solution by linearizing the reaction rates in the vicinity of $n={\cal O}(1)$ breaks down in the presence of bursty reproduction \cite{BA2016}. Fortunately, the momentum-space approach comes to rescue. Here, as in section \ref{extinctionA}, we can match the bulk solution and the boundary-layer solution in their joint region of validity, see Fig.~\ref{momentum-phi}. This yields the MTE including the pre-exponential factor~\cite{BA2016}:
\begin{equation}\label{tauburstfull}
\tau=\frac{1}{(B-1)(e^{-p_f}-1)}\sqrt{\frac{2\pi}{N q_a'(p_f)}}e^{N\Delta S},
\end{equation}
where $q_a(p)=e^p F(p)-1/B$ is the zero-energy solution of the Hamiltonian~(\ref{Hamburst}), $p_f$ is the root of the equation $q_a(p_f)=0$, and $\Delta S$ is given by Eq.~(\ref{tauburst}). Figure~\ref{burstytau} shows comparisons between the analytical and numerical results for $\tau$ for four different burst size distributions.

Close to the transcritical bifurcation of the deterministic rate equation, $\delta \equiv B-1\ll 1$, but still at $N\delta^2 \gg 1$, Eq.~(\ref{tauburstfull}) simplifies to
\begin{equation}
\tau\simeq \frac{\sqrt{2\pi[1+F'(0)]}}{\sqrt{N}\delta^2}\,\exp\left\{\frac{N \delta^2}{2[1+F'(0)]}\right\}.
\label{tauburstclose}
\end{equation}
Let us compare this result with Eq.~(\ref{tausingleAclose}) for the MTE of the ``standard" Verhulst model. As $F'(0)>0$, Eq.~(\ref{tauburstclose}) shows that bursty reproduction can reduce the MTE by an exponentially large factor~\cite{BA2016}.

Bursty deaths, where a random number of individuals dies (for example, due to competition), can be treated in a similar manner. Overall, bursty reactions can either increase or decrease the probability of rare events to occur compared to the non-bursty reactions~\cite{BA2016a}.

\section{Which Stationary WKB Method is Better?}

The stationary real-space and momentum-space WKB methods, in conjunction with additional approximations employing the same large parameter $N\gg 1$, yield accurate large-deviation results for a broad class of problems of single stochastic populations. Such accuracy is impossible to achieve with the more traditional
van Kampen system size expansion which approximates the exact master equation by a
Fokker-Planck equation and, as a result, usually holds only for typical fluctuations.  How does the momentum-space WKB method compare with the real-space WKB method?
From our experience, every problem with ``non-bursty" reactions, which includes a large parameter $N\gg 1$
and can be approximately solved with the momentum-space WKB, can also be approximately solved with the real-space WKB, and these approximate results coincide. For
populations exhibiting escape scenario B  (the Allee effect), the momentum space representation encounters significant difficulties \cite{AMS2010}.

On the other hand, for problems exhibiting bursty reactions, the momentum-space approach is, as of today, the only method that enables one to calculate the preexponential corrections, see Sec.~\ref{bursty} and Refs.~\cite{BAA2016,BA2016}. In this case even the leading-order WKB calculations are more conveniently done with the momentum-space approach. Here the master equation, which contains an infinite sum corresponding to the infinite reaction ensemble, transforms into a single evolution equation for the probability generating function, $G(p,t)$, where the bursty nature of the reactions is manifested by a nontrivial $p$-dependence of this equation. By contrast, the Hamiltonian obtained via the real-space approach contains an infinite number of terms, which makes theory more complicated~\cite{BAA2016,BA2016}.

\section{Extinction due to Demographic Noise and Environmental Variations}
\label{demenv}

\subsection{Periodic Modulation of the Environment}

A periodic rate modulation can greatly reduce the MTE. The authors of Refs. \cite{EscuderoRodriguez,AKM2008}
studied this effect on the prototypical example of the branching-annihilation process $A\to 2A$ and $2A\to \emptyset$. The transition  $2A\to \emptyset$ can be viewed as an extreme variant of competition when two competitors are both killed upon encounter. The transition rate coefficients of the branching and annihilation processes are $\lambda$ and $\mu_0$, respectively. Let $\lambda$ oscillate with time,   $\lambda=\lambda_0(1+\epsilon \,\cos \omega t)$,
so as to describe \textit{e.g.} seasonal effects.  The master equation reads, for $n\geq 1$,
\begin{equation}
\label{masterperiodic} \frac{dP_{n}(t)}{dt}=\frac{\mu_0}{2}\left[(n+2)(n+1)
P_{n+2}(t)-n(n-1)P_{n}(t)\right]
+\lambda(t)\left[(n-1)P_{n-1}(t)-nP_{n}(t)\right].
\end{equation}
Using this equation, one obtains the following exact evolution equation for $G(p,t)$:
\begin{equation}\label{Gdotperiodic}
\frac{\partial G}{\partial t} =
\frac{\mu_0}{2}(1-p^2)\frac{\partial^2 G}{\partial p^2}+
\lambda(t)p(p-1)\frac{\partial G}{\partial p}\,.
\end{equation}
In the absence of modulation, $\epsilon=0$,  the MTE is the following \cite{turner,KS2007,AM2007}:
\begin{equation}\label{MTEnoperiodic}
\tau\simeq \frac{2\sqrt{\pi}\mu^{1/2}}{\lambda_0^{3/2}}e^{\frac{2\lambda_0}{\mu}\left(1-\ln 2\right)}
=\frac{2\sqrt{\pi}}{\mu N^{3/2}}e^{2N(1-\ln 2)},
\end{equation}
where $N=\lambda_0/\mu\gg 1$. This result (which includes an important pre-exponential factor) was obtained by three different methods, among them the real-space WKB method in the leading and subleading WKB order \cite{KS2007}.
For the time-periodic $\lambda$, such a high accuracy can only be achieved in the limit when the modulation frequency is very low, see below. In other regimes one settles for the leading-order WKB accuracy.  Let us use  the non-stationary WKB ansatz in the momentum space $G(p,t)=\exp[-S(p,t)]$ (here and in subsection B we absorb $N$ in the definition of $S$). Defining $q=\partial S/\partial p$, neglecting $\partial ^2 S/\partial p^2$ and shifting the momentum, $p\to p-1$,
we arrive at an effective one-dimensional classical mechanics with the Hamiltonian
\begin{equation}\label{H1}
H(q,p)=\left[\lambda(t)(1+p)-\frac{\mu}{2}(2+p)q\right]\, qp
\end{equation}
that explicitly depends on time. This mechanical problem is in general not solvable analytically. Two existing numerical algorithms will be briefly reviewed in Sec. \ref{extendemic}. Analytical progress requires additional approximations, based on additional small parameters.  Ref. \cite{AKM2008} used three such approximations.

\subsubsection{Weak modulation: linear theory}
\label{weakmodulation}

\begin{figure}
\includegraphics[width=3.3 in,clip=]{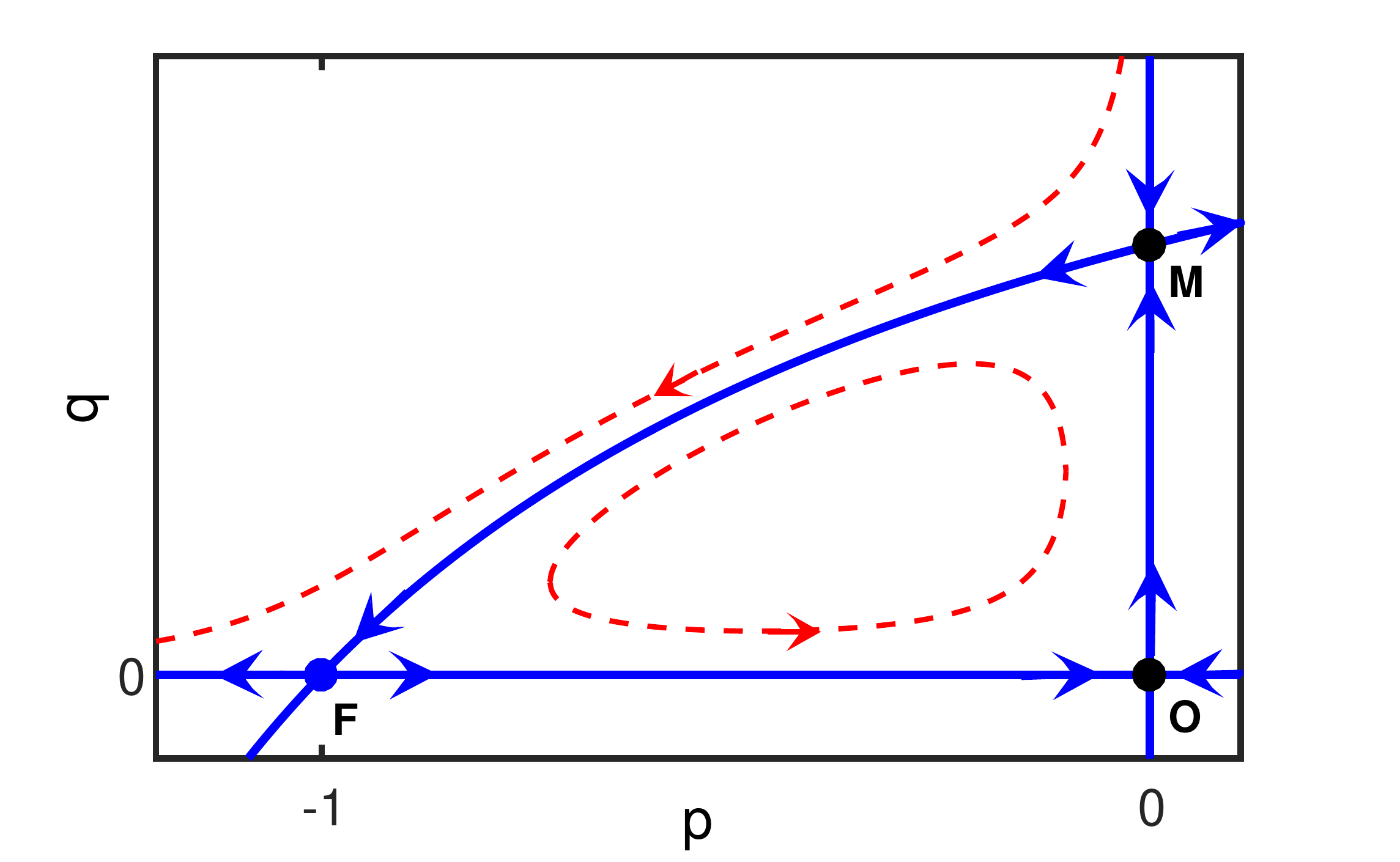}
\caption{The phase plane of the unperturbed Hamiltonian (\ref{h0pert}).
The solid lines denote zero-energy trajectories.} \label{branchingann}
\end{figure}

At sufficiently small, but not too small, $\epsilon$, the relatively small, but important correction to the action is linear with respect to $\epsilon$, and can be calculated perturbatively within the WKB approximation. We split the Hamiltonian into two terms:
\begin{equation}\label{pertHam}
    H(q,p,t)=H_0(q,p) + \epsilon H_1(q,p,t) \,,
\end{equation}
where
\begin{equation}\label{h0pert}
H_0(q,p)=\lambda_0(1+p)pq-\frac{\mu_0}{2}(2+p)p q^2\,,
\end{equation}
and
\begin{equation}\label{h1pert}
H_1(q,p,t)=\lambda_0(1+p)pq \,\cos \omega t\,.
\end{equation}
For the unperturbed Hamiltonian $H_0(q,p)$ the ``extinction instanton" --  the heteroclinic trajectory MF of Fig. \ref{branchingann} -- is~\cite{AM2007}
\begin{equation}\label{q(p)}
q= q_0(p)=\frac{2N(1+p)}{(2+p)}\,.
\end{equation}
One can find explicit formulas for $q(t)$ and $p(t)$:
\begin{equation}\label{unpert}
q_0(t-t_0)=\frac{2N}{2+e^{\lambda(t-t_0)}}\;,\;\;\;p_0(t-t_0)=-\frac{1}{1+e^{-\lambda(t-t_0)}}\,,
\end{equation}
where $t_0$ is an arbitrary time shift. Using Eq.~(\ref{q(p)}), we obtain the unperturbed action $S_0$ along the instanton,
\begin{equation}\label{action1}
S_0 = -\int_0^{-1}
q_0(p)\,dp=2N(1-\ln 2)
\end{equation}
which yields the expression in the exponent of Eq.~(\ref{MTEnoperiodic}).

A small correction to the action $\Delta S$ can be evaluated from \cite{AKM2008}
\begin{equation}
\Delta S=\min_{t_0}\left\{ -\varepsilon
\int_{-\infty}^{\infty}H_1\left[q_0(t-t_0),p_0(t-t_0),t\right]\,
dt\right\}\,, \label{S1}
\end{equation}
where the integration is performed along the \textit{unperturbed} instanton (\ref{unpert}).
After the integration over time and minimization over $t_0$, the perturbed action $S=S_0+\Delta S$ satisfies
\begin{equation}
\frac{S}{S_0}=1+\frac{\Delta S}{S_0}\simeq 1-\frac{\pi\varepsilon}{(1-\ln 2)\sinh(\pi\alpha)}
\left\{\left[\sin(\alpha\ln
2)-\alpha\right]^2+\left[\cos(\alpha\ln
2)-1\right]^2\right\}^{1/2}\,,\label{minact}
\end{equation}
where $\alpha=\omega/\lambda_0$.
The maximum
effect of the rate modulation is obtained at
$\alpha \to 0$. Recall that the perturbed action yields (in the leading order in $N$) minus the natural logarithm of the MTE.

The minimization over $t_0$ reflects an important effect of synchronization between the optimal path of the population to extinction
and the periodic rate modulation. The synchronization lifts the degeneracy of the unperturbed instanton
with respect to an arbitrary time shift present in Eq.~(\ref{unpert}). Synchronization of this nature was also observed in large deviations of continuous stochastic systems subject to a parameter modulation \cite{Smel1,Smel2}.

\begin{figure}
\includegraphics[scale=0.6]{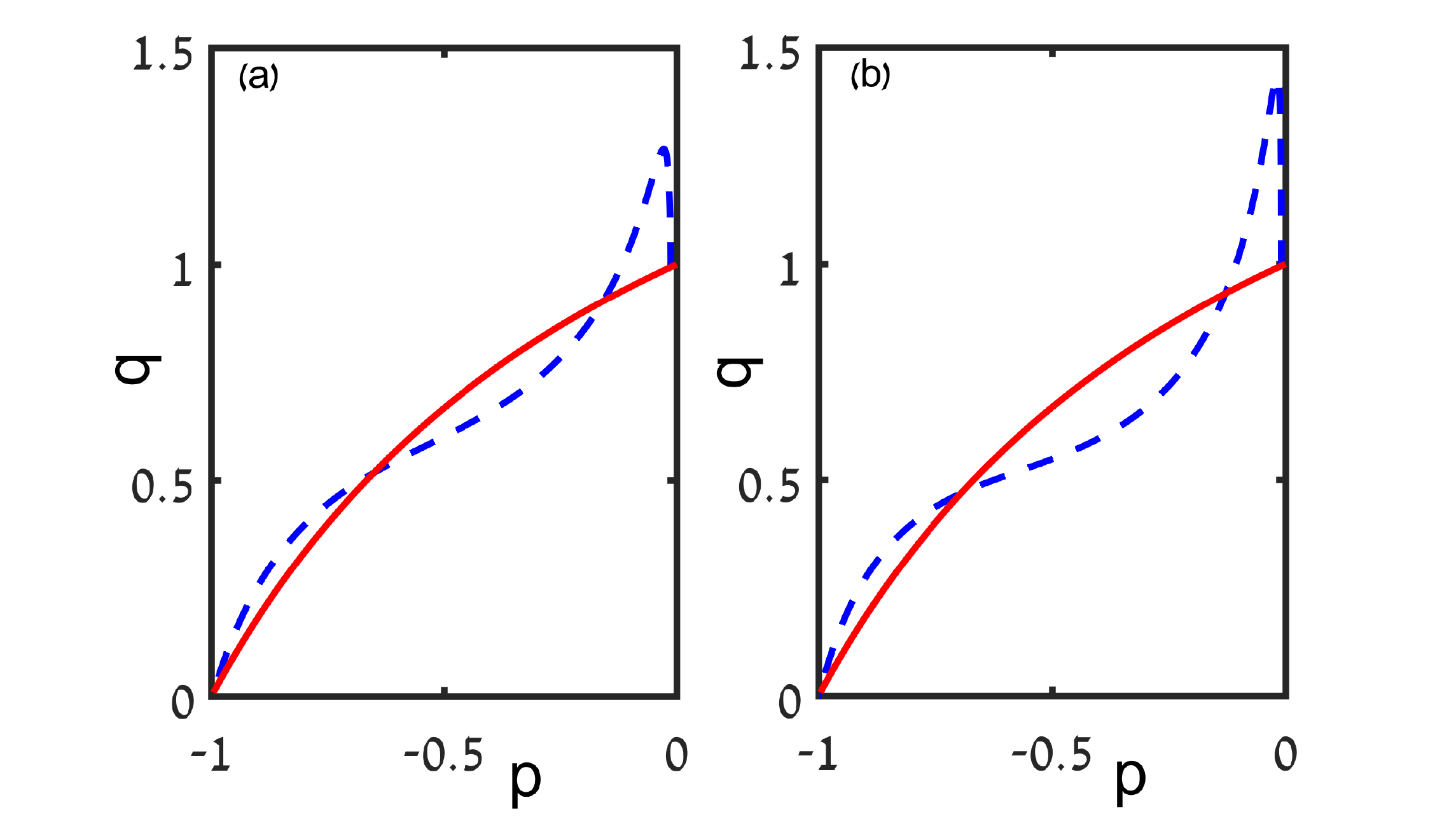}
\caption{Projections on the $(q,p)$ plane of instantons of the Hamiltonian (\ref{H1}), found numerically via the shooting method, for two sets of parameters (the dashed lines). Here $q$ is measured in units of $N$, $\alpha=1$, and $\varepsilon=0.6$ such that ${\cal S}/N=0.467$ in (a) and $\varepsilon=0.9$ such that ${\cal S}/N=0.398$ in (b). The solid line denotes the unperturbed instanton, whose unperturbed action is ${\cal S}_0/N=2(1-\ln 2)\simeq 0.614$.}\label{timedependent-action}
\end{figure}

\subsubsection{Fast modulation: the Kapitsa's pendulum}

In the the high-frequency limit, $\omega \gg \lambda_0$,  the system resembles the Kapitsa's pendulum \cite{LL}:
 a rigid pendulum with a rapidly vibrating
pivot. Here one can also calculate an important correction to the unperturbed action.
Because of the high modulation frequency,  this correction
is small even if $\varepsilon$ is of order 1. To determine this correction Assaf et al \cite{AKM2008}
developed a Hamiltonian version of the asymptotic method for the Kapitsa's pendulum \cite{LL}.
As a first step, one calculates small
high-frequency corrections to the unperturbed ``coordinate" and
``momentum" of the system. Using these corrections, one then constructs a
canonical transformation which, by means of a time averaging (essentially, by
rectifying the high-frequency component of the motion), transforms
the original time-dependent Hamiltonian into an effective
\textit{time-independent} Hamiltonian. The effective Hamiltonian includes a
correction coming from the rectified high-frequency perturbation.
Finally, one finds the perturbed instanton, emerging from this effective time-independent Hamiltonian,
and evaluates the action along this instanton.
The action along the effective instanton turns out to be~\cite{AKM2008}
\begin{equation}\label{Kapitsaaction}
\frac{S}{S_0}\simeq 1-
\frac{K\epsilon^2}{\alpha^2}\,,
\end{equation}
where
\begin{equation*}
K=\frac{6\ln 2-49/12}{1-\ln 2}= 0.2462\dots\,.
\end{equation*}

\subsubsection{Slow modulation: adiabatic approximation}

When the modulation period $2\pi/\omega$ is much
longer than the characteristic relaxation time of the system $t_r=1/\lambda_0$ but much shorter than
the expected MTE, the extinction probability as a function of time can be written, at long times, as \cite{AKM2008}
\begin{equation}\label{extprob10}
    P_0(t)\simeq 1- e^{- \int_0^t r_{\text{ext}}(t) \,dt} .
\end{equation}
Here $r_{\text{ext}}(t)=r_{\text{ext}} [\lambda=\lambda(t)]$ is the
\textit{instantaneous value} of the slowly time-dependent extinction rate. It is obtained by replacing
$\lambda_0$ in Eq.~(\ref{MTEnoperiodic}) by $\lambda(t)=\lambda_0(1+\epsilon \cos \omega t)$.
The \textit{average}, or effective extinction rate $\bar{r}_{\text{ext}}$ during a sufficiently long time $T$ is defined via the
relation
\begin{equation}\label{extprob20}
P_0(T)=1- e^{- \bar{r}_{\text{ext}} T}\,.
\end{equation}
Comparing Eqs.~(\ref{extprob10}) and (\ref{extprob20}), we obtain for our periodic rate modulation
\begin{equation}
  \bar{r}_{\text{ext}}=\frac{2\pi}{\omega} \int_0^{2 \pi/\omega} r_{\text{ext}}(t) \,dt =\frac{N^{3/2}}{4\pi^{3/2}}\int_{0}^{2\pi}(1+\varepsilon\cos
\tau)^{3/2}e^{-S_0(1+\varepsilon\cos \tau)}d\tau\,. \label{extrate}
\end{equation}
The MTE is equal, in the adiabatic approximation, to $1/\bar{r}_{\text{ext}}$. In Ref. \cite{AKM2008} the integral over $\tau$ in Eq.~(\ref{extrate}) was evaluated
analytically in two limits. For $|\epsilon| N \gg 1$, a saddle point evaluation
yields
\begin{equation}\label{specadia}
\bar{r}_{\text{ext}}=\frac{N (1-|\epsilon|)^{3/2}}{4\pi \sqrt{|\epsilon| (1-\ln
2)}}e^{-S_0(1-|\epsilon|)}\,.
\end{equation}
The leading term $S_0(1-|\varepsilon|)$ coincides with the
zero-frequency limit of the linear theory, as can be seen from
Eq.~(\ref{minact}) for $\alpha \to 0$. The
physical meaning of this term is evident: by virtue of the adiabatically slow
rate modulation, the ``activation barrier" to extinction
$S_0(1-|\varepsilon|)$ is determined simply by the minimum value of
$\lambda(t)=\lambda_0 (1+\varepsilon \cos \omega t)$ which is equal
to $\lambda_0 (1-|\varepsilon|)$. It is not surprising, therefore, that the same leading order result also follows from
the leading-order WKB theory \cite{AKM2008}.  However, Eq.~(\ref{specadia})
also includes an important prefactor missed by the leading-order WKB.  Notice that this prefactor scales as
$N$, not as
$N^{3/2}$ as observed without modulation, see Eq.~(\ref{MTEnoperiodic}).

In the opposite limit, $|\varepsilon| N \ll 1$, the integral
(\ref{extrate}) can be calculated via a Taylor expansion of the integrand
in $\varepsilon N$ \cite{AKM2008}.

\begin{figure}
\includegraphics[width=10.0cm,height=7.0cm,clip=]{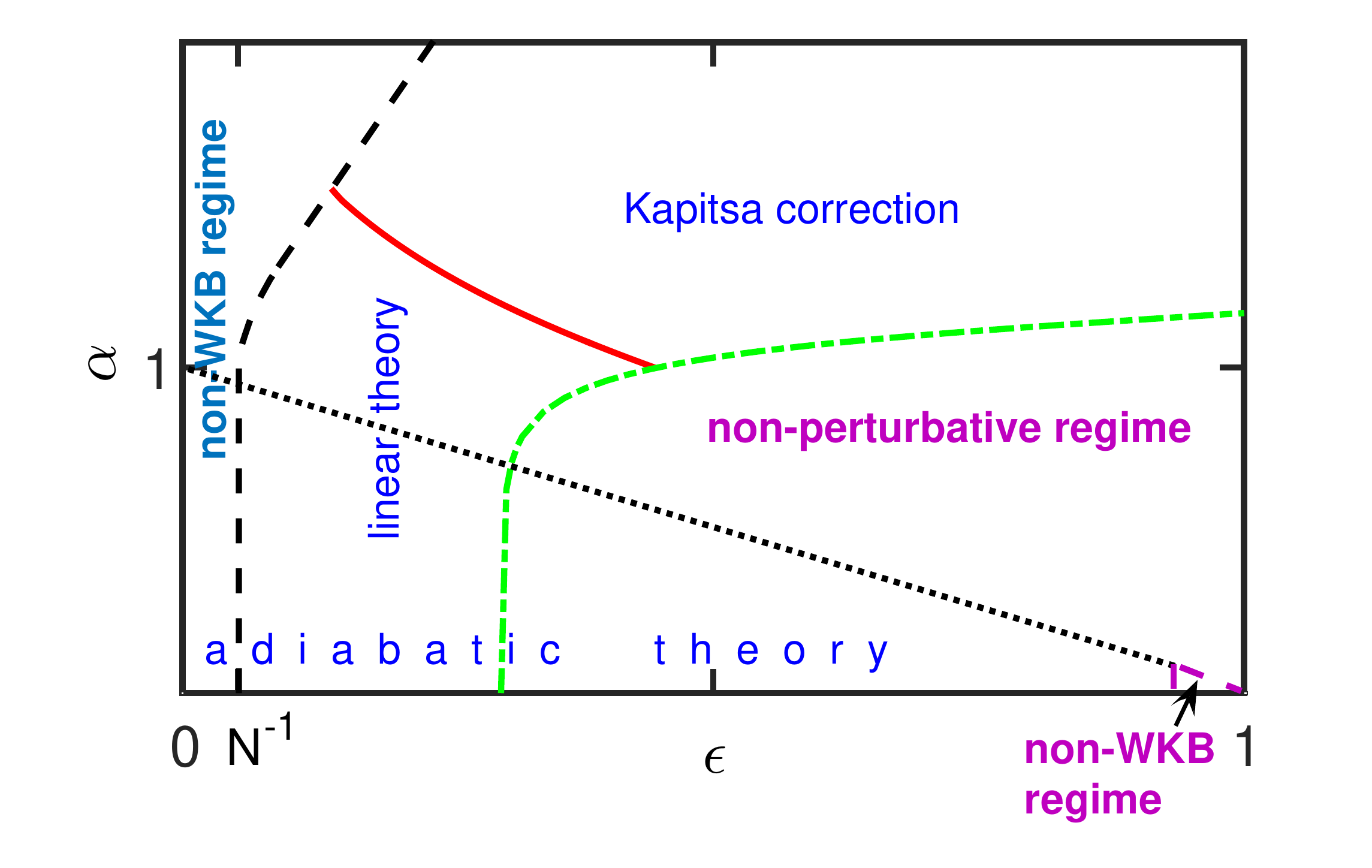}
\caption{Schematic phase diagram of different extinction regimes
in the parameter plane  $(\epsilon, \alpha=\omega/\lambda_0)$ of the
branching-annihilation process with modulation of the branching rate.
Either the linear theory, or the hamiltonian Kapitza's method
is applicable well to the left of and above the dash-dotted line, and
well to the right of the dashed line.
The position of the solid line is calculated in Ref. \cite{AKM2008}.
The adiabatic approximation holds
well below the dotted line. The horizontal size of the non-eikonal (that is, non-WKB)
regions at $\alpha \to 0$ is of order $1/N$, where $N=\lambda_0/\mu\gg 1$. There is no analytic theory
as of today in the  non-perturbative regime.} \label{phasediagram}
\end{figure}

The non-perturbative regime can only be studied numerically. The existing numerical methods
are essentially the same as for multi-population problems, and we postpone their brief discussion
until Sec. \ref{extendemic}. Here we only show, in Fig. \ref{timedependent-action}, two examples
of extinction instantons in this system, found numerically in Ref. \cite{AKM2008}. A schematic ``phase diagram" of the
system, on the plane $(\varepsilon, \alpha=\omega/\lambda_0)$,
is shown in Fig. \ref{phasediagram}.

\subsection{Catastrophic Events}\label{catastrophe}

\begin{figure}
\includegraphics[width=7.0cm,clip=]{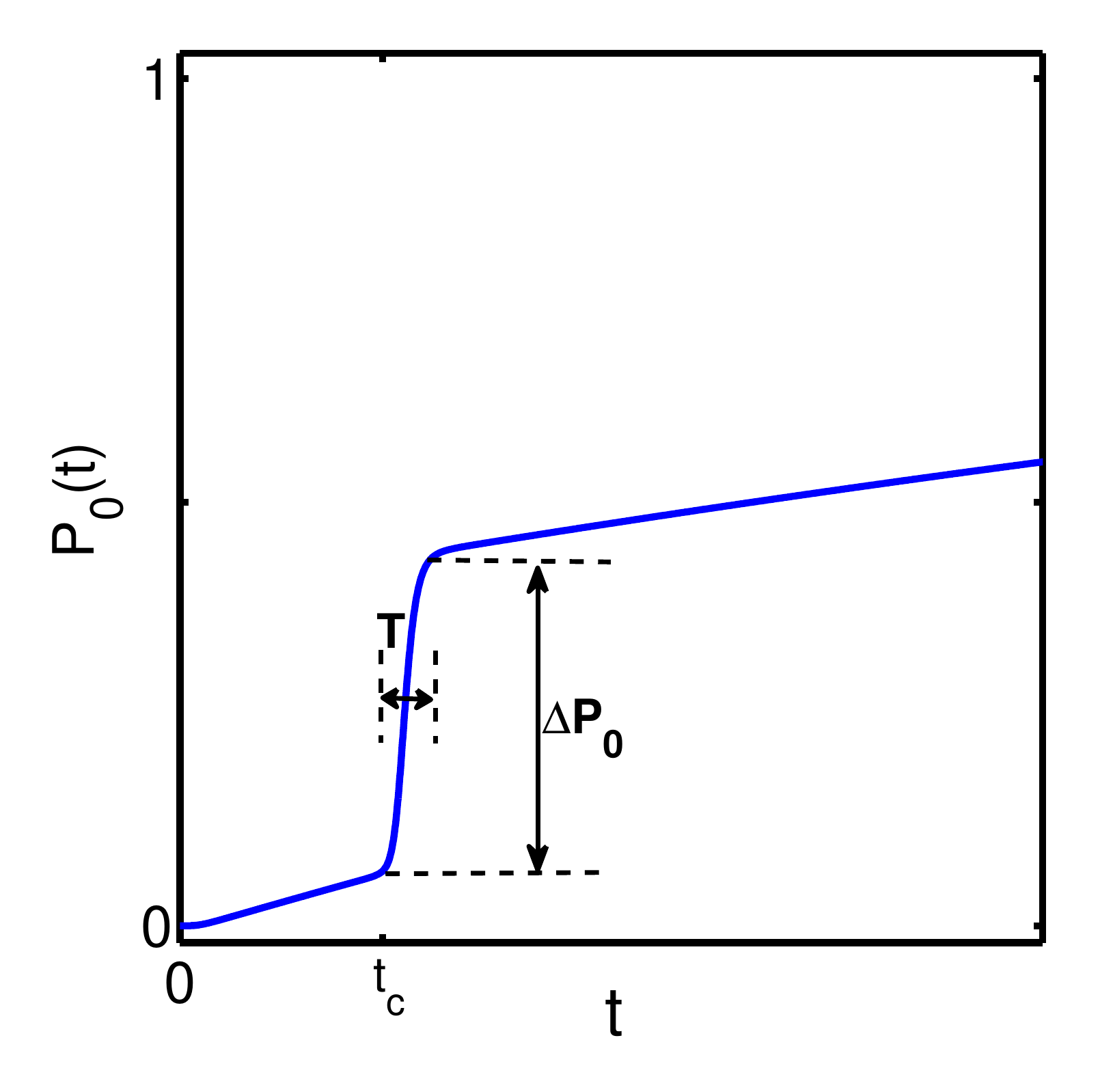}
\caption{A schematic plot of the time-dependent
extinction probability $P_0(t)$, showing the effect of a
catastrophe with characteristic duration $T$. The increase of the
extinction probability $\Delta P_0$ can be evaluated using the WKB approximation.} \label{sketch}
\end{figure}

A catastrophic event in population dynamics can be modeled by a strong temporary decrease of the population birth rate, or increase of the death rate. How can one evaluate $P_0(t)$ -- the probability that the population goes extinct by time $t$ -- at the time when the catastrophic
event is over, and the environmental conditions return to normal? As shown in Ref. \cite{AKM2009}, a catastrophe can lead to an exponentially
large increase  $\Delta P_0$ in the extinction probability $P_0(t)$ compared to the extinction probability without the catastrophe, see Fig. \ref{sketch}. This increase $\Delta P_0$ can be evaluated using the WKB approximation.

\begin{figure}
\includegraphics[width=8.0cm,clip=]{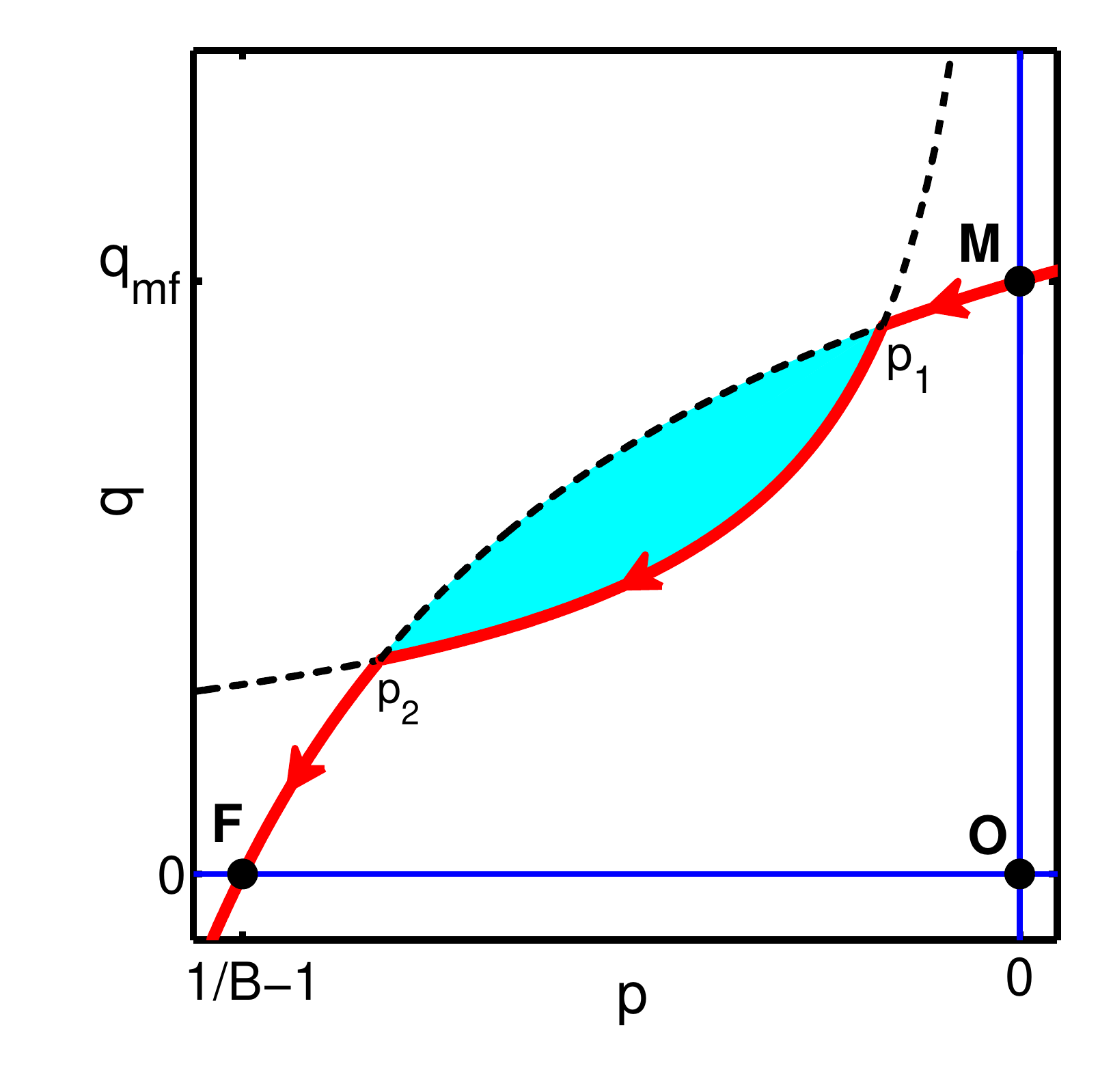}
\caption{Projection on the $(p,q)$ plane of the
``extinction instanton" (the thick solid line going from $M$ to $F$)
for the catastrophic event described by Eqs.~(\ref{cat1}) and (\ref{f}). Points $p_1$
and $p_2$ correspond to times $t=0$ and $t=T$ where the catastrophic
event begins and ends, respectively. At $t<0$ and $t>T$ the
instanton follows the unperturbed zero-energy heteroclinic trajectory
$q=q_0(p)$, whereas at $0<t<T$ it follows a trajectory with a
nonzero energy which depends on
the duration $T$ of the catastrophe. The increase $\Delta P_0$
of the extinction probability due to
the catastrophe is $\sim \exp[-S(T)]$, where $S(T)$ is area of the shaded region.}
\label{phasespacecatast}
\end{figure}

The authors of Ref. \cite{AKM2009} considered the Verhulst model with the birth and death rates
given by Eq.~(\ref{ratesVer}). The catastrophic event was modeled by introducing a time-dependent factor $f(t)$, so that $f(\pm \infty)=1$,
into the birth rate:
\begin{equation}\label{cat1}
W_{+1}(n,t)=B f(t) n .
\end{equation}
The evolution equation for the probability generating function can be written as
$\partial G/\partial t=\hat{\mathcal H}G$,
with the operator
\begin{equation}\label{hatH}
\hat{\mathcal H} =
\frac{B}{N}\left(1-p\right)p\,\frac{\partial^2}{\partial p^2}+
(p-1)\left[B f(t) p-1\right]\frac{\partial}{\partial p} .
\end{equation}
Similarly to the problem of rate modulation considered in the previous subsection, the non-stationary WKB ansatz in the momentum space leads, in the leading order, to a one-dimensional classical
Hamiltonian flow with the time-dependent Hamiltonian
\begin{equation}\label{H1time}
H(p,q,t)=p\left[-\frac{B}{N}(p+1)q +B f(t) (p  + 1)-1 \right] q\,,
\end{equation}
where the new momentum $p$ is shifted by $1$, $p-1\to p$.  Because of the explicit time dependence of the Hamiltonian, only numerical solution is in general available. One analytically solvable
case is when the reproduction rate drops instantaneously
(for simplicity, to zero) at $t = 0$ and recovers to the pre-catastrophe
value, again instantaneously, after time $T$
has elapsed:
\begin{equation}\label{f}
f(t)=\left\{ \begin{array}{ll}
1 & \;\;\;\mbox{if $\;t<0$ or $t>T$}\,, \\
0 & \;\;\;\mbox{if $\;0<t<T$\,.}
\end{array}
\right.
\end{equation}
Because of the special shape of $f(t)$, there are now two distinct
Hamiltonians to consider: the unperturbed
Hamiltonian [Eq.~(\ref{H1time}) with $f(t)=1$] before and after the catastrophe
and the zero-birth-rate Hamiltonian during the catastrophe:
\begin{equation}
H_c(p,q)=-p\left[\frac{B}{N}(p+1)q + 1 \right] q\,. \label{Hc}
\end{equation}
Each of the two Hamiltonians is an integral of motion on the
corresponding time interval. The instanton can be found by matching
three separate trajectory segments: the pre-catastrophe, catastrophe
and post-catastrophe segments. Figure \ref{phasespacecatast} shows a
projection of the instanton on the $(p,q)$ plane. The
instanton must exit, at $t=-\infty$, the  deterministic fixed point $M$ and
enter, at $t=\infty$, the fluctuational fixed point $F$. The matching
conditions at times $t=0$ and $t=T$ are provided by the continuity
of the functions $q(t)$ and $p(t)$. The pre- and post-catastrophe
segments must have a zero energy, $E=0$, so they are parts of the
original zero-energy trajectory of the unperturbed problem, $q_0(p)=N-N/[B(1+p)]$.  For the catastrophe segment, $q_c(p)$, however, the energy
$E=E_c$ is non-zero and \textit{a priori} unknown. It can be found from the demand that the duration of this segment be $T$~\cite{AKM2009}.

The solution simplifies
considerably
at $B-1\ll 1$, that is close to the bifurcation point of the model.  In this limit the pre- and post-catastrophe Hamiltonian reduces
to the ``universal" Hamiltonian
\begin{equation}\label{H1univ}
H(p,q)=p\left(-\frac{q}{N} +p+ B-1\right)\,q\,,
\end{equation}
introduced in Ref. \cite{Kamenev2}. Furthermore, the zero-birth-rate Hamiltonian during the
catastrophe simplifies dramatically:
\begin{equation}
H_c(p,q)\simeq -pq\,. \label{Hc1}
\end{equation}
After some algebra, see Ref. \cite{AKM2009}, one obtains
\begin{equation}
\label{deltaSsimple}  S(T) \simeq \frac{2
S_0}{1+e^{T}}\,,
\end{equation}
where
\begin{equation}\label{area}
S_0 \simeq \frac{N (B-1)^2}{2}\,,
\end{equation}
is the action of the ``universal" model in the absence of a catastrophe. The increase $\Delta P_0$
of the extinction probability due to
the catastrophe is $\sim \exp[-S(T)]$. As one can see, for a prolonged catastrophe, $T\gg 1$, the increase of the
extinction probability becomes very large.

\subsection{Environmental Noise}

Multiple and concurrent environmental factors that affect the birth or death rates of populations can be modeled
as a
rate modulation by \textit{noise}. Early population biology models assumed that the environmental noise is white, that is
delta-correlated in time \cite{Leigh,Lande}. More recent studies focused on the
effect of temporal autocorrelation, or color, of the environmental noise on the population extinction, see e.g. Ref. \cite{OM2010} for a review.  WKB theories of population extinction driven by an interplay of demographic and environmental noise in the absence of an Allee effect and in the presence of it
were developed in Refs. \cite{KMS2008} and \cite{LM2013}, respectively.  Ref. \cite{KMS2008} considered a symmetrized Verhulst model with the birth and death rates
\begin{equation}\label{KMSrates}
    \lambda_n=\frac{n}{2}\, (\mu+r-a n)\,,\;\;\;\;\;\mu_n=\frac{n}{2}\, (\mu-r+a n)\,, \quad r<\mu\,,
\end{equation}
where there is no Allee effect. Ref. \cite{LM2013} considered the set of reactions $2A\to 3A$, $3A\to 2A$ and $A\to \emptyset$, that we dealt with in Sec. \ref{extinctionB},
in the parameter region corresponding to a strong Allee effect. In both papers the environmental noise was introduced via a modulation of a parameter entering the reaction rates. In Ref. \cite{KMS2008} the constant parameter $r$ was replaced by $r-\xi(t)$, where $\xi(t)$ is a ``red" (that is, positively correlated) Gaussian random process -- the Ornstein-Uhlenbeck noise \cite{Gardiner} -- with zero mean, variance $v\ll \mu^2$ and correlation time $t_c$. The same type of noise was adopted in Ref. \cite{LM2013}. Both papers assumed a regime of parameters close to the bifurcation (transcritical and saddle-node, respectively) of the corresponding deterministic equations.  In these cases the master equation for the evolution of the population size distribution \textit{without} environmental noise can be well approximated by a Fokker-Planck equation. The latter is equivalent to a Langevin equation, so that the environmental noise  $\xi(t)$ can be conveniently introduced in the birth-death term \cite{LM2013}. Importantly, the Ornstein-Uhlenbeck noise $\xi(t)$
obeys a Langevin equation of its own,
\begin{equation}\label{Langevin1}
\dot{\xi}(t)=-\frac{\xi(t)}{t_c}+ \sqrt{\frac{2 v}{t_c}}\,\eta(t),
\end{equation}
where $\eta(t)$ is a Gaussian \textit{white} noise with zero mean and $\langle \eta(t_1) \eta(t_2)\rangle =\delta(t_1-t_2)$.  The resulting two-component Langevin equation leads to a two-dimensional Fokker-Planck equation for the joint probability $P(n,\xi,t)$ of observing a certain population size $n$ and a certain value of the noise $\xi$ at time $t$. Applying the WKB approximation to this two-dimensional Fokker-Planck equation, one arrives at a two-dimensional effective mechanical problem where, in the context of population extinction,  one should again look for an instanton solution. In general, this two-dimensional problem can only be solved numerically. Perturbative analytical solutions can be obtained in the limits of short-correlated, long-correlated and weak environmental noise. The details can be found in Refs. \cite{KMS2008,LM2013}. Here we will only summarize the main results of these two works. In both cases the environmental noise causes an exponentially large reduction of the MTE. At fixed variance of the environmental noise, positive correlations quicken extinction.

\begin{figure}[ht]
\includegraphics[width=10.0 cm,clip=]{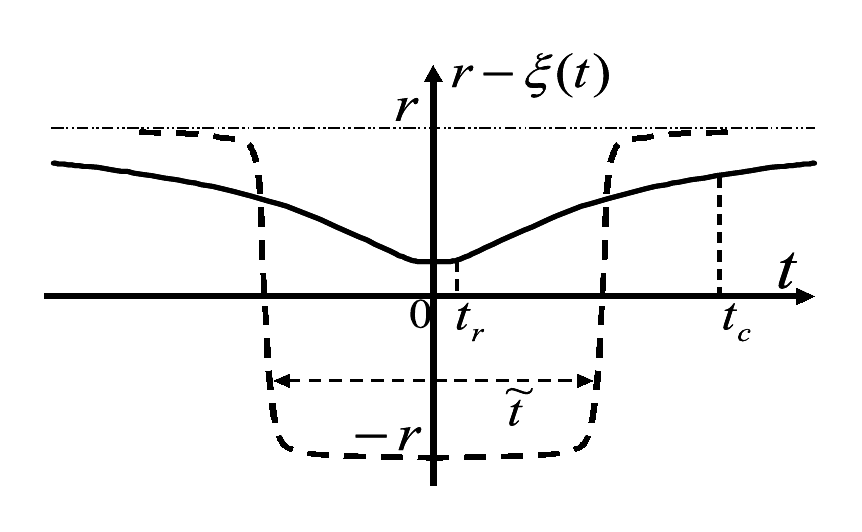}
\caption{Optimal realizations of the environmental noise in the limit of long (the solid line) and short (the dashed line) noise correlations \cite{KMS2008}. $t_r=1/r$ is the characteristic relaxation time of the deterministic theory. The duration of the ``catastrophe'' for the short-correlated noise is $\tilde t\simeq t_r \ln (Kv t_c/\mu)$, where
$K=r/a$.}
\label{fig3KMS2008}
\end{figure}

In the absence of the Allee effect the population-size dependence of the MTE changes from exponential without noise to a power law for a strong short-correlated noise and to almost no dependence for a long-correlated noise \cite{KMS2008}. The power-law dependence of the MTE on the population size for a strong white environmental noise in the absence of the Allee effect was known for quite a while \cite{Leigh,Lande}. The WKB theory also gives the optimal environmental fluctuation -- a non-random function $\xi(t)$ --
that mostly contributes to the population extinction.  The optimal fluctuation looks quite differently in different regimes, see Fig. \ref{fig3KMS2008}.
For a long-correlated environmental noise the effective birth rate slowly goes down with time, and then slowly recovers to its original value. In this case most of the ``extinction current" (the probability current to the absorbing state) is observed around the time when the effective birth rate is at a minimum, as to be expected on the physical grounds.

For short-correlated noise the optimal fluctuation is less intuitive. Surprisingly, it has a form of a catastrophe, when the effective birth rate abruptly drops on a certain time interval (which the theory predicts), stays almost constant and then returns, again abruptly, to its original value \cite{KMS2008}. By predicting the optimal realizations of the environmental noise in different regimes,  the WKB method provides, in addition to the MTE,  an instructive visualization and a better understanding of the rare events where the environmental and demographic noises
``coconspire"  to bring the population to extinction most effectively.

In the presence of a strong Allee effect
the
population-size dependence of the MTE changes from exponential for weak environmental noise to (approximately) no dependence
for strong noise, implying a greatly increased extinction risk \cite{LM2013}.  The exponential-to-power-law crossover of the MTE versus the population size, observed in the absence of the Allee effect,  does not  happen in the presence of the Allee effect.
Here again, at fixed variance of the environmental  noise, the noise correlations quicken extinction.

\begin{figure}[ht]
\includegraphics[scale=1.5]{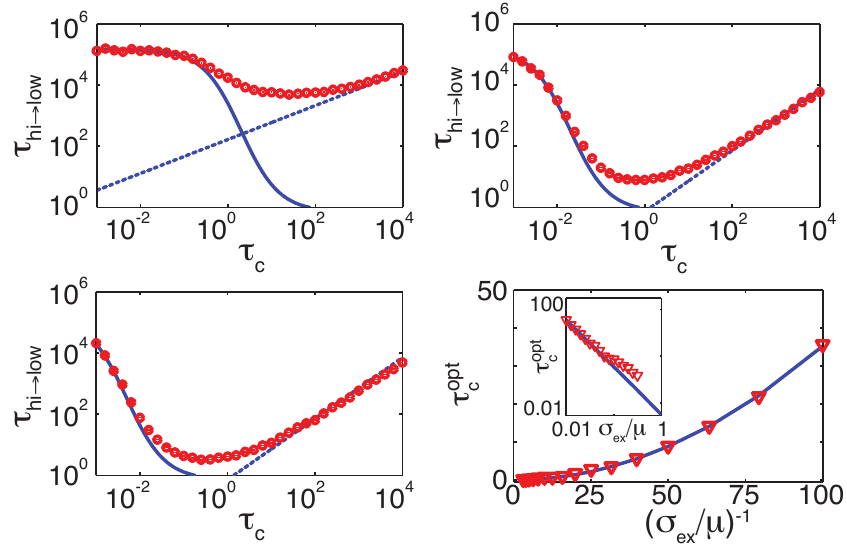}
\caption{The MST versus the correlation time, $\tau_c$, of the extrinsic noise  for different variances of the extrinsic noise (top left and right, bottom left). $\sigma_{\text{ex}}/\mu=0.01$ (top left), $0.1$ (top right) and $0.2$ (bottom left), where $\mu$ and $\sigma_{\text{ex}}$ are the extrinsic noise's mean and standard deviation, respectively, while ``low" and ``hi"  denote two long-lived metastable states, see Sec.~\ref{switchingB}. The solid lines represent the WKB solution,  the dashed lines represent an asymptotic analytical result in the \textit{non}-WKB regime of strong adiabatic extrinsic noise. Here the perturbed action seizes to be large, and the MST is governed by the mean time it takes the Ornstein-Uhlenbeck process to reach such noise magnitude that enables almost instantaneous switching~\cite{ARSG2013}. The bottom right panel shows the optimal correlation time as a function of the coefficient of variation, $\sigma_{\text{ex}}/\mu$ of the extrinsic noise. Parameters are $N=5000$ and $\delta=0.07$, such that the action in the absence of extrinsic noise satisfies $\Delta S_0\simeq N\delta^2/2=12.5$, see Ref.~\cite{ARSG2013} for details.}\label{ENcorrelation}
\end{figure}

Apart from population biology, environmental, or extrinsic noise may play an important role in  gene regulation~\cite{ELSS2002}.  The extrinsic noise results from a multitude of factors such as  variation in the cells' number of ribosomes, transcription factors and polymerases,
fluctuations in the cell division time, as well as environmental fluctuations. The effect of extrinsic noise on the
MST of a model gene regulatory network with positive feedback, see Sec.~\ref{switchingB}, was studied in Ref.~\cite{ARSG2013}.  As in Refs.~\cite{KMS2008,LM2013}, the authors assumed the Ornstein-Uhlenbeck extrinsic noise and worked close to the saddle-node bifurcation of the deterministic model. They derived a two-dimensional Fokker-Planck equation for the joint probability $P(n,\xi,t)$ of observing a protein copy-number $n$ and extrinsic noise magnitude $\xi$ at time $t$.
As Refs.~\cite{KMS2008,LM2013}, they observed that the extrinsic noise causes an exponential decrease of the MST between metastable states. However, going beyond WKB approximation,  they also found that, at fixed variance of the extrinsic noise, there is an optimal correlation time,  $\tau_c^{opt}$, for which the switching occurs at the maximum rate.
A simple calculation, comparing the relative contributions of the short- and long-correlated extrinsic noise to the MST, shows that $\tau_c^{opt}$ is inversely proportional to the extrinsic noise variance~\cite{ARSG2013}, see Fig.~\ref{ENcorrelation}.

The treatment of environmental noise in Refs.~\cite{KMS2008,LM2013,ARSG2013} can be generalized to a non-gaussian noise, as well as to escape from a metastable state when the system is not necessarily close to bifurcation. For a non-gaussian noise the mean, variance and correlation time do not define the environmental noise uniquely.  For gene regulatory networks one often observes reaction rates that are distributed according to a negative binomial distribution or a gamma distribution. From a theoretical viewpoint, such distributions have an advantage over the Gaussian distribution, as for the former distributions the reaction rates cannot reach a non-positive value, while the Gaussian distribution must be trimmed~\cite{RBBSA2015}. As an example, consider a gene regulatory network which is responsible for production of a protein of interest, and where the extrinsic variable is an auxiliary protein that affects the degradation rate of the protein of interest. Such an extrinsic noise variable $\xi$ with, say, a negative-binomial distribution, mean $\langle\xi\rangle=1$, and variance $\sigma_{\text{ex}}^2$ can be generated by defining a discrete variable $k=K\xi$, where $K\gg 1$ is a large integer, and using the following master equation
\begin{eqnarray}\label{auxmasterMain}
\dot{P}_{m,k}(t)&=&\frac{\alpha}{\tau_c}(P_{m-1,k}\!-\!P_{m,k})+\frac{\omega}{\tau_c}[(m+1)P_{m+1,k}\!-\!mP_{m,k}]\nonumber\\
&+&\frac{\omega \beta m}{\tau_c}(P_{m,k-1}\!-\!P_{m,k})+\frac{1}{\tau_c}[(k+1)P_{m,k+1}\!-\!kP_{m,k}]\!.
\end{eqnarray}
Here, $P_{m,k}(t)$ is the probability to create $m$ auxiliary mRNA molecules and $k$ auxiliary proteins at time $t$.  This master equation describes the stochastic dynamics of the auxiliary circuit which includes mRNA transcription at a rate $\alpha/\tau_c$, mRNA degradation at a rate $\omega/\tau_c$, protein translation at a rate $\omega\beta/\tau_c$ and protein degradation at a rate $1/\tau_c$~\cite{PE2000,ShS2008}. In the limit of short-lived mRNA, such that
$\omega\gg 1$, it can be shown that the stationary probability distribution of the auxiliary protein is~\cite{PE2000}
\begin{equation}
P_k=\frac{\Gamma(\alpha+k)}{\Gamma(k+1)\Gamma(\alpha)}\left(\frac{\beta}{\beta+1}\right)^k\,\left(\frac{1}{\beta+1}\right)^{\alpha},
\end{equation}
where $P_k$ is the probability to find $k$ auxiliary proteins. The mean of this distribution is $\alpha\beta$, while its variance is $\alpha\beta(1+\beta)$. As a result, since $\xi=k/K$, by choosing $\alpha=1/(\sigma_{\text{ex}}^2-1/K)$ and $\beta=K\sigma_{\text{ex}}^2-1$ we find that $\langle\xi\rangle=\langle k\rangle/K=1$, and the variance of $\xi$ becomes the variance of $k$ over $K^2$ which equals $\sigma_{\text{ex}}^2$ as required. Finally, the correlation time of $k$ or $\xi$ is given by $\tau_c$ as required.

Using this definition of the extrinsic-noise variable, one can formulate a two-dimensional master equation for $P(n,k,t)$ -- the probability to find $n$ proteins, and extrinsic noise magnitude $\xi=k/K$, at time $t$. This master equation can be analyzed by using a WKB approximation. The resulting Hamilton equations were solved for a short-correlated noise, by finding the optimal environmental fluctuation, and for a long-correlated noise, by adiabatic elimination of the fast variable~\cite{RBBSA2015,AM2008}.

\section{Large Deviations in Multi-Population Systems}

When the demographic noise is weak, the quasi-stationary distribution of an established population is sharply
peaked around an attractor of the corresponding deterministic theory. For example, for a generic two-population system
this can be either an attracting fixed point or a stable limit cycle. The use of the WKB approximation
for the analysis of multi-population systems, initially residing around an attracting fixed point, was pioneered in Ref. \cite{DMR1994} in the context of population switches, and in Refs.
\cite{DSL2008,MS2009} in the context of population extinction.  As of today, there is a large number of works in these two directions,
with applications ranging from population biology and ecology to dynamics of bacterial colonies to intracellular biochemistry.
One way of applying the WKB ansatz is to the master equation describing the evolution
of the probability $P_{m,n,\dots}(t)$ of observing $m$, $n$, $\dots$ individuals of each of the sub-populations at time $t$ \cite{DMR1994,RFRSM2005,DSL2008}. Alternatively, one can first
derive an evolution equation for the probability generating function $G(p_m,p_n, \dots, t)$ -- a linear PDE with multiple arguments -- and then apply the WKB ansatz \cite{KM2008}.
In  both cases the leading-order WKB approximation generates a multi-dimensional effective classical mechanical problem. Again, extinction of one or more of the sub-populations, or a switch from the vicinity of one attractor to another, is encoded
in an instanton-like trajectory.

\subsection{Established Populations Reside in a Vicinity of a Fixed Point}

Here we consider several examples from epidemiology, population dynamics  and cell biochemistry. In each of these examples, in a WKB approximation, the problem of evaluating the mean time to extinction/switch boils down to finding an instanton in the phase space of the corresponding classical mechanics, and computing the action along it. As in one-population problems, the instanton is a  zero-energy heteroclinic trajectory which exits a deterministic fixed point and approaches either a non-trivial ``fluctuational" fixed point  or another deterministic fixed point. For a generic multiple-population system the instanton can only be found numerically, and we briefly review two existing numerical methods that have been proved useful. We also show how one can use additional small parameters to obtain approximate analytical solutions.

\subsubsection{Extinction of endemic disease}

\label{extendemic}

Our first example deals with ``endemic fadeout": spontaneous extinction of an infectious disease from a population
after the disease has become endemic. The disease extinction ultimately occurs if the infectives recover, leave or die, while no new
infectives are introduced into the susceptible population. The WKB approximation enables one to
evaluate the MTE of the disease. Following Ref. \cite{KM2008}, we will consider the
SI (Susceptible-Infected) model of epidemiology with renewal of susceptibles \cite{SI,Grasman,vH,KM2008,MS2009}. In this
model the population is divided into two dynamic sub-populations: susceptible (S) and infected (I).  The set of processes and their rates are given in Table \ref{tableKM}.
\begin{table}[ht]
\centering 
\begin{tabular}{|c c c|} 
\hline\hline 
Process & Transition & Rate \\ [0.5ex] 
\hline 
Infection & $S\to S-1, \, I\to I+1$ &  $(\beta/N) SI$\\
Renewal of susceptible & $S\to S+1$ & $\mu N$ \\
Removal of susceptible & $S\to S-1$ & $\mu S$ \\
Removal of infected & $I\to I-1$ & $\mu_I I$ \\[1ex] 
\hline 
\end{tabular}
\caption{Stochastic SI model with renewal of susceptibles}
\label{tableKM} 
\end{table}

The deterministic rate equations for the SI model can be written as
\begin{eqnarray}
\dot S&=& \mu N-\mu S-(\beta/N) S I\,, \label{Sdot}\\
\dot I&=&-\mu_I I+(\beta/N) S I, \label{Idot}
\end{eqnarray}
For a sufficiently high infection rate,
$\beta > \mu_I$, there is a stable fixed point on the $S,I$ plane,
\begin{equation}\label{stablefp}
    \bar{S} = \frac{\mu_I}{\beta}\,N\,,\;\;\;\;\;\;\bar{I} = \frac{\mu (\beta-\mu_I)}{\beta\mu_I}\,N,
\end{equation}
which describes an endemic state, and an unstable fixed point $\bar{S}=N, \,\bar{I}=0$ (a saddle) which describes an infection-free steady-state  population. At $\mu < 4 \,(\beta-\mu_I) (\mu_I/\beta)^2$
the stable fixed point is a focus;
for the opposite inequality it is a node. The inverse of the real part of the eigenvalues (for the focus), or the inverse of the smaller of the eigenvalues (for the node) yields the characteristic relaxation time $t_r$ towards the ``endemic fixed point".

The master equation for the probability $P_{n,m}(t)$ of finding $n$ susceptible and $m$ infected individuals at time $t$ has the form
\begin{eqnarray}
\label{masterSI}
\dot{P}_{n,m}\!\!&=& \sum\limits_{n^{\prime}, \, m^{\prime}} M_{n, \, m;\, n^{\prime}, \, m^{\prime}}\, P_{n^{\prime}, \, m^{\prime}}(t)\nonumber \\
&=&\!\!\mu \left[N(P_{n-1,m}-P_{n,m})
+(n+1)P_{n+1,m}-n P_{n,m}\right]+\mu_I\left[(m+1)P_{n,m+1}-mP_{n,m}\right]\nonumber\\
&+&\!\! (\beta/N) \left[(n+1) (m-1)P_{n+1,m-1}-nm P_{n,m}\right],\;\;\;\;\mbox{for}\; m>0,
\end{eqnarray}
and
\begin{equation}\label{master0SI}
\dot{P}_{n,0} =\mu_I P_{n,1} .
\end{equation}
Following Ref. \cite{KM2008}, we use Eq.~(\ref{masterSI}) to obtain an exact evolution equation for the probability generating function $G(p_S,p_I,t)=\sum_{n,m=0}^{\infty}p_S^{n}p_I^{m}P_{n,m}(t)$. This equation reads $\partial_t G=\hat{\mathcal H}G$ with the effective Hamiltonian operator
\begin{equation}\label{hamiltonian}
\hat{\mathcal H}=\ \mu (p_S-1)(N-\partial_{p_S})-\mu_I (p_I-1)\partial_{p_I}
-(\beta/N) (p_S -p_I)p_I\partial^2_{p_S p_I} \,.
\end{equation}
Using the WKB ansatz $G(p_S,p_I,t)=\exp[-\mathcal{S}(p_S,p_I,t)]$ and neglecting the second derivatives of the action $\mathcal{S}$ with respect to $p_S$ and $p_I$, we arrive at a Hamilton-Jacobi equation $\partial_t{\cal S}+H=0$ in the $p$-space
with the classical Hamiltonian
\begin{equation}\label{hamiltonian1}
H(S,I,p_S,p_I)= \mu (p_S-1)(N-S) - \mu_I (p_I-1)I -(\beta/N) (p_S-p_I)p_I SI\,,
\end{equation}
where $S=-\partial_{p_S} {\cal S}$  and $I=-\partial_{p_I} {\cal S}$. As $H$ does not depend explicitly on time, it is an integral of motion: $H(S,I,p_S,p_I)=E=const$. The deterministic trajectories, described by Eqs.~(\ref{Sdot}) and (\ref{Idot}), lie in the zero-energy, $E=0$, two-dimensional plane $p_S=p_I=1$. We are interested in extinction starting from the long-lived, quasi-stationary endemic state. This requires an instanton: a non-deterministic activation  trajectory. The stable fixed point (\ref{stablefp}) of the deterministic theory becomes
a hyperbolic point $M_2=[\bar{S},\bar{I},1,1]$ in the four-dimensional phase space with two stable and two unstable eigenvalues (the sum of which is zero) and respective eigenvectors. There are two more zero-energy fixed points that describe an infection-free population: the point $M_1=[N,0,1,1]$, which is also present in the deterministic description, and  the ``fluctuational" fixed point $F=[N,0,1,\mu_I/\beta]$.

Let us sum up Eq.~(\ref{master0SI}) over $n$. We obtain
\begin{equation}\label{mastersumSI}
\frac{d}{dt} \sum_{n=0}^{\infty}P_{n,0} =\mu_I \sum_{n=0}^{\infty} P_{n,1} .
\end{equation}
The left hand side in Eq.~(\ref{mastersumSI}) describes the growth rate of the probability of disease extinction with time, whereas
the right hand side is equal to $\mu_I$ times the probability of observing exactly one infected individual (at any number of
susceptibles). At times much longer than the characteristic relaxation times of the deterministic theory, the probability distribution $P_{n,0}$ is sharply peaked close to the deterministic fixed point $(N,0)$ corresponding to the infection-free steady-state population. Essentially, by using the leading-order WKB approximation, one replaces the sums
$\sum_n P_{n,0}$ and $\sum_n P_{n,1}$ by their maximum terms $P_{N,0}$ and  $P_{N,1}$, respectively,  for the purpose of calculating the probability flux to the infection-free absorbing state. Correspondingly, the instanton -- the optimal path of the disease extinction -- is a heteroclinic trajectory that exits, at $t=-\infty$, the ``endemic" fixed point $M_2$  along its two-dimensional  unstable zero-energy manifold in the four-dimensional phase space and enters, at $t=\infty$, the fluctuational extinction point $F$, along its two-dimensional  stable manifold \cite{KM2008,DSL2008}.

Up to a pre-exponent, the MTE of the disease is $\tau\propto \exp(\mathcal{S}_0)$, where
\begin{equation}\label{action1a}
   \mathcal{S}_0 = \int_{-\infty}^\infty (p_S \dot{S}+p_I \dot{I})\,dt\,,
\end{equation}
and the integration is performed along the instanton going from $M_2$ to $F$. Following Ref. \cite{KM2008} we introduce  new ``coordinates" $x=S/N-1$ and $y=I/N$, time $\tilde{t}=\mu t$, new momenta $p_{x,y}=p_{S,I}-1$ and  the bifurcation parameter $\delta=1-\mu_I/\beta$, $0<\delta<1$. The action (\ref{action1a}) can now be rewritten as
${\cal S}_0=N  \sigma $, where
$ \sigma(K,\delta)$ is the action along the proper instanton of the rescaled Hamiltonian
\begin{equation}\label{hamiltonian2}
\tilde{H}= -p_x x -K \left[(1-\delta) p_y+(p_x-p_y)(p_y+1)(x+1)\right]y
\end{equation}
and $K=\beta/\mu>1$. The fixed points $M_1$, $M_2$ and $F$ become
$$
[0,0,0,0],\;\;\left[-\delta,\;\; \frac{\delta}{K(1-\delta)},0,0\right]\;\;\mbox{and}\;\;[0,0,0,-\delta],
$$
respectively.

\begin{figure}[ht]
  \includegraphics[scale=0.68]{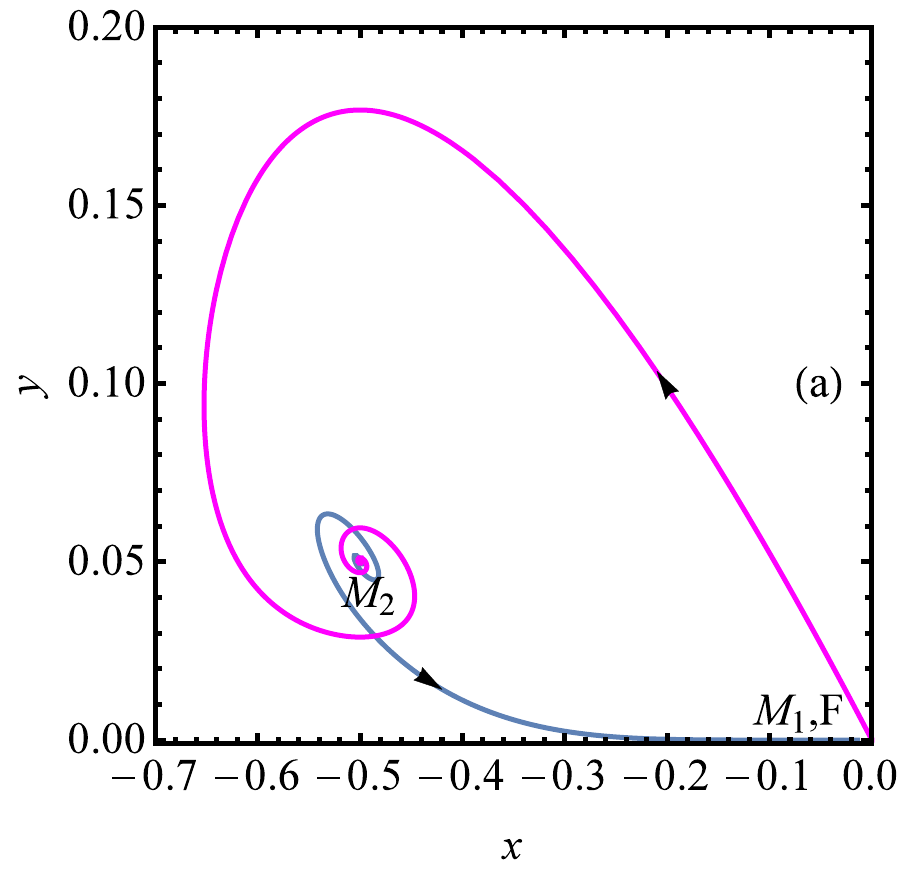}
  \includegraphics[scale=0.70]{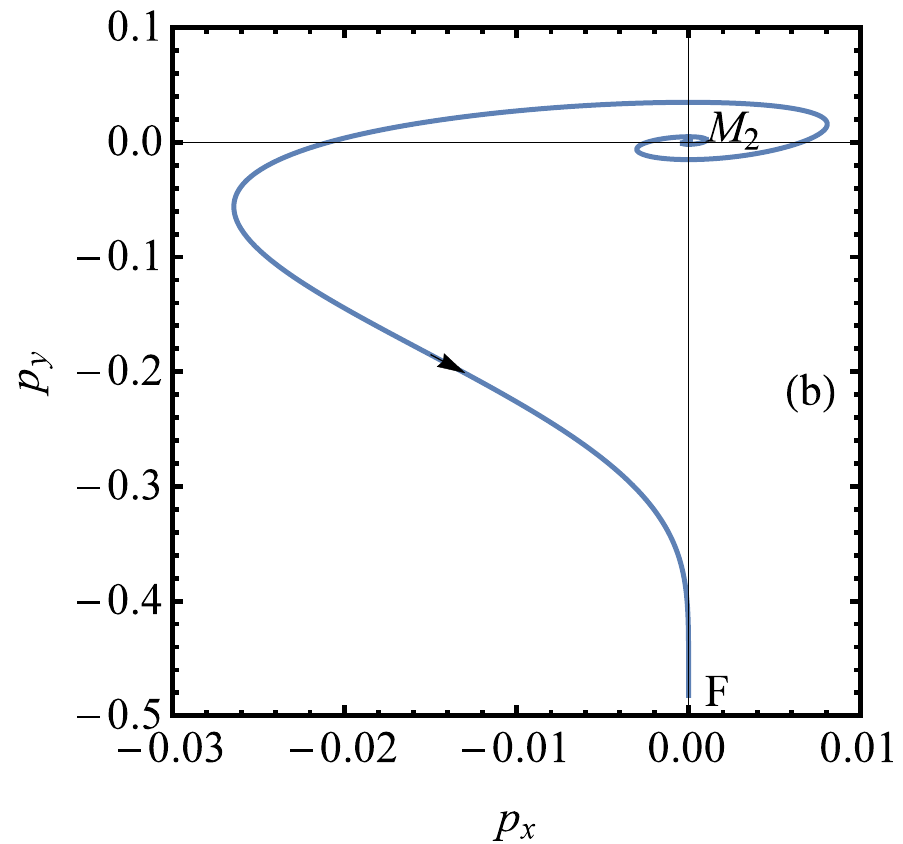}
\caption{(a) The $x,y$ projection of the four-dimensional optimal path in the phase space of the SI model with population turnover  (the lower curve) and the deterministic trajectory ($p_x=p_y=0$) describing an epidemic outbreak (the upper curve). (b) The $p_x,p_y$ projection of the optimal path.
$x=S/N-1$, $y=I/N$;  $\,\,\,K=20$ and $\delta=0.5$.} \label{typical1}
\end{figure}

The endemic extinction instanton can be found numerically: either by the shooting method \cite{RFRSM2005,DSL2008,KM2008}, or by the iteration method \cite{ChernykhStepanov}. The shooting method works as follows. One
linearizes the Hamilton equations near the ``endemic" fixed point and finds the two unstable eigenvectors.
The instanton is then found by looking for the correct linear combination of the two unstable eigenvectors of a
fixed (and very small) norm. An example of an instanton, found by the shooting method in Ref. \cite{KM2008}, is shown in Fig.~\ref{typical1}. In this case $4 K \delta (1-\delta)^2>1$, and the  endemic fixed point $M_2$ is a focus. For comparison, Fig.~\ref{typical1}a also shows the deterministic heteroclinic trajectory ($p_x=p_y=0$) connecting the fixed points $M_1$ and $M_2$. It describes, in deterministic terms, an epidemic outbreak following an introduction of a small number of infectives into a steady-state susceptible population. Deterministically, this outbreak always leads to an endemic state. In contrast to equilibrium systems, the $x,y$ projection of the optimal path of a large fluctuation is different from the corresponding time-reversed relaxation path.  As one can see, the difference between the ``activation spiral" and the time-reversed ``relaxation spiral" in this example is striking.

The shooting method  usually works well for well-mixed two-population systems (and for one-population systems with time-dependent rates, see Fig.~\ref{timedependent-action} in Sec. \ref{demenv}). For three-population systems it becomes much less convenient, as now one needs to do shooting with respect to two parameters. The shooting method of course becomes impractical for spatially explicit systems (which, after discretization, represent multi-dimensional systems with a very large number of ``dimensions"). Here the back-and-forth iteration method comes to rescue. The iteration method was suggested by Chernykh and Stepanov \cite{ChernykhStepanov}  in the context of the instanton theory of the ``Burgers turbulence" \cite{Migdal,Balkovsky}. Then, starting from Ref. \cite{EK2004},  it was adopted for population dynamics and for a host of other stochastic systems where a WKB approximation is used, and the ensuing Hamilton's equations need to be solved. The  back-and-forth iteration method is applicable if the boundary conditions in time involve the knowledge of $x$ and $y$ at the initial time and  $p_x$ and $p_y$ at the final time. For zero-energy instantons the initial time is at $-\infty$, and the final time is at $+\infty$. The iterations proceed as follows. One solves the Hamilton equations for $\dot{x}$ and $\dot{y}$ forward in time (with the initial conditions for $x$ and $y$ and with $p_x(t)$ and $p_y(t)$ from the previous iteration), and equations for $\dot{p}_x$ and $\dot{p}_y$ backward in time (with the ``initial" conditions for $p_x$ and $p_y$ and with $x(t)$ and $y(t)$ from the previous iteration). At the very first iteration reasonable ``seed functions" for $p_x(t)$ and $p_y(t)$ are used. A recent review \cite{Grafke} presents a detailed discussion of the Chernykh-Stepanov algorithm and its modifications and extensions in the context of  hydrodynamic turbulence.

Let us return to the endemic fadeout in the SI model. At $\delta \ll 1$  -- close to the bifurcation point corresponding to the emergence of an endemic state -- the system exhibits time-scale separation,
and the optimal path to extinction and the action  can be calculated analytically \cite{KM2008,DSL2008}. Introducing rescaled  variables $q_1=x/\delta$, $q_2=yK/\delta$,
$p_1=p_x/\delta^2$, and $p_2=p_y/\delta$ and neglecting higher order terms in $\delta$, one obtains the following approximate equations of motion:
\begin{eqnarray}
\left.\begin{array}{rcl}
\dot{q}_1 &=& -q_1-q_2\,,
\\
\dot{p}_1 &=& p_1 - q_2 p_2\,,\\
\end{array}\right.
\left.\begin{array}{rcl}
\dot{q}_2 &=& K \delta \, q_2(1+q_1+2 p_2)\,,\\
\dot{p}_2 &=& K \delta \,(p_1-p_2-p_2^2-q_1p_2)\,.
\end{array}\right.
\label{4equations}
\end{eqnarray}
The fixed points become $M_1=[0,0,0,0]$, $M_2=[-1,1,0,0]$, and $F=[0,0,0,-1]$. For $K\delta\ll 1$ the subsystem $(q_1,p_1)$ is fast, whereas  $(q_2,p_2)$ is slow. On the fast time scale (that is, the time scale $\mu^{-1}$ in the original, dimensional variables) the fast subsystem approaches the state $q_1\simeq -q_2$ and $p_1 \simeq q_2 p_2$ which then slowly evolves according to the equations
\begin{equation}\label{2equations}
    \dot{q}_2 \simeq K \delta \, q_2(1-q_2+2 p_2)\,,\;\;\;\dot{p}_2 \simeq K \delta \,p_2(2q_2 -1-p_2)
\end{equation}
that are Hamiltonian, as they follow from the reduced Hamiltonian $ H_r(q_2,p_2)=K\delta\;q_2 p_2 (1-q_2+p_2)$.
This universal Hamiltonian  appears in the theory of a whole class of \textit{single}-species models in the vicinity of the transcritical bifurcation point \cite{Kamenev2}, see also Sec. \ref{catastrophe}.
As $H_r(q_2,p_2)$ is independent of time, it is an integral of motion.  The optimal extinction path goes along the zero-energy trajectory $1-q_2+p_2=0$. Evaluating the action (\ref{action1a}) along this trajectory, we find in the leading order:
${\cal S}_0\simeq \left[N \delta^3/(K\delta)\right]\int_1^0 p_2dq_2= N\delta^2/(2K)$. For the MTE of the disease we obtain
\begin{equation}\label{tauSI}
 \ln (\tau)/N \simeq \delta^2/(2 K) = [\mu/(2 \beta)]\left(1-\mu_I/\beta\right)^2;
\end{equation}
this asymptote is valid when ${\cal S}_0\gg 1$. Many other stochastic multi-population models exhibit a similar time-scale separation close to the bifurcation, defining a universality class \cite{Kamenev2,DSL2008}.

\subsubsection{Speeding up endemic disease extinction with a limited vaccination}

A common way of fighting epidemics is vaccination. Mathematical modeling of different aspects of vaccination has attracted considerable attention from mathematicians and, more recently, from physicists \cite{D'Onofrioreview}. Here we will briefly consider only one aspect of vaccination, following Ref. \cite{KDM2010}. If there is enough vaccine (and the susceptible individuals are willing to be vaccinated), the infection may be often eradicated ``deterministically". The amount of available vaccine, however,
can be insufficient. The vaccine can be expensive, or dangerous to store in large amounts, or it can be effectively short-lived because of mutations of the infection agent.
Finally (and unfortunately), many people nowadays opt out of vaccination programs for reasons unrelated to science. In some cases spontaneous extinction of an endemic disease from a population can still be greatly accelerated even if only a fraction of susceptible individuals are vaccinated \cite{DSL2008,KDM2010}. By reducing
the number of susceptible individuals, the vaccination perturbs the instanton (which describes the optimal path toward the disease extinction in the absence of vaccination), and leads to a reduction of the ``entropic barrier" to extinction. The mathematical solution of this problem boils down to an optimization problem where one maximizes
the reduction of the classical action for given constraints on the vaccine \cite{KDM2010}.  It turns out that, if the available amount of vaccine is constrained by a given average vaccination rate, the optimal vaccination protocol turns out to be model-independent for a whole class of epidemiological models (including the SI model with population turnover, considered in the previous subsubsection). Furthermore, if a vaccination protocol is periodic in time,  the optimal protocol
represents a periodic sequence of delta-like pulses. The disease extinction rate can strongly depend on the period of this sequence, and display
sharp peaks when the vaccination frequency is close to the characteristic frequency of the oscillatory deterministic dynamics of the epidemic
in the absence of demographic fluctuations, or to its sub-harmonics \cite{KDM2010}.

\subsubsection{Minimizing the population extinction risk by migration}
\label{minmigration}

In our next example we will follow Ref. \cite{KMKS2012} and evaluate the MTE of a metapopopulation which consists of several local populations occupying separate
``patches". A local population is prone to extinction due to the demographic noise.
A migrating population from another patch can dramatically delay the extinction. What is the
optimal migration rate that maximizes the MTE of the whole population? This question
was addressed in Ref. \cite{KMKS2012}. The authors considered $N$ local populations of individuals $A$ located on a connected network of patches $i=1,2,...,N$ with different carrying
capacities. The individuals
undergo branching
$A\to 2A$ with rate constant $1$ on each patch and annihilation $2A\to \emptyset$ with rate constant  $1/(\kappa_i K)$ on patch $i$.  It is assumed that $K\gg 1$. The parameters $\kappa_i=O(1)$, $i=1,2,\dots,N$, describe the difference among the local carrying capacities $\kappa_i K$. Each individual can also migrate between connected patches $i$ and $j$ with rate constant $\mu_{ij}=\mu_{ji}$. It is assumed that $\mu_{ij}  = \mu M_{ij}$, where elements
of $M_{ij}$ are of order unity.

The MTE of the meta-population, $\tau$,  is exponentially large in $K$ but finite.
How does $\tau$ depend on the characteristic migration rate $\mu$? At $\mu=0$ each local population goes extinct separately, and $\tau_{\mu=0}$ is determined by the patch with the greatest carrying capacity, $K_m=K \max_i\{\kappa_i\}$:
\begin{equation}\label{muequal00}
   \ln \tau_{\mu=0}/K \simeq 2(1-\ln 2) \,\max_i\{\kappa_i\}
\end{equation}
see Eq.~(\ref{MTEnoperiodic}). At very fast migration, $\mu \to \infty$, the local populations are fully synchronized: both at the level of the expected local carrying capacities, and at the level of large fluctuations leading to population extinction. The total carrying capacity of the meta-population, as derived from the deterministic rate equation for this model \cite{KMKS2012}, becomes $\bar{\kappa} K$, where
\begin{equation}\label{barkappa}
\bar{\kappa}=N^2 /\sum(\kappa_i^{-1}).
\end{equation}
As a result,
at $\mu\to \infty$ the meta-population  goes extinct as if it were occupying a single effective patch with the total rescaled carrying capacity $\bar{\kappa}$, that is
\begin{equation}\label{fastgen}
\ln \tau_{\mu\to \infty}/K \simeq 2(1-\ln 2)\, \bar{\kappa}.
\end{equation}
The main result of Ref. \cite{KMKS2012} is that, for unequal $\kappa_i$, $\tau$ reaches its maximum  at a finite (and very small) value of the migration rate. This fact is intimately related to synchronization of the most probable local extinction events that occurs already at very small but positive
migration rates. The synchronization makes $\tau$ close to that for a single patch with the \textit{combined} carrying capacity $K \sum_i\kappa_i$:
\begin{eqnarray}
  &&\ln \tau_{\mu \to +0}/K \simeq  2 (1-\ln 2) \sum_i \kappa_i . \label{S1N}
\end{eqnarray}
Now let us inspect the MTE in the cases of $\mu=0$, $\mu=\infty$ and very small but finite $\mu$, as described by Eqs.~(\ref{muequal00}), (\ref{fastgen}) and (\ref{S1N}), respectively.
As $\sum_i\kappa_i\geq\max_i\{\kappa_i \}$ and $\sum_i\kappa_i \ge \bar{\kappa}$ for any $\kappa_i$, the MTE must indeed reach a maximum at a finite value $\mu=\mu_*$, unless all the patches have the same carrying capacity. As was found in Ref. \cite{KMKS2012},
$\mu_*\ll 1$ and scales as $1/K$.

Now we expose the theory of Ref. \cite{KMKS2012} in some detail. For simplicity, we will limit ourselves to a system of only two patches. The deterministic rate equations are:
\begin{equation}\label{eq:mf}
\dot x= x-x^2-\mu x+\mu y ,\quad \dot y=y-\frac{y^2}{\kappa}+\mu x-\mu y,
\end{equation}
where $x$ and $y$ are the local population sizes rescaled by $\kappa_1 K$, and $\kappa=\kappa_2/\kappa_1$.  Equations (\ref{eq:mf}) have two fixed points: the unstable point $x_0=y_0=0$ that describes an empty system, and a stable point $[x_*(\kappa,\mu)>0,\, y_*(\kappa,\mu)>0]$ that describes an established meta-population.
At $\mu= 0$ one has $x_*=1$ and $y_*=\kappa$, whereas for  infinitely fast migration, $\mu \to \infty$,
\begin{equation}\label{harmonicmean}
    x_*=y_*=2 \kappa/(1+\kappa).
\end{equation}
The characteristic time $t_r$ of population establishment is determined by the smaller of the two eigenvalues  of the linear stability matrix of Eqs.~(\ref{eq:mf}) at the fixed point $(x_*,y_*)$.

In the stochastic formulation, the probability $P_{m,n}(t)$ to find $m$ individuals in patch 1 and $n$ individuals in patch 2 evolves in time according to the master equation
\begin{eqnarray}
\label{eq:master_master}
\dot{P}_{m,n}(t)=\hat{H} P_{m,n}&\equiv& (m-1)P_{m-1,n}+(n-1)P_{m,n-1} \nonumber \\
&+&\frac{(m+1)(m+2)}{2 K}P_{m+2,n}+\frac{(n+1)(n+2)}{2 \kappa K}P_{m,n+2} \nonumber \\
&+&\mu (m+1)P_{m+1,n-1}+\mu (n+1)P_{m-1,n+1} \nonumber \\
&-&\left[(1+\mu)(m+n)+ \frac{m(m-1)}{2K}+\frac{n(n-1)}{2\kappa K}\right] P_{m,n}.
\end{eqnarray}
The probability $P_{0,0}$ that the meta-population
goes extinct by time $t$ obeys the equation
\begin{equation}\label{probext}
    \dot{P}_{0,0}(t)=\frac{1}{K} P_{2,0}+\frac{1}{\kappa K} P_{0,2}.
\end{equation}

At $t\gtrsim t_r$,  $P_{m,n}(t)$ is sharply peaked at the local carrying capacities $m_*=Kx_*$ and $n_*=Ky_*$, corresponding to the stable fixed point $(x_*, y_*)$ of the deterministic theory. The subsequent slow decay of $P_{m,n}$ in time is determined by the lowest excited eigenmode $\pi_{m,n}$ of the master equation operator $\hat{H}$: $P_{m,n} (t) \simeq \pi_{m,n} \exp(-t/\tau)$. Simultaneously, a probability peak at $m=n=0$ grows with time: $P_{0,0}(t) \simeq 1- \exp(-t/\tau)$.
The inverse eigenvalue $\tau$ is an accurate approximation to the MTE. Since it is exponentially large with respect to $K\gg 1$, one can
neglect the right-hand-side of the eigenvalue problem $\hat{H} \pi_{m,n}=-\pi_{m,n}/\tau$ and consider
the quasi-stationary equation $\hat{H} \pi_{m,n}\simeq 0$. Once $\pi_{m,n}$ is found, the MTE can be determined from Eq.~(\ref{probext}):
\begin{equation}\label{tau}
    \tau= [\pi_{2,0}/K+\pi_{0,2}/(\kappa K)]^{-1}
\end{equation}
The WKB ansatz for $\pi_{m,n}$ is
\begin{equation}\label{WKB}
    \pi_{m,n} = \exp[-KS(x,y)],
\end{equation}
where $x=m/K$ and $y=n/K$. In the leading order in $1/K\ll 1$ one obtains a zero-energy Hamilton-Jacobi equation $H(x, y, \partial_x S, \partial_y S) = 0$ with  classical Hamiltonian
\begin{eqnarray}
H(x,y,p_x,p_y) &=& x\left(e^{p_x}-1\right)+\frac{x^2}{2} \left(e^{-2p_x}-1\right)
+y\left(e^{p_y}-1\right)+\frac{y^2}{2\kappa} \left(e^{-2p_y}-1\right)\nonumber \\
&+&\mu x\left(e^{-p_x+p_y}-1\right)+\mu y\left(e^{p_x-p_y}-1\right). \label{H}
\end{eqnarray}
The established population corresponds to the fixed point $M=(x_*, y_*, 0,0)$ of the Hamiltonian flow. Up to a pre-exponent, $\tau \sim \exp (K S)$,
where $S$ is the action along the instanton that exits, at time $t=-\infty$, the fixed point $M$ and approaches the fluctuational extinction point $F$ that, for the two-patch branching-annihilation model, is $(0,0,-\infty,-\infty)$. In general, the instanton and the action along it can be found only numerically. Analytical results are possible in the limits of small and large $\mu$. When $\mu\to +0$ the Hamiltonian (\ref{H}) becomes separable, and the extinction instanton can be easily found analytically leading to the MTE~(\ref{S1N}) with $N=2$. This MTE is exponentially large compared to the one obtained if one neglects migration completely, see Eq.~(\ref{muequal00}) with $\max \{\kappa_i\}=1$. The sharp increase of $\tau$ once slow migration is allowed results from synchronization of the most probable local extinction paths of the local populations.

\begin{figure}[ht]
\includegraphics[width=4.0 in,clip=]{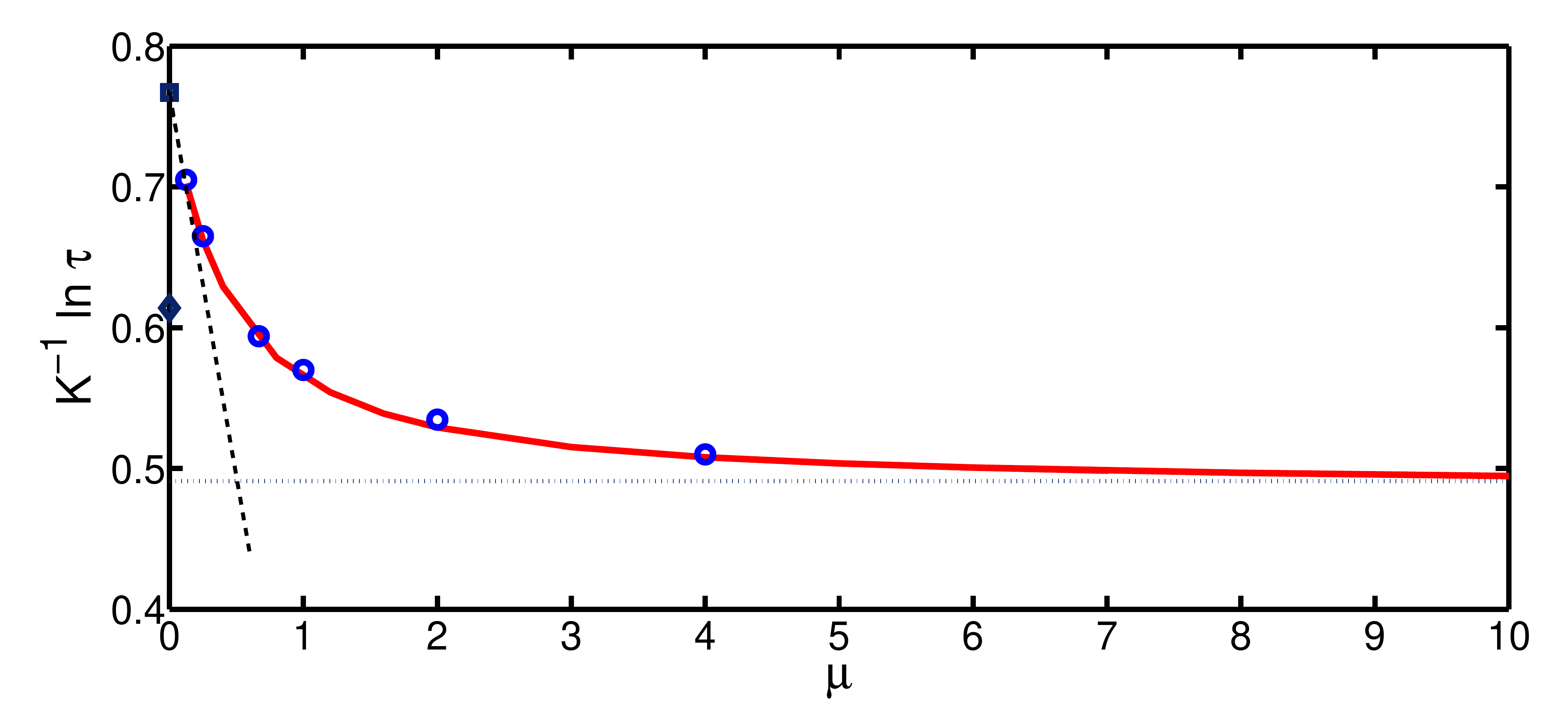}
\caption{$K^{-1}\ln \tau$ versus the migration rate $\mu$ for a two-patch metapopulation and $\kappa=0.25$ \cite{KMKS2012}. Circles: numerical WKB solutions.  Diamond and square: predictions of Eqs.~(\ref{muequal00}) and ~(\ref{fastgen}), respectively. Dashed line: prediction of linear theory for $\mu\ll 1$. Dotted line: prediction of the theory for $\mu\gg 1$. The solid line was obtained from a numerical solution of the master equation (\ref{eq:master_master}) for $K=220$.}
\label{II}
\end{figure}

The first order correction in $\mu$ to the action can be calculated perturbatively, similarly to the case of a weak environmental modulation, see subsubsection \ref{weakmodulation},  by integrating the terms proportional to $\mu$ of the classical Hamiltonian (\ref{H}) over the unperturbed ($\mu \to +0$) $x$- and $y$-instantons. In the opposite limit, $\mu \to \infty$, the total population size  $Q=x+y$  varies  slowly in comparison with the fast migration. The fast variables
$x$ and $y$ rapidly adjust to the slow dynamics of $Q$, staying close to their stationary values for the
instantaneous value of $Q$. Transforming to $Q$ and $q=x$ and associated conjugate momenta as a new set of canonical variables, one arrives at an effective \textit{one-dimensional}
Hamiltonian. The action along its instanton can be calculated analytically, and it yields Eqs.~(\ref{barkappa}) and (\ref{fastgen}) with $N=2$. It would be interesting to also calculate the subleading correction in $1/\mu$. Figure~\ref{II} shows the numerically found $S$ for $\kappa=0.25$ and different $\mu$. The numerical instantons were found in Ref. \cite{KMKS2012} by using both the shooting method and the iteration method, with very close results.

At $\mu\ll 1$ and $\mu\gg 1$ the numerical results agree with analytical predictions. Figure~\ref{II} also compares the WKB results with those of a numerical solution of (a truncated version of) the full master equation
(\ref{eq:master_master}).

\begin{figure}[ht]
\includegraphics[width=5.0 in,clip=]{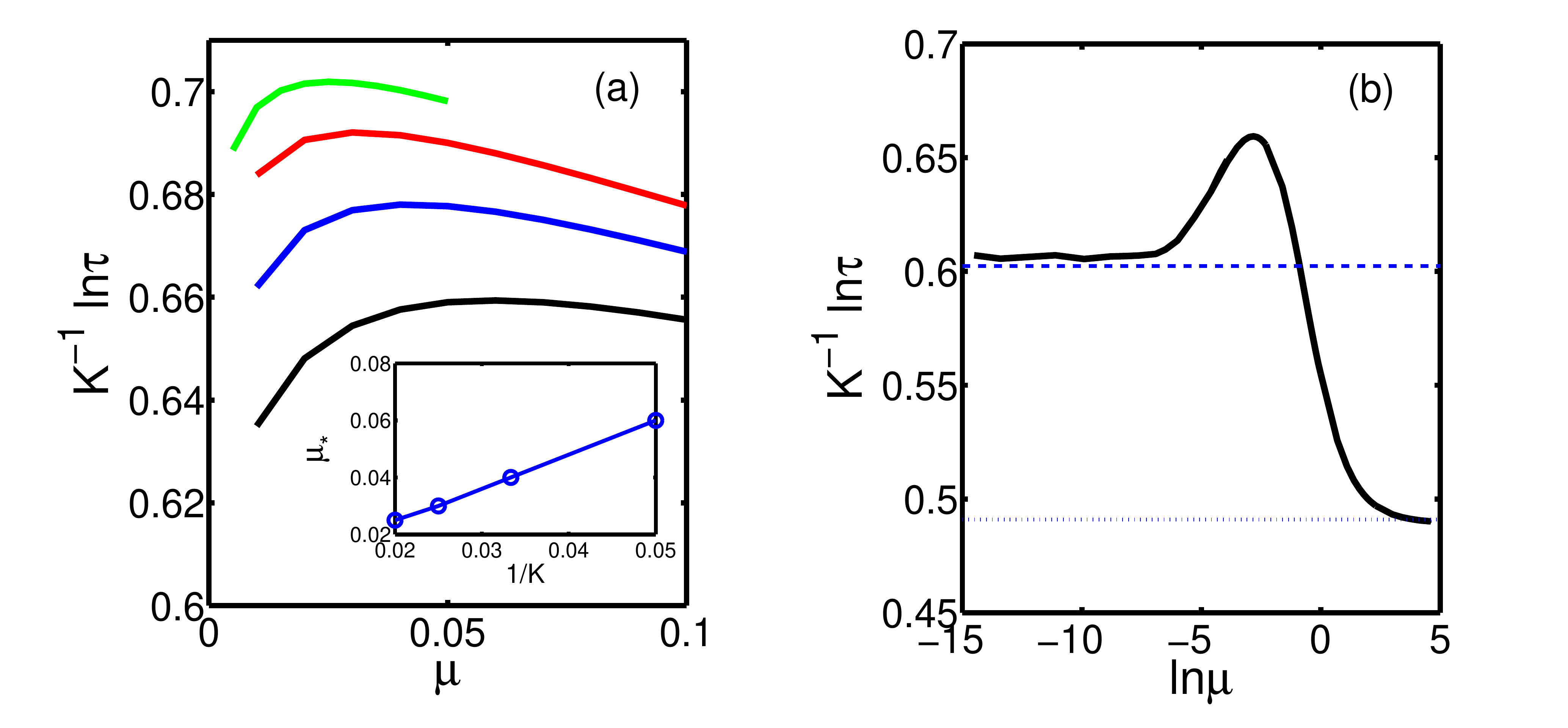}
\caption{$K^{-1}\ln \tau$ versus $\mu$ (a) and $\ln \mu$ (b)
for a two-patch meta-population from a numerical solution of the master equation and stochastic simulations. (a) $\kappa=0.25$ and $K=20, 30, 40$ and $50$ (bottom to top).  Inset: the migration rate $\mu_{*}$, at which the maximum of the MTE is observed, versus $K$. (b) $\kappa = 0.25$ and $K = 20$; dashed line: Eq.~(\ref{muequal00}) with $N=2$,
dotted line: Eqs.~(\ref{barkappa}) and (\ref{fastgen}) with $N=2$.}
\label{III}
\end{figure}

To evaluate the maximum MTE and the optimal migration rate, one needs to resolve the jump of $(\ln \tau)/K$ at $\mu=0$ predicted by the WKB theory, see Eqs.~(\ref{muequal00}) and~(\ref{S1N}). The authors of Ref. \cite{KMKS2012} determined the MTE for very small $\mu$ by  numerically solving the master equation (\ref{eq:master_master}) and by performing stochastic simulations.
The resulting $\mu$-dependence of the MTE, at $\kappa=0.25$ and different $K$, is shown in Fig.~\ref{III}. The maximum of $\tau$ is observed at a small migration rate $\mu_*$ that apparently scales as $K^{-1}$. This regime is beyond the WKB
approximation, and an analytical theory of this regime is currently unavailable.

\subsubsection{Multiple routes to extinction}
\label{multipleroutes}

In some population models extinction of a population may occur via more than one scenario. This situation
was considered in Ref. \cite{GM2012} on the example of a simple predator-prey model, see Table 2. This model
generalizes the celebrated Lotka-Volterra model \cite{LV}
by taking into account competition among prey.

\begin{table}[ht]
\centering 
\begin{tabular}{|c c c|} 
\hline\hline 
Process & Transition & Rate \\ [0.5ex] 
\hline 
Birth of rabbits & $R \rightarrow 2R$ 				& $aR$ \\ 
Predation and birth of foxes & $F+R \rightarrow 2F$	& $\frac{RF}{\Gamma N}$ \\ 
Death of foxes & $F \rightarrow \emptyset$ 					& $F$ \\ 
Death of rabbits & $R\to \emptyset$ &  $b R$\\
Competition among rabbits & $2R \rightarrow R$ 			& $\frac{R\left(R+1\right)}{2N}$  \\ [1ex] 
\hline 
\end{tabular}
\caption{Stochastic predator-prey model \cite{GM2012}}
\label{modelGM} 
\end{table}

By  re-interpreting the rabbits (R) as susceptibles (S) and the foxes (F) as infected (I), this model also describes extinction of an endemic disease in an isolated community \cite{GM2012}. As in the SI model  with population turnover, a susceptible individual can become infected upon contact with another infected, and the susceptibles and infectives are removed with constant per-capita rates. In the conventional SI model with population renewal, see subsubsection 1, the susceptibles arrive from outside. Instead, in the modified SI model, the susceptibles reproduce by division.  Finally, the susceptibles compete for resources, $2S\to S$, so their population size remains bounded.

\begin{figure}
\includegraphics[width=2.5 in,clip=]{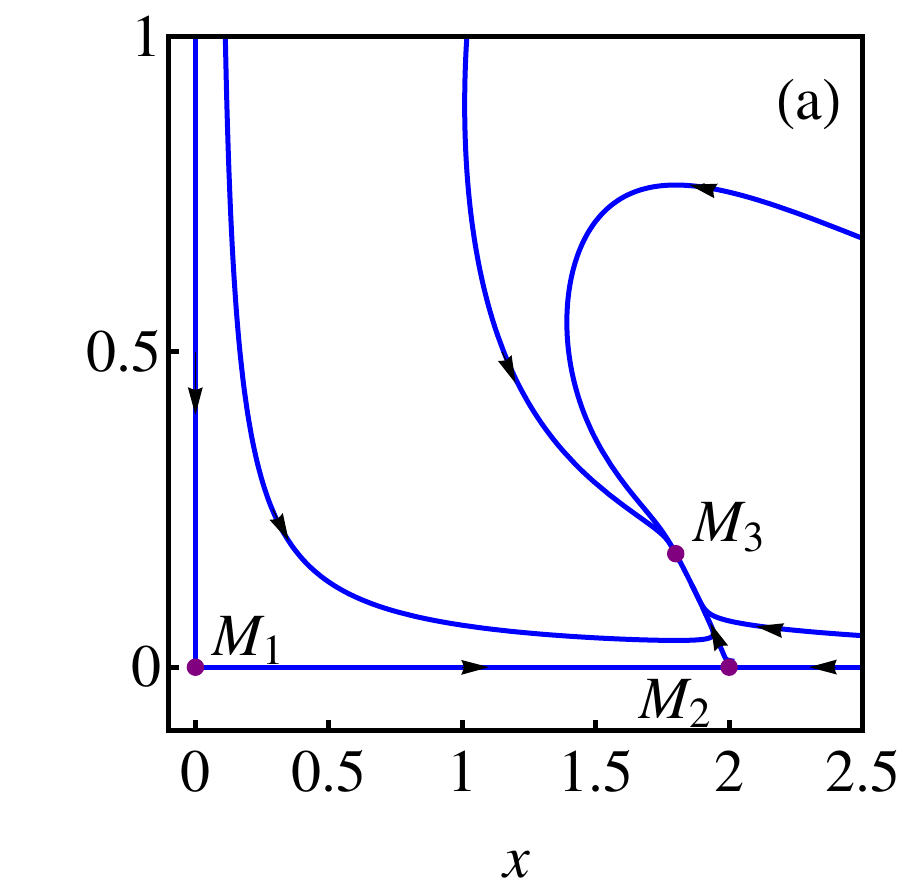}
\includegraphics[width=2.5 in,clip=]{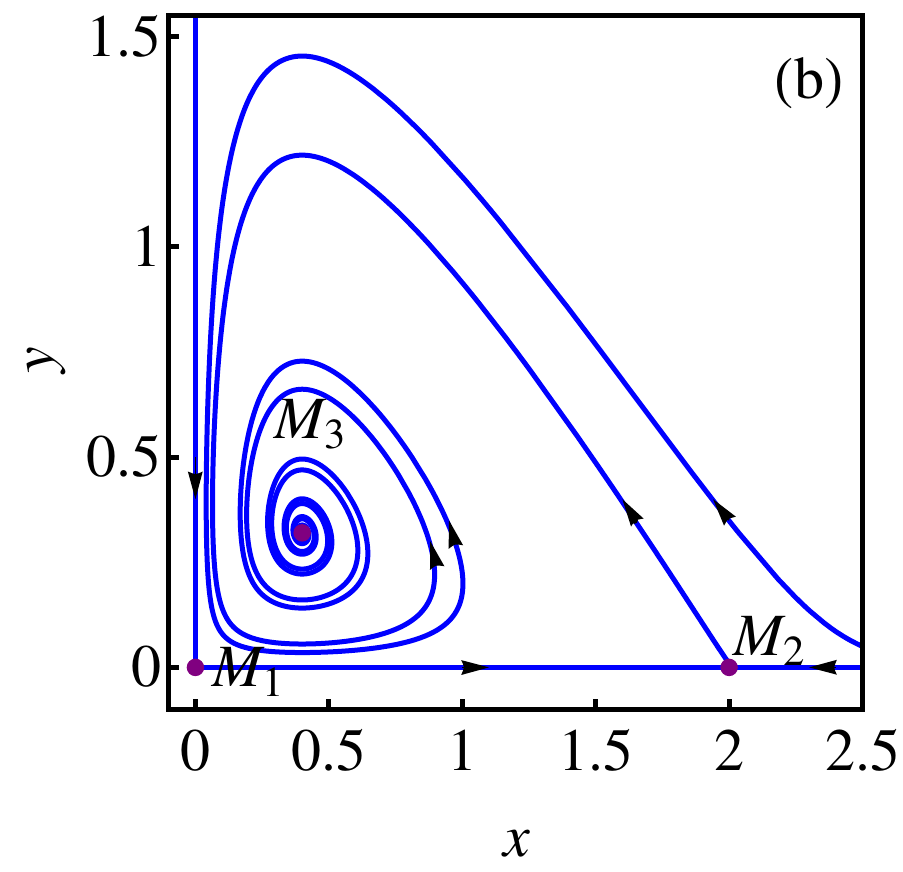}
\caption{Deterministic trajectories of the predator-prey model (see Table \ref{modelGM}) for $a=2$, $b=1$ and two different values of $\Gamma$. (a) $\Gamma=1.8$: $M_3$ is a stable node. (b) $\Gamma=0.4$:  $M_3$ is a stable focus.}
\label{MeanFieldPhase}
\end{figure}

Introducing the rescaled population sizes $x=R/N$ and $y=F/N$, one can write the deterministic rate equations as
\begin{equation}\label{xydot}
    \dot{x} = (a-b)x-\frac{xy}{\Gamma}-\frac{x^2}{2}\,,\;\;\;\;\;\dot{y}=\frac{xy}{\Gamma}-y\,.
\end{equation}
These equations are fully characterized by two parameters, $a-b$ and $\Gamma$. For $a>b$ and  $\Gamma<\Gamma_*\equiv 2(a-b)$, Eqs.~(\ref{xydot}) have three fixed points corresponding to non-negative population sizes. The fixed point $M_1$ ($\bar{x}_1=0$, $\bar{y}_1=0$) describes an empty system. It is a saddle point: attracting in the $y$-direction (no rabbits), and repelling in the $x$-direction. The fixed point $M_2$ [$\bar{x}_2=\Gamma_*$, $\bar{y}_2=0$] describes an established population of rabbits in the absence of foxes. It is also a saddle: attracting in the $x$-direction (no foxes), and repelling in a direction corresponding to the introduction of a small number of foxes into the system. The third fixed point $M_3$ [$\bar{x}_3=\Gamma$, $\bar{y}_3=\Gamma(\Gamma_*-\Gamma)/2$] is attracting and describes the ``coexistence" state. It is either a stable node (for $\Gamma>\Gamma_0=4(\sqrt{1+a-b}-1)$), or a stable focus (for $\Gamma<\Gamma_0$).
Note that $\bar{y}_3$ is a non-monotonic function of $\Gamma$. It vanishes at  $\Gamma=0$ and $\Gamma=\Gamma_*$, and reaches a maximum, $\Gamma_*^2/8$, at $\Gamma=\Gamma_*/2$ corresponding to the optimal predation rate.  Figure \ref{MeanFieldPhase} shows two examples of deterministic trajectories: for $\Gamma_0<\Gamma<\Gamma_*$ (a) and for $\Gamma<\Gamma_0$ (b).

Reintroducing the stochasticity, both populations ultimately go extinct, but now there are two distinct extinction routes. In the first, \textit{sequential} route the predators (or infectives) go extinct first, whereas the prey (or susceptibles) typically thrive for a long time and only then go extinct. In the second, \textit{parallel} route the prey (or susceptibles) go
extinct first, causing a rapid extinction of the predators (or infected).

\begin{figure}
\includegraphics[width=4 in,clip=]{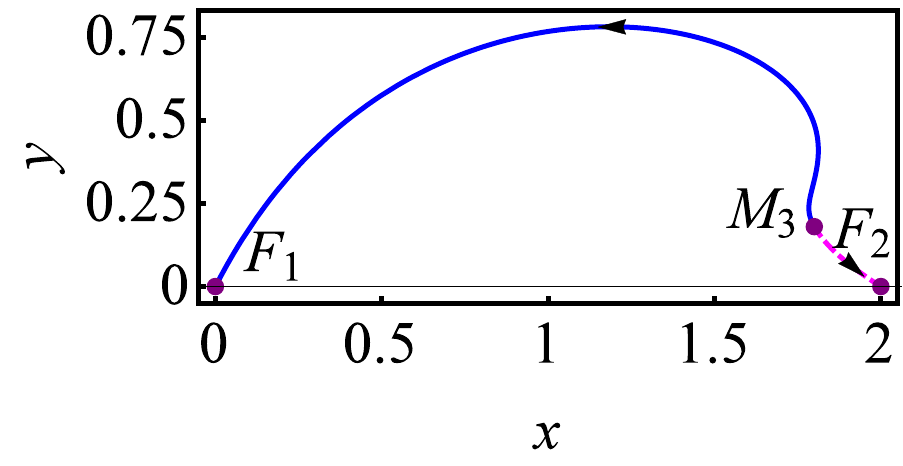}
\caption{Instantons $M_3F_1$ and $M_3F_2$ for $a=2,\;\;b=1$ and $\Gamma=1.8$, when the fixed point $M_3$ is a stable node \cite{GM2012}. Shown are
the $x,y$-projections of the four-dimensional instantons.}
\label{nodeinst}
\end{figure}

\begin{figure}
\includegraphics[width=2.0in,height=1.5in,clip=]{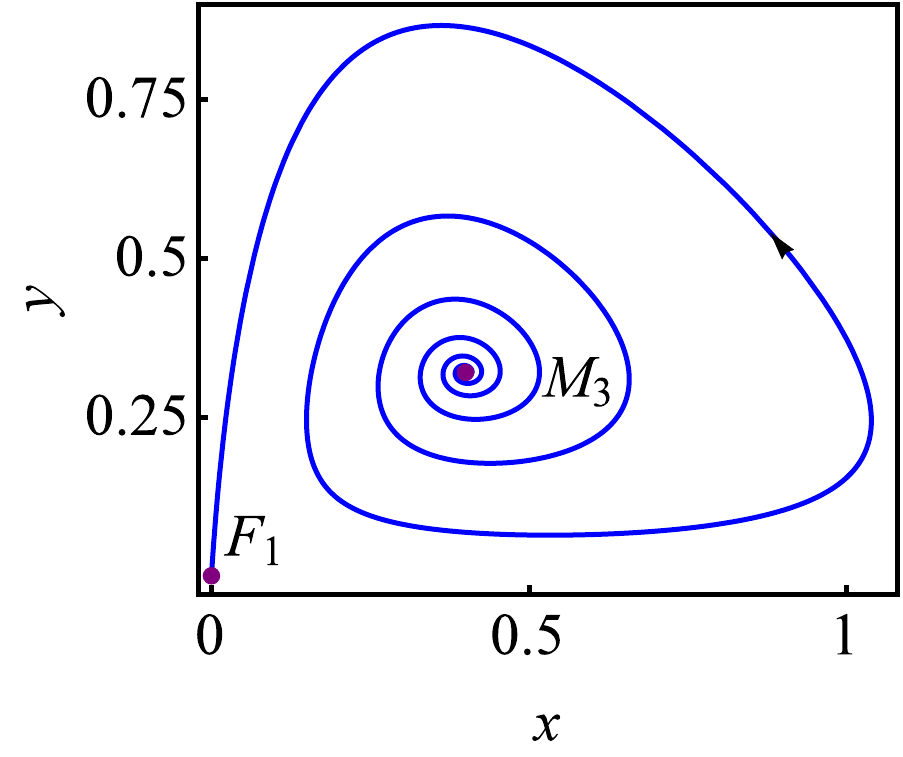}
\includegraphics[width=2.5in,height=1.5in,clip=]{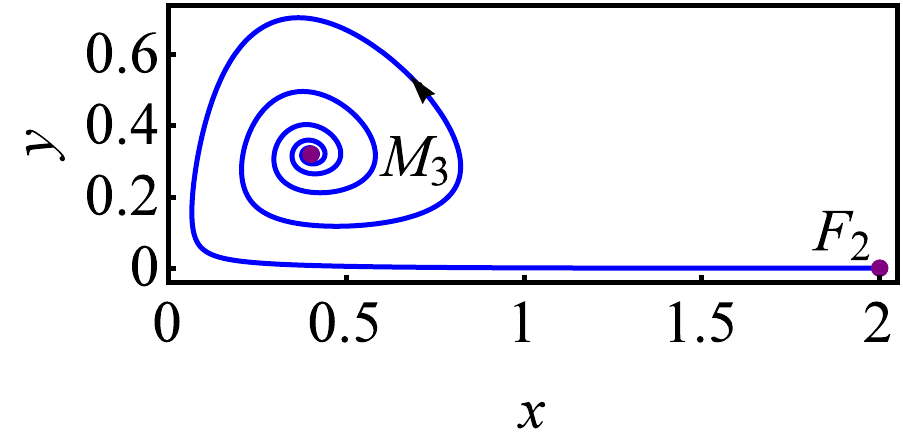}
\caption{(left) Instanton $M_3F_1$ and (right) instanton $M_3F_2$, for $a=2,\;\;b=1$ and $\Gamma=0.4$, when the fixed point $M_3$ is a stable focus \cite{GM2012}. Shown is the $x,y$-projection of the four-dimensional instantons.}
\label{focinst31}
\end{figure}

Let $P_{m,n}(t)$ be the probability to observe, at time $t$, $m$ rabbits and $n$ foxes, where $m,n=0,1,2 \dots$. The evolution of
$P_{m,n}(t)$ is described by the master equation
\begin{eqnarray}
\label{MasterEquationGM}
   \dot{P}_{m,n}&=&\hat{H} P_{m,n}=a[(m-1)P_{m-1,n}-mP_{m,n}] \nonumber \\
   &+&(1/\Gamma N)[(m+1)(n-1)P_{m+1,n-1}-mnP_{m,n}] \nonumber \\
 &+&b[(m+1)P_{m+1,n}-mP_{m,n}]+(n+1)P_{m,n+1}-nP_{m,n}\nonumber \\
 &+&(1/2N)[(m+1)m P_{m+1,n}-m(m-1)P_{m,n}],
\end{eqnarray}
where $P_{i,j}=0$ when any of the indices is negative. Before delving into the WKB analysis of this master equation, let us discuss the probability flow in this system.
At large $N$ and at sufficiently long times, the probability distribution $P_{m,0}(t)$ with $m>0$, which describes the dynamics of rabbits conditional on prior extinction of the foxes, is a one-dimensional distribution peaked at the fixed point $M_2$ of the deterministic theory. The probability distribution  $P_{m,n}(t)$ with
$m,n>0$ (which describes the long-time dynamics of the coexisting rabbits and foxes) is a two-dimensional distribution peaked
at the fixed point $M_3$. Finally, the extinction probability of both sub-populations $P_{0,0}(t)$ corresponds to a Kronecker-delta probability distribution.  Not only the  structure, but the long-time dynamics of these three distributions are different. To emphasize this point, the authors of Ref. \cite{GM2012} defined the total ``probability
contents" of the vicinities of each of the fixed points $M_1$, $M_2$ and $M_3$:
\begin{eqnarray}
  \mathcal{P}_1 (t) &=& P_{0,0} (t)\,, \label{calP1} \\
  \mathcal{P}_2 (t) &=& \sum_{m=1}^{\infty}P_{m,0} (t)\,, \label{calP2} \\
  \mathcal{P}_3 (t) &=& \sum_{m=1}^{\infty} \sum_{n=1}^{\infty} P_{m,n} (t)\,. \label{calP3}
\end{eqnarray}
At $N\gg 1$ and $t\gg t_r$ the sums in Eqs.~(\ref{calP2}) and (\ref{calP3}) are mostly contributed to by close vicinities of the points $M_2$ and $M_3$, respectively. The long-time evolution of $\mathcal{P}_1(t)$,  $\mathcal{P}_2(t)$ and  $\mathcal{P}_3(t)$ is described by the effective \textit{three-state} master equation:
\begin{equation}
\label{efrateeqn}
  \dot{\mathcal{P}}_1 = r_{21}  \mathcal{P}_2 + r_{31} \mathcal{P}_3,\quad
  \dot{\mathcal{P}}_2=  -r_{21}  \mathcal{P}_2 + r_{32} \mathcal{P}_3, \quad
  \dot{\mathcal{P}}_3= -(r_{31} + r_{32}) \mathcal{P}_3,
\end{equation}
where $r_{ij}$ is the (yet unknown) transition rate from the vicinity of the fixed point $i$ to the vicinity of the fixed point $j$. Let the initial condition correspond to the coexistence state around $M_3$:
\begin{equation}\label{incond}
   [\mathcal{P}_1 (0),\,\mathcal{P}_2 (0),\,\mathcal{P}_3 (0)]=(0,0,1)\,.
\end{equation}
Then the solution of Eqs. ~(\ref{efrateeqn}) is
\begin{eqnarray}
\mathcal{P}_1 (t) &=& 1+\frac{r_{32}\,e^{-r_{21}t}+(r_{31}-r_{21})\,e^{-(r_{31}+r_{32})t}}{r_{21}-r_{31}-r_{32}},\label{P1}\\
\mathcal{P}_2 (t) &=&\frac{r_{32}\,[e^{-(r_{31}+r_{32})t}-e^{-r_{21}t}]}{r_{21}-r_{31}-r_{32}},  \label{P2} \\
\mathcal{P}_3 (t)  &=& e^{-(r_{31}+r_{32})t}.\label{P3}
\end{eqnarray}
Once the transition rates $r_{31}$, $r_{32}$ and $r_{21}$ are known, Eqs.~(\ref{P1})-(\ref{P3}) provide a useful ``coarse-grained" description of this system in terms of the long-time
evolution of the probabilities to observe the coexistence state around $M_3$, the fox-free state around $M_2$ and the extinction state at $M_1$. In particular, the MTE of foxes is
$\tau_{F} \simeq (r_{31}+r_{32})^{-1}$, whereas the MTE of both populations is
\begin{equation}\label{MTEtotal}
    \tau_{RF}=\int_0^{\infty} dt\,t\, \dot{\mathcal{P}}_1(t)=\frac{r_{21}+r_{32}}{r_{21}(r_{31}+r_{32})}.
\end{equation}
The transition rate $r_{21}$  comes from solving the one-population problem $R \rightleftarrows 2R$ and $R\to \emptyset$; it is known with a high accuracy  \cite{KS2007,AM2010}. To evaluate the transition rates $r_{31}$ and $r_{32}$, the authors of Ref. \cite{GM2012} used the real-space WKB theory. The classical Hamiltonian takes the form
\begin{equation}\label{HamiltonianGM}
H(x,y,p_x,p_y)=ax\left(e^{p_x}-1\right)+bx\left(e^{-p_x}-1\right)
+\frac{xy}{\Gamma}\left(e^{p_y-p_x}-1\right)+y\left(e^{-p_y}-1\right)+\frac{x^2}{2}\left(e^{-p_x}-1\right).
\end{equation}
The Hamiltonian flow, generated by this Hamiltonian, has five zero-energy fixed points:
\begin{eqnarray}\label{FixedPoints}
    M_1&=&(0,0,0,0)\,;\;\;\;\;\;\;\;\;\;\;\;M_2=(\Gamma_*,0,0,0)\,;\;\;\;\;\;\;\;\;\;\;\;
    M_3=[\Gamma,\Gamma (\Gamma_*-\Gamma)/2,0,0]\,;\nonumber \\
    F_1&=&[0,0,\ln(b/a),0]\,;\;\;\;\;F_2=[\Gamma_*,0,0,-\ln(\Gamma_*/\Gamma)]\,.
\end{eqnarray}
The zero-momentum fixed points $M_1$, $M_2$ and $M_3$ correspond to the three fixed points of the deterministic equations (\ref{xydot}). The two other fixed points, $F_1$ and $F_2$, are fluctuational fixed points describing a fox-free state at a non-zero number of rabbits ($F_2$) and an empty system ($F_1$).
The two different routes to extinction -- the sequential and the parallel -- are encoded in two different instantons. Both start from the fixed point $M_3$ which describes the established populations, but end in a different fluctuational fixed point, $F_1$ or $F_2$.  The instantons and the action along each of the instantons can be found numerically  \cite{GM2012}. Figure \ref{nodeinst} shows typical examples of the numerically found instantons $M_3F_1$ and $M_3F_2$ in the case when $M_3$ is a node. Figure \ref{focinst31} refers to the case when $M_3$ is a focus. The actions $S_{31}$ and $S_{32}$ along these (zero-energy) instantons,
$$
S_{31}=\int_{M_3}^{F_1} p_x dx + p_y dy\;\;\mbox{and}\;\;S_{32}=\int_{M_3}^{F_2} p_x dx + p_y dy,
$$
yield, with exponential accuracy, the transitions rates $r_{31}$ and $r_{32}$:
\begin{equation}\label{ratesGM}
  r_{31} \sim  \exp(-NS_{31}), \quad r_{32} \sim  \exp(-NS_{32}).
\end{equation}
As the actions along the different instantons are different, and $N\gg 1$, there is usually a great (exponential) disparity between the transition rates: $r_{21}\ll r_{31}\ll r_{32}$, implying that the sequential extinction route (foxes first, rabbits second) is usually much more likely than the parallel extinction route. In other words (and somewhat surprisingly), the predators are usually more prone to extinction than their prey.

\subsubsection{Epidemic fadeout}

An infectious disease can disappear from a population immediately after the first infection outbreak \cite{Bartlett,vH}.
Such an ``epidemic fade-out" happens if the epidemic dynamics
are oscillatory at the level of the deterministic theory, and the number of infected individuals at the end of the first outbreak of the disease
is relatively small so that fluctuations in the disease transmission and in the recovery or removal of the infected can ``switch off" the disease.
Epidemic fade-out has been mostly addressed by epidemiologists via stochastic simulations. (One exception is Ref. \cite{vH} which reported
important early analytical results.) To determine the probability of the epidemic fadeout, Ref. \cite{MS2009} considered the SI model with population turnover,
see Table \ref{tableKM},  therefore the master equation coincides with Eq.~(\ref{masterSI}).  The initial condition, $P_{n,m}(t=0) = \delta_{n,N} \delta_{m,m_0}$, used
in Ref. \cite{MS2009}, describes introduction of $m_0$ infected individual into a steady-state susceptible population. One boundary condition reflects the fact that $m=0$ is, for any $n$, an absorbing state. Being interested in epidemic fade-out, one should exclude from consideration all stochastic trajectories that
do not reach the extinction boundary $m=0$ immediately after the first outbreak.  This can be achieved by introducing an artificial
absorbing boundary \cite{Gardiner,vH,MS2009}.

In contrast to the endemic fadeout, which is studied assuming quasi-stationarity of $P_{n,m}$,
the epidemic fadeout occurs on a fast time scale determined by the deterministic equations. Still, a stationary formulation can be obtained
if one uses the master equation (\ref{masterSI}) to derive an exact stationary  equation for the \textit{mean residence time} of the population
in a certain state $(n,m)$, where $m>0$:  $T_{n,m}=\int\limits_0^\infty P_{n,m}(t)\, dt$ \cite{MS2009}. (Alternatively, one could
write a recursive equation for the extinction probability starting from the state $n,m$, as was done in Sec. \ref{nonest}.)
The
accumulated extinction probability ${\cal P}_n$ from the state $(n,1)$ becomes
${\cal P}_n =\mu_I \, T_{n,1}$, and the total extinction probability
is ${\cal P}=\sum_n{\cal P}_n$. Integrating Eq.~(\ref{masterSI}) over $t$ from $0$ to $\infty$ and using the
equality $P_{n,m}(t=\infty)=0$ and the initial condition, one obtains for $T_{n,m>0}$:
\begin{equation}
\sum\limits_{n^{\prime}, m^{\prime}>0} M_{n,  m;\, n^{\prime},  m^{\prime}}\, T_{n^{\prime}, m^{\prime}}
+\delta_{n, n_0}\, \delta_{m, m_0}=0\,,
\label{foi_760}
\end{equation}
where  $M_{n,m;n',m'}$ was defined in Eq.~(\ref{masterSI}).
For $N\gg 1$ this stationary equation can be approximately solved by the WKB ansatz $T_{n, m}=a(x,y)\, e^{-N S(x,y)}$,
where $a$ and $S$ are smooth functions of $x=n/N-1$ and $y=m/N$~\cite{MS2009}.  Remarkably, the ensuing classical Hamiltonian
is the same as in Eq.~(\ref{hamiltonian}) up to the canonical transformation from the momentum space to the real space.

\begin{figure}[ht]
\includegraphics[width=4.0in,clip=]{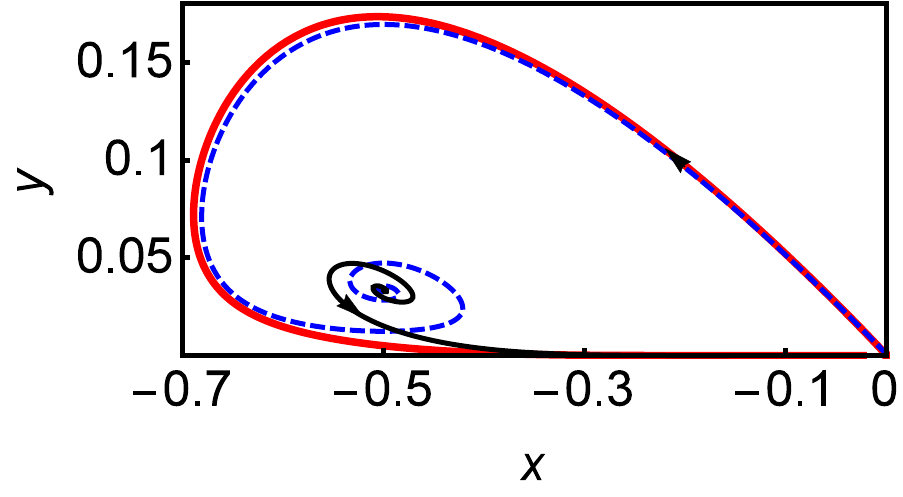}
\caption{An epidemic outbreak on the $xy$-plane as predicted
by the deterministic rate equations for the SI model (dashed line)
and the epidemic fade-out instanton  (thick solid line). For comparison, the thin solid line
shows the \textit{endemic} fade-out instanton considered in section \ref{extendemic}.
The rescaled parameters are $K=\beta/\mu=30$ and $\delta=1-\mu_I/\beta=0.5$. For $K\delta\gg 1$
the epidemic fade-out instanton closely follows the deterministic trajectory and departs from it at $y\ll 1$.}
\label{far}
\end{figure}

If only a few infected individuals are introduced into an infection-free population then
one has to find an instanton that exits, at $t=-\infty$, the deterministic fixed point corresponding
to the infection-free steady state
$[S=N,I=0,p_S=0,p_I=0]$ and enters, at $t=\infty$, the fluctuational fixed point for which $S=N$, $I=0$ and $p_S=0$, but $p_I\neq 0$ \cite{MS2009}.
As was found in Ref. \cite{MS2009}, such an instanton exists if and only
if the endemic fixed point, predicted by the rate equations,  is a focus rather than a node. In fact,
there exist \textit{multiple} heteroclinic trajectories between the same two points in this case. They can be classified by whether their
$x,y$-projections exhibit a single loop, two loops, three loops, \textit{etc}. \cite{MS2009}.
A single-loop instanton
corresponds to a disease fade-out immediately
after the first outbreak.   A two-loop instanton  corresponds to a fade-out immediately
after the second outbreak, \textit{etc}. An example of the $x,y$-projection of the one-loop instanton is shown in Fig.~\ref{far}, where $x=S/N$ and $y=I/N$.
In the leading WKB order one has ${\cal P}\sim {\cal P}_{n=N}$, therefore one again needs to calculate the accumulated action ${\cal S}_0$
along the instanton. The numerics can be done by shooting. It is convenient to introduce the same rescaled parameters  $K=\beta/\mu$ and $\delta=1-\mu_I/\beta$ as
in Sec. \ref{extendemic}. As shown in Ref. \cite{MS2009}, the regime of $K\delta \gg 1$ can be investigated analytically.
In this regime the epidemic fadeout instanton closely follows the deterministic trajectory, and departs from it only when $y=I/N$ becomes much smaller than unity,
see Fig. \ref{far}.  Further,  for $K\delta \gg 1$ the fluctuations of the number of susceptibles can be neglected, and one can Taylor-expand the Hamiltonian in $p_x \ll 1$ and truncate
the expansion at first order. Under these conditions the main contribution to the action comes from
a narrow region of $x$ where the instanton quickly departs from the deterministic orbit.
The resulting action looks especially simple if, in addition to the inequality  $K\delta \gg 1$, one also demands $\delta \ll 1$. Here one obtains
$-\ln \mathcal{P}/N \simeq {\cal S}_0\simeq (2 \delta^5/\pi e^4 K)^{1/2} (e/2)^{-K \delta}$. We refer the reader to Ref. \cite{MS2009} for details.
We also note that, slightly above the node-focus transition of the deterministic theory, the epidemic fadeout instanton first closely follows the deterministic trajectory,
and then rapidly approaches the \textit{endemic} fadeout instanton considered in section \ref{extendemic}. The intimate connection between the three heteroclinic trajectories in the four-dimensional phase space is fascinating. It deserves a further study and is likely to be universal for a whole class of models.

\subsubsection{Extinction: two more examples}

It was realized in Ref.~\cite{KD2009}  that the population extinction rate in multi-population systems may exhibit \textit{fragility}, where the WKB action (that determines the effective barrier to extinction)
depends on some of the elemental transition rates non-analytically. As an example, the authors of Ref.~\cite{KD2009}
considered the classic SIS (susceptible-infectious-susceptible) model and its extension accounting for population turnover.
The classic SIS model includes the following basic reactions \cite{Andersson,Nasellbook,Nasellbook2}. Upon interaction with an infected individual, a susceptible can be infected, while an infected individual can recover:
\begin{equation}
S+I\stackrel{\beta}{\rightarrow}I+I,\;\;\;\;\;I\stackrel{\gamma}{\rightarrow}S.
\label{classicSIS}
\end{equation}
The SIS model with population turnover includes, in addition to these transitions, the renewal of susceptibles, $\emptyset \stackrel{\mu} {\rightarrow}S$. The authors of Ref. \cite{KD2009} observed that the WKB action, corresponding to extinction
of the endemic disease from the population, experiences a jump when the population renewal is allowed, even in the limit of $\mu \to 0$.  To better understand the extinction rate fragility, the subsequent work \cite{KhasinMeersonSasorov} introduced the concept of time-resolved extinction rate. When $\mu$ is small, there is a time-scale separation in the system which enables one to define a short-time quasistationary extinction rate $W_1$ and a long-time quasistationary extinction rate $W_2$, and to follow
the transition between $W_1$ and $W_2$ as it develops in time. Importantly, $W_1$ and $W_2$ coincide with the
extinction rates when the population turnover is absent and present but very slow, respectively. The extinction rate fragility, discovered in Ref. \cite{KD2009}, manifests itself in the exponentially
large disparity between $W_1$ and $W_2$.

An interesting variant of the problem of extinction of a stochastic metapopulation was considered in Ref. \cite{Mehlig}.
The authors considered $N\gg 1$ patches supporting a local birth-death-competition population dynamics. In addition, an individual of each patch can migrate
into a pool of migrants, and the reverse transition -- from the pool to the patch -- also occurs. The authors studied
a long-lived quasi-stationary state of the system and employed the real-space WKB method to evaluate the MTE of the whole population. They also found the conditions under which their individual-based model reduces to several metapopulation models, previously used  \textit{ad hoc} by ecologists.

\subsubsection{Switching in a two-population system}\label{Switch2D}
\begin{figure}[ht]
\includegraphics[scale=1.2]{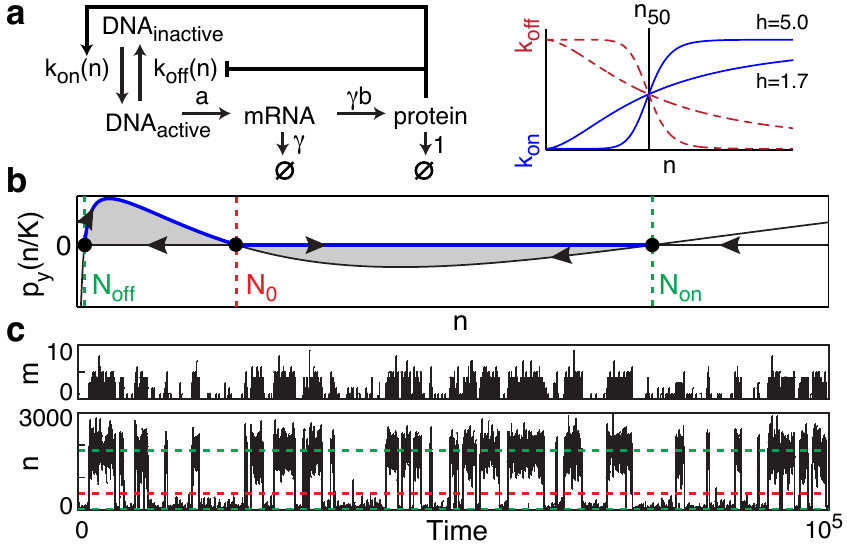}
\caption{(a) A model for a two-state positive feedback network. Transcription of mRNA and translation of proteins are modeled as first-order processes with rates $a$ and $b\gamma$, respectively. The mRNA and protein molecules also undergo first-order degradation with rates $\gamma$ and $1$ (all the rates are rescaled by the protein decay rate). The feedback functions $k_{\text{on}}(n)$ and $k_{\text{off}}(n)$ control promoter transitions. (b) The momentum $p_y$, obtained from solving $H_r^{(2)}(y,p_y)=0$ [where $H$ is given by Eq.~(\ref{hred2})] versus protein copy number $n$. The thick line indicates the off$\to$on instanton, the thin line indicates the on$\to$off instanton, while the shaded areas correspond to the switching actions. (c) Time-dependent fluctuations in mRNA (denoted by $m$), and protein (denoted by $n$) copy-numbers, in a typical Monte Carlo trajectory undergoing switching. In (b) and (c) $K =ab=2400$, $b=22.5$, $h_1=h_2=2$, $n_{50}=1000$, $k_0^{\text{min}}=k_1^{\text{min}}=a/100$, $k_0^{\text{max}}=k_1^{\text{max}}=a$, and $\gamma=50$.}\label{twostate}
\end{figure}

The calculations of the MST in the model of a self-regulating gene, presented in Sec.~\ref{switchingB}, can be generalized by explicitly accounting for mRNA dynamics, and this generalization may dramatically affect the MST \cite{MMW2008,ARS2011}, see below. Here we follow Ref.~\cite{VLACB2013},
where this generalization was made on an example of a one-state DNA model with positive feedback (see also Ref. \cite{MMW2008}). The fact that the DNA has only one state means that the mRNA molecules are being constantly transcribed, as opposed to multi-state models~\cite{ARS2011}, see Fig.~\ref{twostate}(a). Regardless of the number of DNA states, we assume that the protein of interest positively regulates the transcription of the mRNA molecules that are responsible for the protein's production, giving rise to a positive feedback loop and a genetic switch.

The deterministic rate equations for the average copy numbers of mRNA, $m$, and proteins, $n$, are~\cite{VLACB2013}:
\begin{eqnarray}\label{REgene}
\dot m&=& f(n)-\gamma m, \nonumber\\
\dot n&=&b\gamma m - n.
\end{eqnarray}
Here transcription of mRNA occurs at a rate $f(n)$ which depends on the current protein number, $\gamma$ is the mRNA degradation rate, $b\gamma$  is the protein translation rate, and all rates are rescaled by the protein's degradation rate or cell division rate~\cite{ShS2008}. Furthermore, in the limit of short-lived mRNA, $\gamma\gg 1$, $b$ is the mean number of proteins made per mRNA~\cite{MMW2008}, see below. We assume that the nonlinear function $f(n)$ introduces a positive feedback leading to three deterministic fixed points:  $n_1\equiv n_{\text{off}}<n_2<n_3\equiv n_{\text{on}}$, where $n_{\text{off}}$  and $n_{\text{on}}$  are attracting fixed points corresponding to the off and on phenotypic states, see Sec.~\ref{switchingB}, while $n_2$  is a repelling fixed point. Let us also denote by $N\equiv n_{\text{on}}\gg 1$  the protein abundance in the
on state, see below.

Being interested in the MST between, say, the on and off states, we turn  to the master equation for $P_{m,n}(t)$,  the probability to find $m$ mRNA molecules and $n$  proteins:
\begin{eqnarray}\label{mastergene}
\dot{P}_{m,n}=\hat{H} P_{m,n}&=& f(n)[P_{m-1,n}-P_{m,n}]+\gamma[(m+1)P_{m+1,n}-mP_{m,n}]\nonumber\\
&+&b\gamma m (P_{m,n-1}-P_{m,n})+(n+1)P_{m,n+1}-nP_{m,n}.
\end{eqnarray}
To evaluate the MST, we set
$P_{m,n}=\pi_{m,n}e^{-t/\tau}$ and employ the WKB ansatz for the QSD, $\pi_{m,n}=\pi(x,y)\sim e^{-NS(x,y)}$, where $x=m/N$ and $y=n/N$. In the leading WKB order the problem reduces to finding the proper instanton of the Hamiltonian~\cite{VLACB2013}
\begin{equation}\label{Hamgene}
H=y(e^{-p_y}-1)+b\gamma x(e^{p_y}-1)+\gamma x (e^{-p_x}-1)+\tilde{f}(y)(e^{p_x}-1),
\end{equation}
where $\tilde{f}(y)=f(n)/N$. The instanton  can be found numerically, \textit{e.g.} via the shooting method~\cite{RFRSM2005}. Then one computes the action $S(x,y)=\int p_x dx + p_y dy=\int^t (p_x\dot{x}+p_y\dot{y})dt$ along the optimal path, and evaluates the MST.

\begin{figure}[ht]
\includegraphics[scale=1.8]{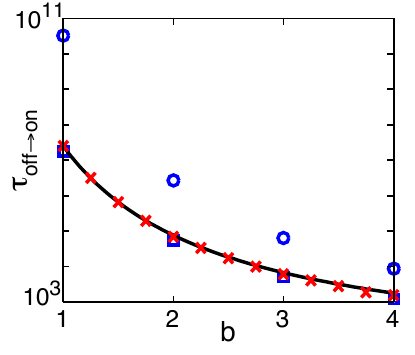}
\caption{(a) The MST $\tau_{\text{off}\to \text{on}}$ as a function of the mean burst size $b$ for the two-state model presented in Fig.~\ref{twostate}. Here $K=ab=2400$, $h_1=h_2=2$, $n_{50}=720$, $k_0^{\text{min}}=k_1^{\text{min}}=a/100$, $k_0^{\text{max}}=k_1^{\text{max}}=a$ and $\gamma=50$. Shown are results of Monte-Carlo simulations, see Fig.~\ref{twostate}(c), of the full two-component model (x's), simulations of the protein-only model with geometrically-distributed bursts with mean $b$ ($\Box$) and simulations of the protein-only model with constant burst size $b$ ($\circ$). The line is the theoretical prediction. It is evident that, while incorporating the protein production via geometrically-distributed bursts effectively accounts for the mRNA noise, ignoring the mRNA noise exponentially increases the MST.}\label{mrnanoise}
\end{figure}

One regime where analytical progress is possible is that of short-lived mRNA, $\gamma\gg 1$, when there is time-scale separation between the mRNA and proteins~\cite{MMW2008,ShS2008}. Here $x(t)$ and $p_x(t)$ equilibrate rapidly, and one can express $x$ and $p_x$ via the protein concentration and momentum (see also Ref.~\cite{AM2008}). In this way one finds~\cite{VLACB2013} $e^{-p_x}=b(1-e^{p_y})+1$, and $x=[\tilde{f}(y)/\gamma][b(1-e^{p_y})+1]^{-2}$, which gives rise to a \textit{reduced} protein-only Hamiltonian~\cite{VLACB2013}
\begin{equation}\label{redHamgene}
H_{\text{red}}(y,p_y)=y(e^{-p_y}-1)+\tilde{f}(y)\frac{b(e^{p_y}-1)}{b(1-e^{p_y})+1}.
\end{equation}
At this point we pause and examine the effect of explicitly incorporating the mRNA species into the gene-expression dynamics. Despite adiabatic elimination of the mRNA species, the reduced Hamiltonian~(\ref{redHamgene}) differs from the 1D Hamiltonian~(\ref{hamilswitch}). In the former, mRNA noise is manifested through the nontrivial dependence of the protein production term on the momentum~\cite{ARS2011,VLACB2013}, which effectively accounts for the fact that the proteins are produced in geometrically distributed bursts with mean $b$~\cite{PE2000,ARS2011,BAA2016}, see Fig.~\ref{mrnanoise}. As expected, the Hamiltonians coincide in the limit of $b\to 0$ (here $b\tilde{f}$ is equivalent to $\tilde{f}$ in Sec.~\ref{switchingB}), which corresponds to a constant burst size~\cite{ARS2011,BAA2016}.

Going back to our reduced Hamiltonian, the instanton is~\cite{VLACB2013}
\begin{equation}\label{onestatemom}
p_y(y)=\ln\frac{(b+1)y}{b[y+\tilde{f}(y)]}.
\end{equation}
Calculating the reduced action $S(y)=\int^y p_y(y')dy'$, which can be done explicitly for any function $\tilde{f}(y)$, gives rise to the protein-only QSD, $\pi(y)\sim e^{-NS(y)}$. [Note that here while calculating the action $S$ we have neglected the contribution from $\int p_xdx$, which scales as ${\cal O}(\gamma^{-1})$ compared to $\int p_ydy$~\cite{ARS2011}.] As a result, the MST, say from the on state to the off state, is~\cite{VLACB2013}
\begin{equation}\label{tauswitch2d}
\tau_{\text{on}\to \text{off}}\sim e^{N[S(y_2)-S(y_{\text{on}})]}.
\end{equation}
The calculation of $\tau_{\text{off}\to \text{on}}$ is done in a similar way.  Here the accumulated action, $S(y_2)-S(y_{\text{off/on}})$, turns out to be smaller than the accumulated action obtained in the protein-only model for any $b>0$, see Sec.~\ref{switchingB}. As a result, an account for mRNA noise exponentially decreases the MST compared to the protein-only case, see Fig.~\ref{mrnanoise}.

The real-space WKB method can be also used in more complicated scenarios of gene regulatory networks in the presence of multiple DNA states, a mechanism which is used to control mRNA production in a more efficient manner. Examples of such analysis were presented in Ref.~\cite{ARS2011} for a two-state system with nonlinear feedback, and in Refs.~\cite{N2012,ERADS2013} for multiple DNA-state systems.

Let us briefly review, following Ref.~\cite{ARS2011}, the case of a two-state positive feedback switch, which was experimentally shown to describe biological switching~\cite{CCFX2008}. We consider a two-state gene-expression model, where transitions between a transcriptionally active and inactive promoter are controlled by the protein copy number $n$ via
positive feedback, see Fig.~\ref{twostate}(a). The transition rates into the active and inactive states, respectively, can be chosen to be
\begin{equation}
\label{choice}
k_{\text{on}}(n)=k_0^{\text{min}}\!+\!(k_0^{\text{max}}\!
-\!k_0^{\text{min}})\frac{n^{h_1}}{n^{h_1}+n_{50}^{h_1}}\;,\;\;\;\;k_{\text{off}}(n)
=k_1^{\text{min}}\!+\!(k_1^{\text{max}}\!-\!k_1^{\text{min}})\frac{n^{h_2}}{n^{h_2}+n_{50}^{h_2}}.
\end{equation}
While the analytical approach holds for generic functions $k_{\text{on}}(n)$ and $k_{\text{off}}(n)$, the rates~(\ref{choice}) were shown to be biologically relevant,
\textit{e.g.}, in the \textit{lac} operon~\cite{RMOBL2011}. As before, we denote the stable fixed points by $N_{\text{off}}$ and $N_{\text{on}}$, while the unstable point is $N_0$.

In the case of the two-state model, the master equation has to be written for $P_{m,n}$ and $Q_{m,n}$ representing the probability distribution functions
of having $m$ mRNAs and $n$ proteins at time $t$ with the promoter in the inactive and active state, respectively, see Ref.~\cite{ARS2011} for details. Using  the WKB machinery, and assuming short-lived mRNA, $\gamma\gg 1$, one can adiabatically eliminate the mRNA species and arrive at a reduced two-state Hamiltonian $H_r^{(2)}(y,p_y)$~\cite{ARS2011}:
\begin{eqnarray}\label{hred2}
H_r^{(2)}=(e^{-p_y}-1)\left\{y+\left[y+\frac{e^{p_y}}{b(e^{p_y}-1)-1}\right]\left[\frac{\tilde{k}_{\text{on}}(y)-y(e^{-p_y}-1)}{\tilde{k}_{\text{off}}(y)}\right]\right\},
\end{eqnarray}
where $\tilde{k}_{\text{on}/\text{off}}(y)=k_{{\text{on}/\text{off}}(n)}/K$, and $K=ab$ is the typical number of proteins in the on state. One finds the instanton $p_y(y)$, see Fig.~\ref{twostate}(b),  from the equation $H=0$, and calculates $S(y)=\int^y p_y(y')dy'$. Setting $\tilde{k}_{\text{off}}(y)=0$, which corresponds to a one-state positive feedback switch, and by properly rescaling $\tilde{k}_{\text{on}}(y)$, one recovers Eq.~(\ref{onestatemom}). Finally, having calculated the action $S(y)$, the MST from the off$\to$on states (and similarly from the on$\to$off) is given by
\begin{equation}
\ln\tau_{\text{off}\to \text{on}}\simeq K\Delta S_{\text{off}},
\end{equation}
where $\Delta S_{\text{off}}=S(y_0)-S(y_{\text{off}})$, $y_{\text{off}}=N_{\text{off}}/K$ and $y_0=N_0/K$. Figure~\ref{mrnanoise} compares the analytical predictions of Ref. \cite{ARS2011} with Monte Carlo simulations [see Fig.~\ref{twostate}(c)], and an excellent agreement is observed. Figure~\ref{mrnanoise} also demonstrates that mRNA noise can exponentially decrease the MST between different phenotypic states, see also Ref.~\cite{MMW2008}.

\subsection{Established Populations Reside in a Vicinity of a Limit Cycle}

Some populations exhibit persistent oscillations in their sizes \cite{LV,Elton1942, Butler1953, Bazykin1998, Odum2004, Murray2008}.  At the level of the deterministic theory, these oscillations can often by described by a stable limit cycle in the space of population sizes. One deterministic model that shows this feature -- an extension of the celebrated Lotka-Volterra model \cite{LV} -- is due to Rosenzweig and MacArthur \cite{RM1963}. Qualitatively similar models are used  in epidemiology -- for a description of the oscillatory dynamics of susceptible and infected populations during an epidemic
\cite{Bartlett, Anderson, Andersson} and the oscillatory dynamics of tumor growth \cite{Kirschner, Donofrio}. Similar models describe oscillatory chemical reactions  \cite{Schloegl, vanKampen}. Population extinction, driven by demographic noise, acquires new features in this case, as has been shown in Ref. \cite{SM2016} on the example of a stochastic version of the Rosenzweig-MacArthur (RMA)  model, see Table \ref{table:reactionsMS}. The model differs from that
used in Ref. \cite{GM2012} (see Table \ref{modelGM}) in two aspects. First, in the absence of the predators, the prey is immortal.
Second, for a very large prey
population, the predation rate saturates so as to describe satiation of the predators.

\begin{table}[ht]
\centering 
\begin{tabular}{|c c c|} 
\hline\hline 
Process & Transition & Rate \\ [0.5ex] 
\hline 
Birth of rabbits & $R \rightarrow 2R$ 				& $aR$ \\ 
Predation and birth of foxes & $F+R \rightarrow 2F$	& $\frac{sRF}{1+s\tau R}$ \\ 
Death of foxes & $F \rightarrow \emptyset$ 					& $F$ \\ 
Competition among rabbits & $2R \rightarrow R$ 			& $\frac{R\left(R+1\right)}{2N}$  \\ [1ex] 
\hline 
\end{tabular}
\caption{Stochastic  Rosenzweig-MacArthur  model \cite{SM2016}.}
\label{table:reactionsMS}
\end{table}

\begin{figure}
\includegraphics[width=2.5 in,clip=]{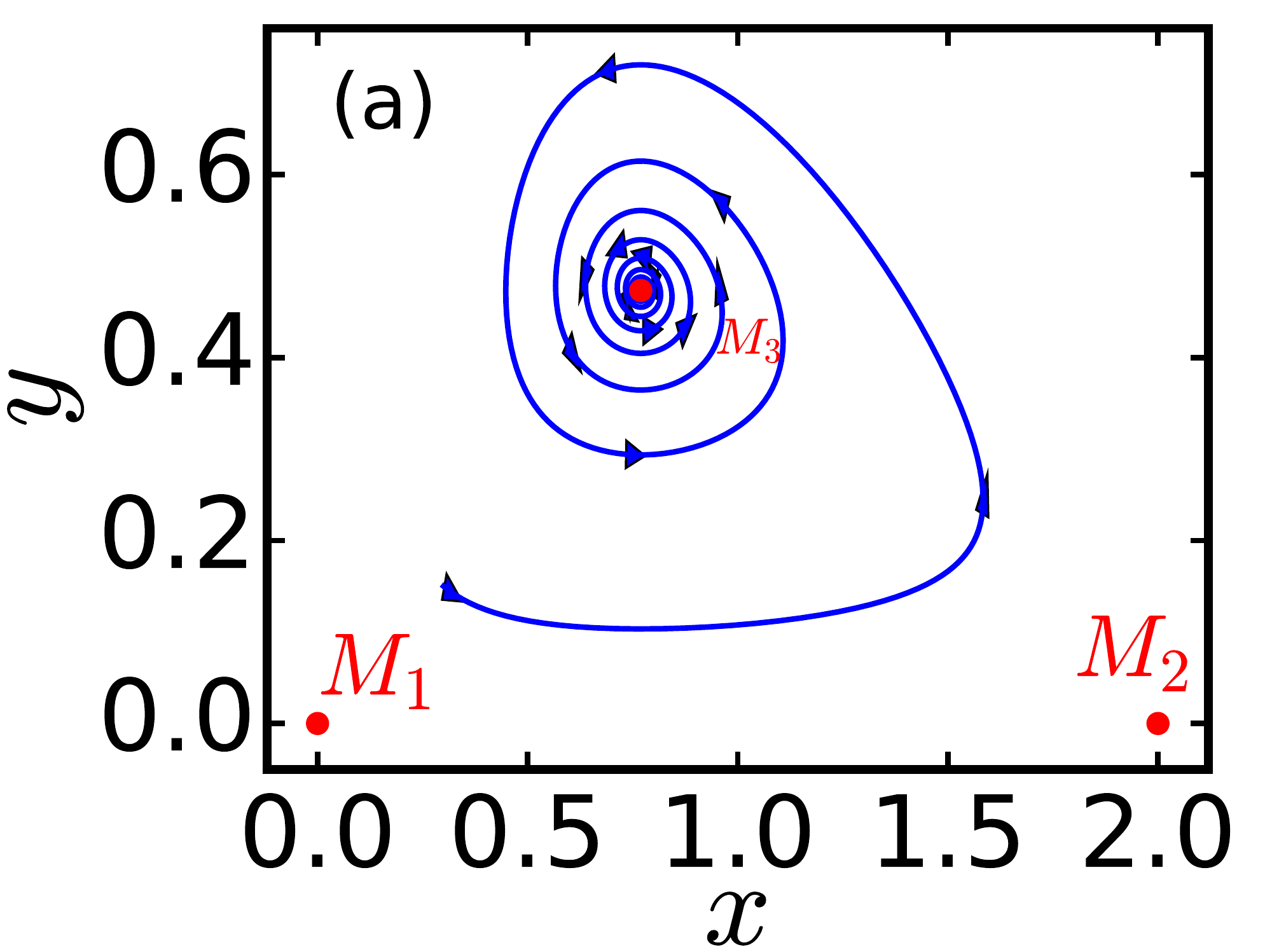}
\includegraphics[width=2.5 in,clip=]{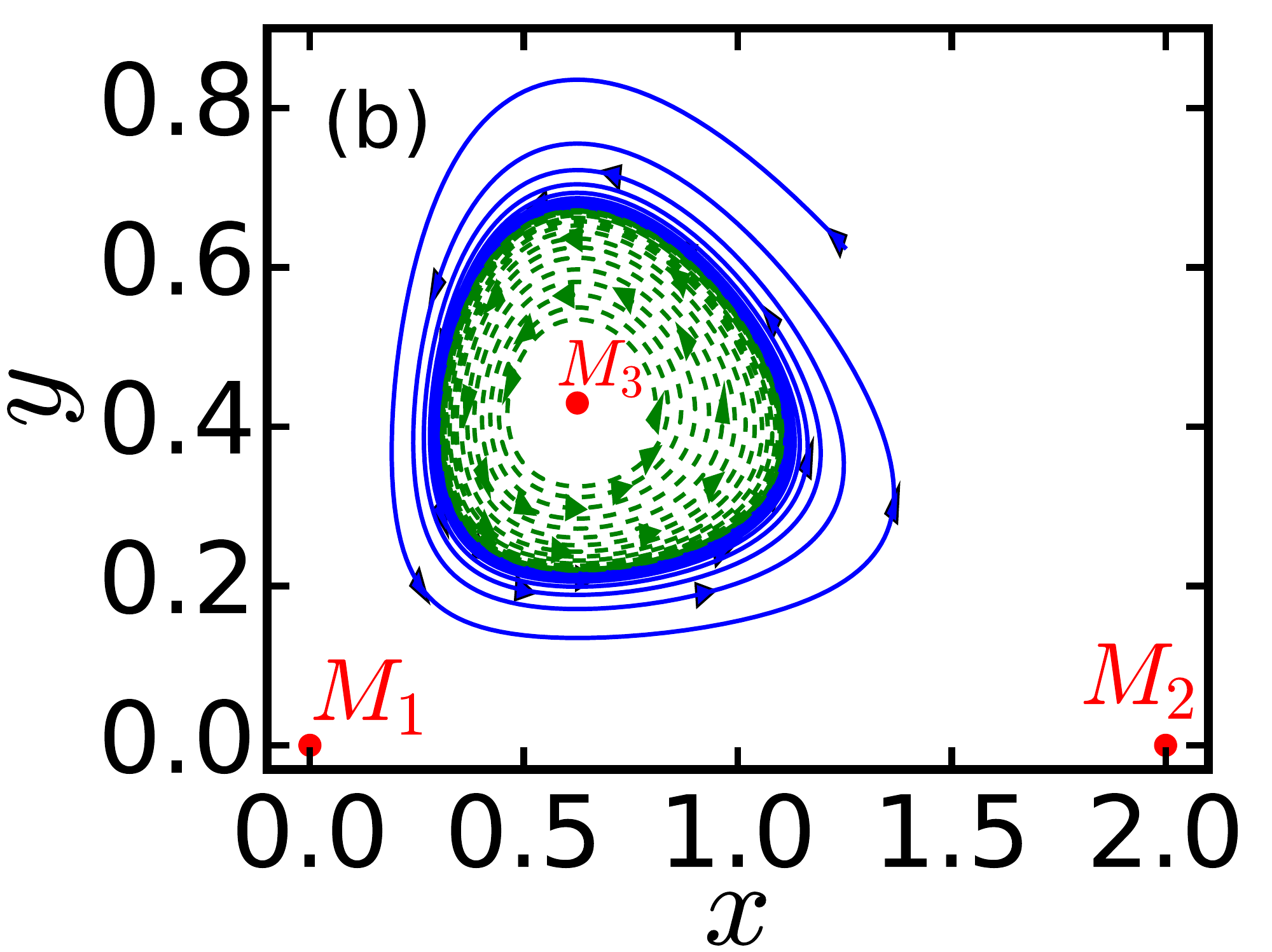}
\caption{Deterministic trajectories of the Rosenzweig-MacArthur model (\ref{eq_xydot_mean_field}) for $a=1$, $\tau=0.5$ and
two different values of $\sigma$: $\sigma=2.6$, where $M_3$ is a stable focus (a), and $\sigma=3.2$, where $M_3$ is an unstable focus, and a stable limit cycle exists (b).}
\label{fig_deterministic_SM}
\end{figure}

The deterministic equations for the RMA model can be written as
\begin{equation}
\label{eq_xydot_mean_field}
\dot{x}=ax-\frac{x^{2}}{2}-\frac{\sigma xy}{1+\sigma\tau x}, \quad \dot{y}=-y+\frac{\sigma xy}{1+\sigma\tau x} ,
\end{equation}
where  $x= R/N$, $y= F/N$, and  $\sigma= sN ={\mathcal O} (1)$. These equations have been extensively studied \cite{RM1963,Cheng1981,HalSmith2008}.
When
\begin{equation}
\label{eq_tau_and_sigma_condition}
0<\tau< 1,\quad  \sigma>\sigma_0=\frac{1}{2a\left(1-\tau\right)} ,
\end{equation}
Eqs.~(\ref{eq_xydot_mean_field}) have three fixed points describing nonnegative population sizes.
The fixed point $M_1$ $\left(\bar{x}_{1}=0,\bar{y}_{1}=0\right)$ describes an empty system. It is a saddle point: attracting in the $y$-direction (when there are no rabbits in the system), and repelling in the $x$-direction. The fixed point $M_2$ $\left(\bar{x}_{2}=2a,\bar{y}_{1}=0\right)$ describes a steady-state population of rabbits in the absence of foxes. It is also a saddle: attracting in the $x$-direction (when there are no foxes), and repelling in a direction corresponding to the introduction of a few foxes into the system. The third fixed point $M_3$ is given by $\left(\bar{x}_{3},\bar{y}_{3}\right)$, where
\begin{equation}
\label{eq_M3_coordinates}
\bar{x}_3=\frac{1}{\sigma\left(1-\tau\right)}, \quad \bar{y}_3=\frac{2a\sigma\left(1-\tau\right)-1}{2\sigma^{2}\left(1-\tau\right)^{2}},
 \end{equation}
and describes a non-oscillatory coexistence of the rabbits and foxes. For
\begin{equation}
\label{eq_sigma_node2focus}
\sigma_0 < \sigma < \bar{\sigma}=\frac{\frac{a\tau\left(1+\tau\right)}{2\left(1-\tau\right)}-1-\sqrt{1+a\frac{1+\tau}{1-\tau}}}{a^{2}\tau^{2}-4a\left(1-\tau\right)} ,
 \end{equation}
$M_3$ is a stable node.  For
\begin{equation}
\label{eq_sigma_star}
\bar{\sigma} < \sigma < \sigma^{*}=\frac{1+\tau}{2a\tau\left(1-\tau\right)}
\end{equation}
it is a stable focus. Finally, for $\sigma > \sigma^{*}$, $M_3$ is unstable, and a stable limit cycle appears around it. A Hopf bifurcation occurs at
$\sigma=\sigma^{*}$. Figure \ref{fig_deterministic_SM} shows the behavior of the deterministic model for $\bar{\sigma} < \sigma < \sigma^{*}$ (a) and for  $\sigma > \sigma^{*}$ (b).

Reintroducing the noise, one can see that, following the population establishment (either around the fixed point, or around the limit cycle),  there are two extinction routes in this model.
In the first (parallel) route all the rabbits are eaten by the foxes, followed by a quick extinction of the foxes. In the second route the foxes go extinct, while the (immortal) rabbits reach a steady state. Ref. \cite{SM2016} considered these two routes for $\sigma>\sigma^*$, when
the established populations are distributed in the vicinity of the stable limit cycle of the deterministic theory.

Again, we denote by  $P_{m,n}\left(t\right)$ the probability to observe $m$ rabbits and  $n$ foxes at time $t$. The master equation is
\begin{eqnarray}
\label{MasterEquationSM}
    \dot{P}_{m,n}&=&\hat{H} P_{m,n}=a\left[\left(m-1\right)P_{m-1,n}-mP_{m,n}\right] \nonumber \\
    &+&\frac{\sigma\left(m+1\right)\left(n-1\right)}{N+\sigma\tau\left(m+1\right)}P_{m+1,n-1}-
    \frac{\sigma mn}{N+\sigma\tau m}P_{m,n}\nonumber \\
    &+&\left(n+1\right)P_{m,n+1}-nP_{m,n}+(1/2N)[(m+1)m P_{m+1,n}-m(m-1)P_{m,n}],
\end{eqnarray}
where $P_{m,n}=0$ when any of the indices is negative. As in Sec. \ref{multipleroutes}, one can define the ``effective probability
contents" of the vicinities of the fixed points $M_1$ and $M_2$ via Eqs.~(\ref{calP1}) and (\ref{calP2}). In its turn,
Eq.~(\ref{calP3}) now defines the effective probability content of the limit cycle.  The long-times dynamics of $\mathcal{P}_1(t)$,  $\mathcal{P}_2(t)$ and  $\mathcal{P}_3(t)$  are given by the three-state master equation \cite{SM2016}
\begin{eqnarray}
\label{efrateeqn_SM}
  \dot{\mathcal{P}}_1 = \mathcal{R}_1 \mathcal{P}_3, \quad
  \dot{\mathcal{P}}_2 =  \mathcal{R}_2 \mathcal{P}_3, \quad
  \dot{\mathcal{P}}_3 = -(\mathcal{R}_1 +\mathcal{R}_2) \mathcal{P}_3,
\end{eqnarray}
where $\mathcal{R}_{1}$ and $\mathcal{R}_{2}$ are the extinction rates along the first and second extinction route, respectively. For the initial condition
$\left[\mathcal{P}_{1}\left(0\right),\,\mathcal{P}_{2}\left(0\right),\,\mathcal{P}_{3}\left(0\right)\right]=\left(0,0,1\right)$
the solution of Eqs. ~(\ref{efrateeqn_SM}) is
\begin{eqnarray}
\mathcal{P}_1 (t) &=& \frac{\mathcal{R}_{1}\,\left[1-e^{-\left(\mathcal{R}_{1}+\mathcal{R}_{2}\right)t}\right]}{\mathcal{R}_{1}+\mathcal{R}_2},\label{P1_SM}\\
\mathcal{P}_2 (t) &=&\frac{\mathcal{R}_{2}\,\left[1-e^{-\left(\mathcal{R}_{1}+\mathcal{R}_{2}\right)t}\right]}{\mathcal{R}_{1}+\mathcal{R}_2},  \label{P2_SM} \\
\mathcal{P}_3 (t)  &=& e^{-(\mathcal{R}_1 +\mathcal{R}_2)t}.\label{P3_SM}
\end{eqnarray}
As one can see from Eqs.~(\ref{P1_SM})-(\ref{P3_SM}), the long-time behavior of the system is determined by the extinction rates $\mathcal{R}_{1}$ and $\mathcal{R}_{2}$. These
can be evaluated in the WKB approximation. The effective Hamiltonian is \cite{SM2016}
\begin{equation}\label{Hamiltonian_SM}
   H\left(x,y,p_{x},p_{y}\right)=ax\left(e^{p_{x}}-1\right)+\frac{\sigma xy}{1+\sigma\tau x}\left(e^{p_{y}-p_{x}}-1\right)+
   y\left(e^{-p_{y}}-1\right)+\frac{x^{2}}{2}\left(e^{-p_{x}}-1\right).
\end{equation}
The three deterministic fixed points become $M_{1}=\left(0,0,0,0\right),\; M_{2}=\left(2a,0,0,0\right)$ and $M_{3}=\left(x^{*},y^{*},0,0\right)$. In its turn, the deterministic limit cycle is an exact time-periodic solution of the Hamilton equations with $p_x=p_y=0$.
There are also two fluctuational fixed points:
\begin{equation}\label{FFpoints}
 F_{1}= \left(0,0,0,-\infty\right), \quad  F_{2}=\left[2a,0,0,\ln\left(\frac{1+2a\tau\sigma}{2a\sigma}\right)\right] .
\end{equation}
The extinction rates $\mathcal{R}_{1}$ and $\mathcal{R}_{2}$ can be evaluated by evaluating the actions $S_1$ and $S_2$ along the instantons in the four-dimensional phase space. These instantons exit, at $t=-\infty$, the deterministic limit cycle and approach, at $t=\infty$, the fluctuational fixed point $F_1$ or $F_2$, respectively:
$$
S_{1}\!=\!\int_{\left(x_{lc},y_{lc}\right)}^{F_{1}}\!\!\!\!\!\!\!\!p_{x}dx+p_{y}dy\quad\mbox{and}\quad S_{2}\!=\!\int_{\left(x_{lc},y_{lc}\right)}^{F_{2}}\!\!\!\!\!\!\!\!p_{x}dx+p_{y}dy,
$$
The extinction rates $\mathcal{R}_{1}$ and $\mathcal{R}_{2}$ are, with exponential accuracy,
\begin{equation}\label{ratesSM}
  \mathcal{R}_{1} \sim  \exp(-NS_{1}), \quad  \mathcal{R}_{2} \sim  \exp(-NS_{2})\,.
\end{equation}

\begin{figure}
\includegraphics[width=4 in,clip=]{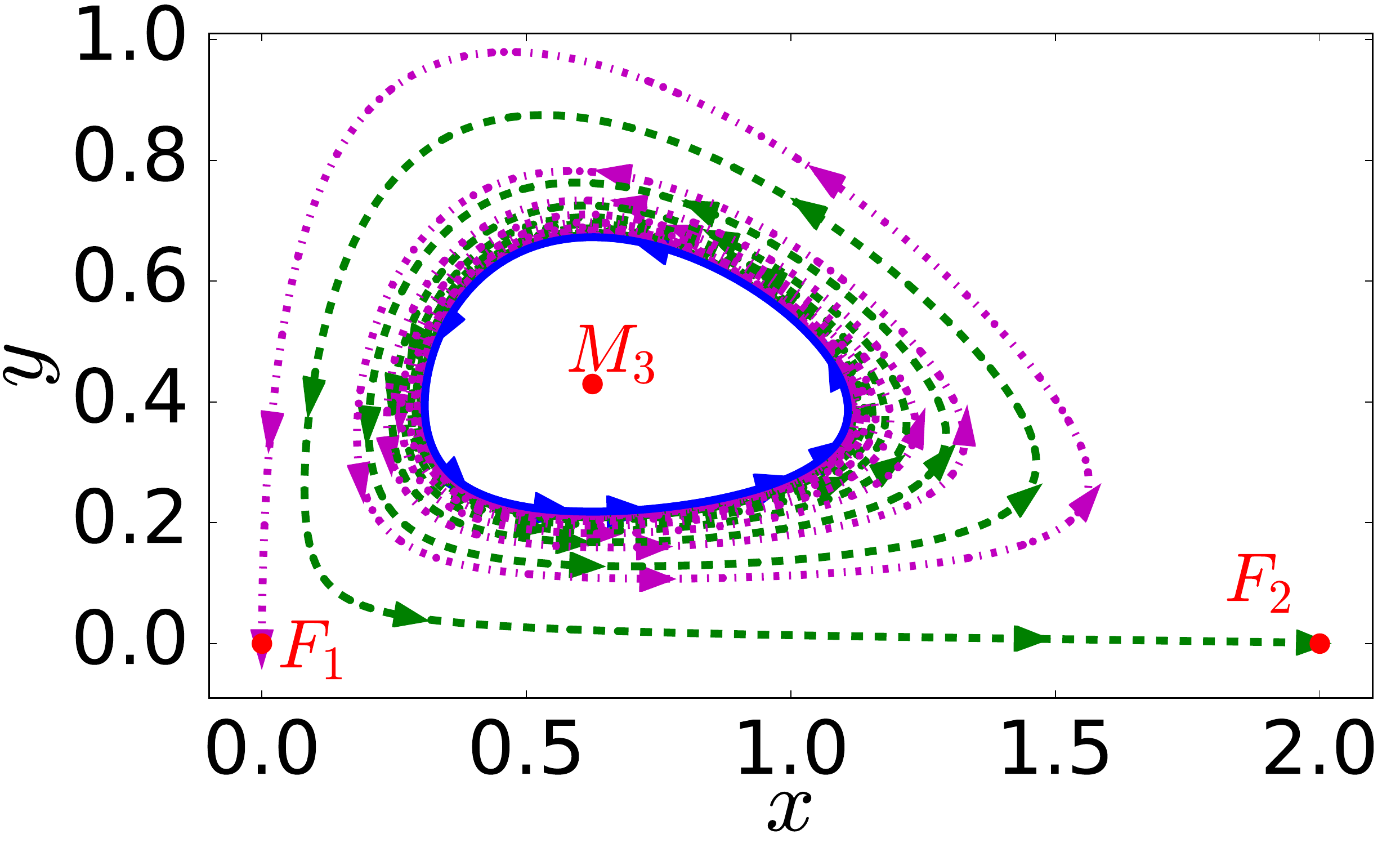}
\caption{Extinction instantons (the $x,y$-projections) going from the stable deterministic limit cycle (solid line) to the fluctuational fixed points $F_1$ (dot-dashed) and $F_2$ (dashed). The parameters are $a=1,\;\sigma=3.2$, and $\tau=0.5$.}
\label{nodeinstSM}
\end{figure}

Until recently, no numerical algorithm existed for finding an instanton that describes a fluctuation-driven escape from a deterministically stable limit cycle. One such algorithm has been recently developed \cite{SM2016}. It is based on Floquet theory  of differential equations with periodic coefficients, see \textit{e.g.} \cite{Ward2010}. The reader is referred to Ref. \cite{SM2016} for a detailed description of the algorithm. Examples of instantons, found with this algorithm, are shown in Fig. \ref{nodeinstSM}.

\begin{figure}
\includegraphics[width=3.5 in,clip=]{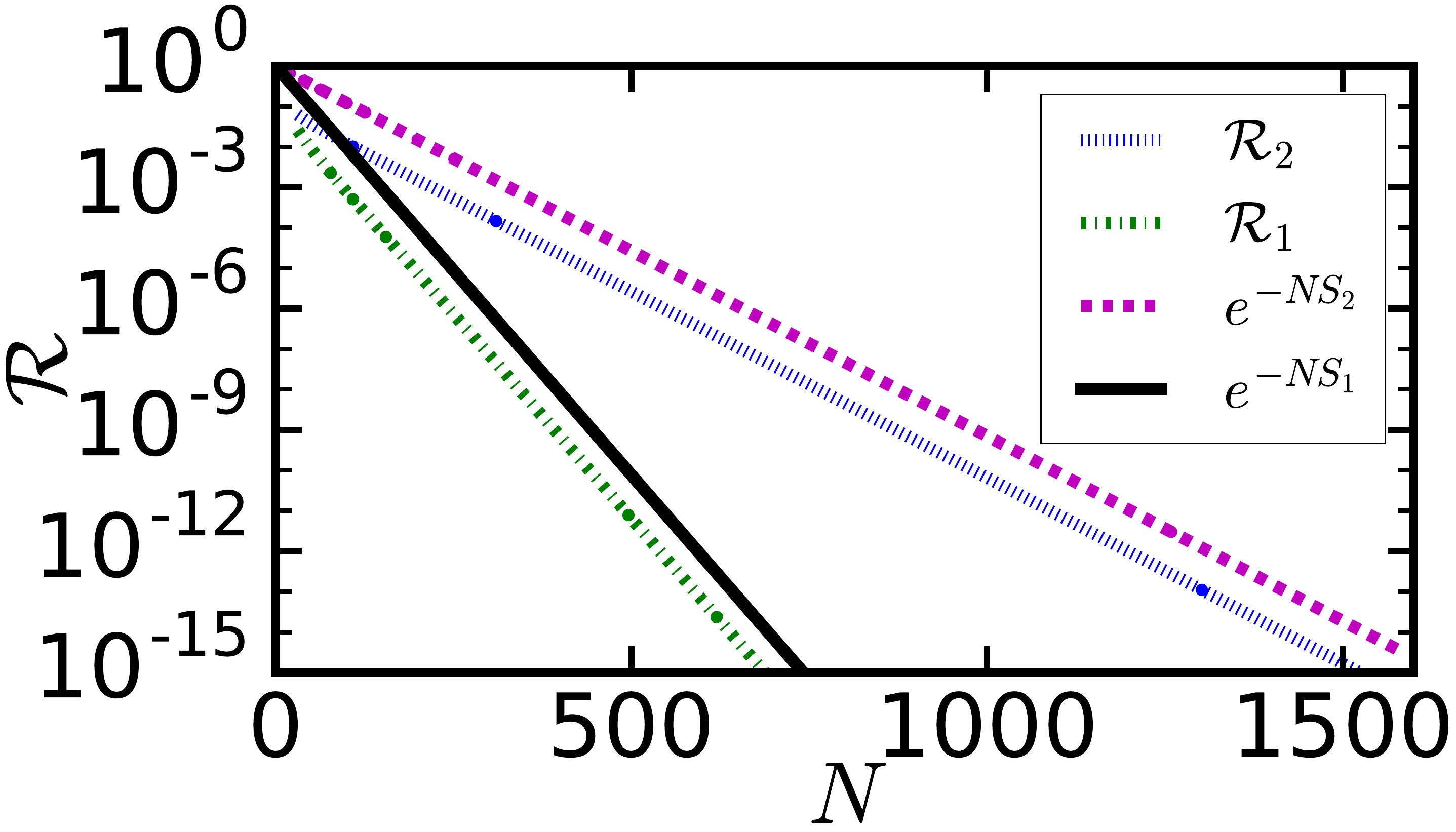}
\caption{Extinction rates  $\mathcal{R}_{i}$, determined by numerically solving the master equation (\ref{MasterEquationSM})
for different $N$, are compared to the extinction rates $e^{-NS_{i}}$ computed in the leading-order WKB approximation. The parameters are $a=1$, $\tau=0.5$, and $\sigma=3.1$. The actions along the instantons are $S_1 \simeq 0.0466$ and $S_2 \simeq 0.0211$. The observed vertical shifts are due to the undetermined WKB prefactors $\mathcal{F}_1$ and $\mathcal{F}_2$. }
\label{fig_extinction_ratesSM}
\end{figure}

\begin{figure}
\includegraphics[width=3.5 in,clip=]{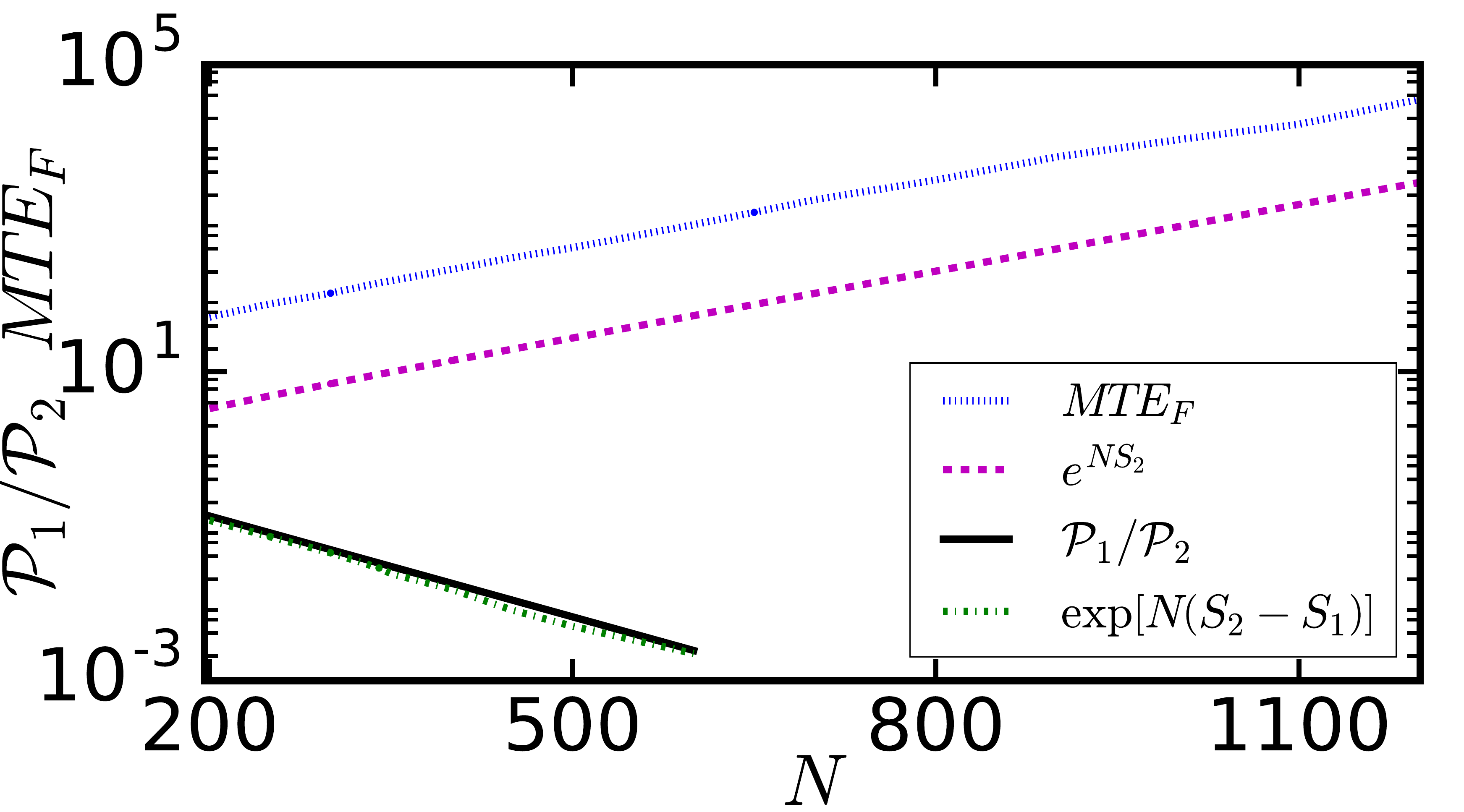}
\caption{$N$-dependence of the MTE of foxes $\text{MTE}_{F}$ and of the ratio of probabilities of the two extinction scenarios $\mathcal{P}_{1}/\mathcal{P}_{2}$ for fixed $a=1$, $\tau=0.5$, and $\sigma=4$.
The results obtained by averaging over many Monte Carlo simulations are compared to the leading-order WKB predictions. The values are plotted on a semi-log scale. The actions are $S_1 \simeq 0.0167$ and $S_2 \simeq 0.00665$.  As $S_2 < S_1$, the MTE of the foxes is approximated in the WKB theory as $\text{MTE}_{F}\simeq e^{NS_{2}}$.
The vertical shifts are due to the undetermined WKB prefactors.}
\label{fig_MTE_and_P1_div_P2SM}
\end{figure}

Figure \ref{fig_extinction_ratesSM} compares the extinction rates $\mathcal{R}_{i}$ obtained by solving numerically the (truncated) master equation (\ref{MasterEquationSM}), with the numerical result of the leading order WKB approximation, $e^{-NS_i}$. In its turn, Fig. \ref{fig_MTE_and_P1_div_P2SM}  shows the MTE of foxes $\text{MTE}_{F}$ and the ratio of the probabilities of the two extinction routes, $\mathcal{P}_{1}/\mathcal{P}_{2}$, obtained by averaging over many Monte-Carlo simulations of the stochastic model. The same figure shows the corresponding leading-order WKB predictions  $\left(e^{-NS_{1}}+e^{-NS_{2}}\right)^{-1}\simeq e^{NS_{2}}$ and $e^{N\left(S_{2}-S_{1}\right)}$, respectively. In both cases a good agreement, up to undetermined WKB prefactors, was observed between the numerical and WKB results \cite{SM2016}.

Ref. \cite{SM2016} also showed that the ``extinction action" changes in a non-analytic way as the system passes through the Hopf bifurcation of the deterministic theory at $\sigma=\sigma^{*}$, when the limit cycle is born. The second derivative of the action with respect to $\sigma$ experiences a jump at $\sigma=\sigma^*$
as in a second-order phase transition. Finally, using the numerical results for the actions $S_1$ and $S_2$ along the instantons, and numerically solving the master equation (\ref{MasterEquationSM}), the authors of Ref. \cite{SM2016} determined the $N$-dependence of the prefactors $\mathcal{F}_{i}$ of the rates, $\mathcal{R}_{i}=\mathcal{F}_{i}e^{-NS_{i}}$, $i=1,2$ at $N\gg 1$, and followed
the change of the $N$-dependence as the system passes through the Hopf bifurcation. At $\sigma<\sigma^*$, the prefactors behave as $\mathcal{F} \propto N^{1/2}$. Above the bifurcation the prefactors are independent of $N$. These findings can be explained by simple normalization arguments \cite{SM2016}.

\section{Population Extinction on Heterogeneous Networks}

Population dynamics on heterogenous networks are very different from those in well-mixed settings because of
the variability in the reaction rates along the network~\cite{PV2001,N2010}. While rare events have been studied on Erd\"{o}s-R\'{e}nyi networks, which are only slightly heterogenous, see \textit{e.g.} Ref.~\cite{LSS2008}, the recent years have witnessed a surge of works
on the long-time population dynamics on degree-heterogenous networks with strong heterogeneity, such as scale-free networks~\cite{BA1999}. Refs.~\cite{ARS2006,SAR2008,AM2012,SMA2017} studied evolutionary models where the network nodes represent either a wild-type individual $A$, or a mutant $B$, and their interactions such as $A+B\to A+A$ or $A+B\to B+B$ determine the evolutionary dynamics. The main questions of interest here are in calculating the fixation probability of each species and the mean fixation time. The authors of Refs.~\cite{ARS2006,SAR2008,AM2012,SMA2017} showed that the non-trivial network topology of heterogenous graphs, and especially of scale-free networks, may cause a dramatic decrease in the mean time of rare events to occur. A striking manifestation of this fact is that,  in contrast to well-mixed systems (or complete graphs), the logarithm of the mean time to fixation may scale sublinearly with the system size $N$, $\;\ln\tau\sim N^{\alpha}$, where $\alpha<1$.

Recently, in Ref.~\cite{HS2016} the authors have investigated the long-time behavior and the MTE of an infectious disease on networks, using the classic SIS model of epidemiology, see Eq.~(\ref{classicSIS}).
They employed the real-space WKB method to analyze the multivariate  master equation describing the stochastic population dynamics on the network. We will now present their
work in some detail.

Following Ref.~\cite{HS2016}, we assume an uncorrelated random
network with a given degree distribution $g_k$, such that $\sum_k g_k=1$, where the
degree $k$ is the number of links of a node. Simple graphs with $N$ nodes can be generated from $g_k$ in several ways, for example, with a configuration model network~\cite{N2010}.
It is useful to represent such networks by the adjacency matrix A, where $A_{ij}$ is $1$ if nodes $i$ and $j$ are linked, and is $0$ otherwise. An infected node is denoted by $\nu_i=1$, and a susceptible node is denoted by $\nu_i=0$. The SIS dynamics involve infection -- a change of $\nu_i$  from $0$ to $1$ -- with probability per unit time $\beta(1-\nu_i)\sum_j A_{ij}\nu_j$, and recovery -- a change of $\nu_i$ from $1$ to $0$ -- with probability per unit time $\alpha\nu_i$, where $\beta$ and $\alpha$ are the infection and recovery rates, respectively~\cite{HS2016}.

To have a closed-form description, the authors of Ref. \cite{HS2016} assumed that there are no correlations between the nodes and replaced $A_{ij}$ by its expectation value in an ensemble of networks $\langle A_{ij}\rangle=k_i k_j/(N\langle k\rangle)$, in the limit of large $N$. This (in general, uncontrolled) assumption is known as the ``annealed" network approximation: a mean-field theory  of heterogeneous networks \cite{ARS2006}. Given this mean-field form, the state of the network can be described by the number of
infected nodes with degree $k$, $I_k$, which has the corresponding
reactions and rates: $I_k\to I_k +1$ with rate of infection
$$
w_k^{inf}(\textbf{I})=\beta k (N_k-I_k)\sum_{k'}k' I_{k'}/(N\langle k \rangle),
$$
and $I_k\to I_k-1$ with recovery rate
$$
w_k^{rec}(\textbf{I})=\alpha I_k,
$$
where $\textbf{I}=(I_1,I_2,\dots,I_{k_{\text{max}}})$, and $N_k=Ng_k$~\cite{HS2016}.

In the stochastic description one writes down a multivariate master equation for the probability $P(\textbf{I},t)$ to find a vector of infected nodes $\textbf{I}$ at time $t$. For networks with a general degree distribution the master equation reads
\begin{eqnarray}\label{masterNET}
\frac{\partial}{\partial t}P(\textbf{I},t)=\sum_{k}&w&_{k}^{inf}(\textbf{I}-\textbf{1}_k)P(\textbf{I}-\textbf{1}_k,t)-w_k^{inf}(\textbf{I})P(\textbf{I},t)\nonumber\\
+&w&_{k}^{rec}(\textbf{I}+\textbf{1}_k)P(\textbf{I}+\textbf{1}_k,t)-w_k^{rec}(\textbf{I})P(\textbf{I},t),
\end{eqnarray}
where $\textbf{1}_k=(0_{k_1},0_{k_2},\dots,1_k,\dots,0_{k_{\text{max}}})$. Essentially, this master equation describes the dynamics of $k_{\text{max}}$ interacting populations.

For large but finite networks,  the population enters a long-lived endemic state. We set $P(\textbf{I},t)=\pi(\textbf{I})e^{-t/\tau}$, where $\tau$ is the MTE of the disease, and use the WKB ansatz for the QSD, $\pi(\textbf{x})=e^{-NS(\textbf{x})}$, where $\textbf{x}=\textbf{I}/N$. In the leading order of $N\gg 1$ this leads to a stationary Hamilton-Jacobi equation $H(\textbf{x},\textbf{p})=0$ with the Hamiltonian~\cite{HS2016}
\begin{equation}\label{HamiltonNET}
H(\textbf{x},\textbf{p})=\sum_{k}\left[\beta k(g_k-x_k)(e^{p_k}-1)\sum_{k'}\frac{k'x_{k'}}{\langle k\rangle}+\alpha x_k(e^{-p_k}-1)\right],
\end{equation}
where $\textbf{p}=\partial_{\textbf{x}}S$ are the momenta.
The problem boils down to finding the instanton: the zero-energy activation trajectory of the Hamilton's equations
\begin{eqnarray}
\dot{y}_k&=&\tilde{\beta}k(1-y_k)e^{p_k}\sum_{k'}\frac{k'g_{k'}}{\langle k \rangle}y_{k'}-y_k e^{-p_k},\nonumber\\
\dot{p}_k&=&\tilde{\beta}k\sum_{k'}\frac{k'g_{k'}}{\langle k \rangle}\left[y_{k'}(e^{p_k}-1)-(1-y_{k'})(e^{p_k'}-1)\right]-e^{-p_k}+1.
\end{eqnarray}
Here $y_k=x_k/g_k$ is the fraction of
each degree class infected, $\tilde{\beta}=\beta/\alpha$, and time is rescaled: $t\to\alpha t$. The activation trajectory
must exit, at $t=-\infty$, the endemic fixed point  $y_k^*=[1+1/(Y\tilde{\beta}k)]^{-1}$ and $p_k=0$ and enter, at $t=\infty$,
the extinction fixed point $y_k=0$ and $p_k^*=-\ln[1+\tilde{\beta}k(1-P)]$. The functions $Y$ and $P$ satisfy the conditions $(Y,P)\to (0,1)$ when $R_0\to 1$, and $(Y,P)\to (1,0)$ as $R_0\to\infty$, where $R_0\equiv \tilde{\beta}\langle k^2\rangle/\langle k\rangle$ is the basic reproduction number. The MTE can be evaluated as  $\tau\sim e^{S}$, where  $S=\sum_k g_k\int p_k \dot{y}_k dt$ is the action along the activation trajectory. In Ref.~\cite{HS2016} this problem was solved numerically, whereas analytical results were obtained close to bifuraction, $R_0-1\ll 1$, and in the strong infection limit, $R_0\gg 1$.

Future work will have to account for correlations between nodes and incorporate more general statistics of adjacency matrix  $A_{ij}$.

\section{Extinction of Spatial Populations}
\label{spatialextinction}

It has been long known \cite{Skellam}, that migration of individuals plays a crucial role in  many environments of relevance to population biology and epidemiology \cite{Murray2008}.
In Secs. \ref{minmigration}  we reviewed some recent work on extinction of populations whose individuals can migrate on a network of ``patches".
Here we will focus on  populations whose individuals migrate in space. For simplicity, we will limit ourselves
to one spatial dimension. Following Refs. \cite{EK2004,MS2011}, let us consider a single population residing on a one-dimensional lattice of $N\gg 1$ sites (or habitat patches) labeled by index $i=1,2,\dots, N$. The population size $n_i$ at each site  varies in time as a result of two types of Markov processes. The first set of processes involves on-site stochastic dynamics of birth-death type, with
birth and death rates $\lambda(n_i)$ and $\mu(n_i)$, respectively, where $\mu(0)=0$. As there is no creation of new individuals ``from vacuum", one also has $\lambda(0)=0$. The second process is random and independent migration of each individual between neighboring sites with migration rate coefficient $D_0$. What happens at the edges of the refuge, $i=1$ and $i=N$, needs to be specified separately.

Assuming $n_i\gg 1$ and neglecting fluctuations, one obtains a set of coupled deterministic rate equations
\begin{equation}\label{rateeq1}
   \dot{n}_i=\lambda(n_i) -  \mu(n_i)+ D_0 (n_{i-1}+n_{i+1}-2 n_i)\,.
\end{equation}
Let the characteristic population size on a single site, predicted by a non-trivial steady-state solution of Eq.~(\ref{rateeq1}), scale as $K\gg 1$.  This implies \cite{Doering,AM2010} that, in the leading order of $K$, we can represent the birth and death rates as
\begin{equation}\label{rates}
    \lambda(n_i)=\mu_0 K \bar{\lambda}(q_i)\;\;\;\;\;\mbox{and} \;\;\;\;\;\mu(n_i)=\mu_0 K \bar{\mu}(q_i),
\end{equation}
where $q_i=n_i/K$ is the rescaled population size at site $i$, $\bar{\lambda}(q_i)\sim \bar{\mu}(q_i) ={\cal O}(1)$, and $\mu_0$ is a characteristic rate coefficient. Now Eq.~(\ref{rateeq1}) can be rewritten as
\begin{equation}\label{rateeq2}
   \dot{q}_i=\mu_0 f(q_i)+ D_0 (q_{i-1}+q_{i+1}-2 q_i)\,,
\end{equation}
where $f(q_i)=\bar{\lambda}(q_i) - \bar{\mu}(q_i)$. When  $D_0\gg \mu_0$, one can introduce a continuous spatial coordinate $x$ instead of the discrete index $i$ and replace the discrete Laplacian in Eq.~(\ref{rateeq2}) by the continuous one \cite{MS2011}. This leads to the reaction-diffusion equation
\begin{equation}\label{rateeq3}
   \partial_t q=\mu_0 f(q)+ D \partial_x^2 q\,,
\end{equation}
where $D=D_0 h^2$ is the diffusivity, and $h$ is the lattice spacing. The system size becomes $L=N h$. Equation~(\ref{rateeq3}) has been the subject of numerous studies, see \textit{e.g.} \cite{Murray2008,Mikhailov}. Ref. \cite{MS2011} considered periodic, $q(x+L)=q(x)$, and absorbing, $q(0)=q(L)=0$, boundary conditions. The absorbing boundaries model extremely harsh conditions outside of the refuge \cite{EK2004,Skellam}. Results for still another type of boundaries -- reflecting walls at $x=0$ and $x=L$ -- can be easily obtained from the results for periodic boundaries.

Spatial deterministic profiles of established populations are described by stable steady-state solutions $q=q(x)>0$ of Eq.~(\ref{rateeq3}).  They satisfy the ODE
\begin{equation}\label{steady}
   D q^{\prime\prime}(x)+\mu_0 f(q)=0
\end{equation}
subject to the chosen spatial boundary conditions. The first integral of this equation,
\begin{equation}\label{energy}
    \frac{D}{2\mu_0}\left(q^{\prime}\right)^2+V(q)=const\,,
\end{equation}
with effective potential $V(q)=\int_0^q f(\xi) \,d \xi$, makes the problem solvable in quadratures and readily yields the phase portrait of possible steady states on the plane $(q,q^{\prime})$.
Importantly, the reaction-diffusion Eq.~(\ref{rateeq3}) is a gradient flow, $\partial_t q =- \delta {\cal F}/\delta q$, where
\begin{equation}\label{freeenergy}
    {\cal F}[q(x,t)]=\int_0^L \,dx\,\left[-\mu_0 V(q)+(1/2) \,D(\partial_x q)^2\right].
\end{equation}
Therefore, it describes a deterministic flow towards a minimum of the
Ginzburg-Landau free energy ${\cal F}[q]$. This helps identify linearly stable and unstable $x$-dependent solutions, as they correspond to local minima and maxima of ${\cal F}[q]$, respectively \cite{Mikhailov}.

Going back to the lattice formulation and reintroducing the stochasticity of the on-site processes and of the migration, one
should deal with a master equation for the multivariate distribution of population sizes on different patches \cite{EK2004,MS2011}. Being interested in
extinction of established populations in a finite region of space, one arrives at a \textit{quasi-stationary} master equation \cite{MS2011}. Applying the WKB ansatz to the quasi-stationary equation, and assuming a high migration rate, one finally arrives at the following Hamilton's equations in the real space \cite{MS2011}:
\begin{eqnarray}
\partial_t q&=&\mu_0\left[\bar{\lambda}(q)e^p - \bar{\mu}(q)e^{-p}\right]
+D \left[\partial_x^2q -2\partial_x\left(q\partial_x p\right)\right] ,
\label{p100}\\
\partial_t p&=& -\mu_0 \left[\bar{\lambda}^{\prime}(q)(e^p-1) + \bar{\mu}^{\prime}(q)(e^{-p}-1)\right]
-D\left[\partial_x^2 p +\left(\partial_x p\right)^2\right] ,
\label{p110}
\end{eqnarray}
Equivalent equations, up to a canonical transformation, were obtained in Ref.~\cite{EK2004}.  Without migration, these equations coincide with those for well-mixed populations, see Eq.~(\ref{hamil}).  Without births and deaths,
they describe a simple particular case --  non-interacting random walkers -- of the so called Macroscopic Fluctuation Theory: a WKB theory of large deviations in diffusive lattice gases \cite{MFT}.

The boundary conditions in space are problem-dependent, see above. For periodic or no-flux boundaries $p(x,t)$ is also periodic or no-flux at the boundaries. For the absorbing boundaries $p$ vanishes at the boundary (up to corrections linear in the lattice constant) \cite{MS2011}.  As shown in Ref. \cite{MS2011}, the optimal path of a quasi-stationary population to extinction is again an instanton: a ``heteroclinic trajectory" between two ``fixed points" in the infinite-dimensional functional phase space of the system. The first ``fixed point" corresponds to the long-lived quasi-stationary distribution of the population size in space.  In the absence of the Allee effect, the second ``fixed point" corresponds to a zero-population-size state with a nontrivial spatial profile of the conjugate momentum $p$. This extinction instanton can be found numerically by the Chernykh-Stepanov iteration algorithm \cite{ChernykhStepanov}, and in some limits analytically, see Refs. \cite{EK2004,MS2011}.

\begin{figure}[ht]
\includegraphics[width=2.5 in,clip=]{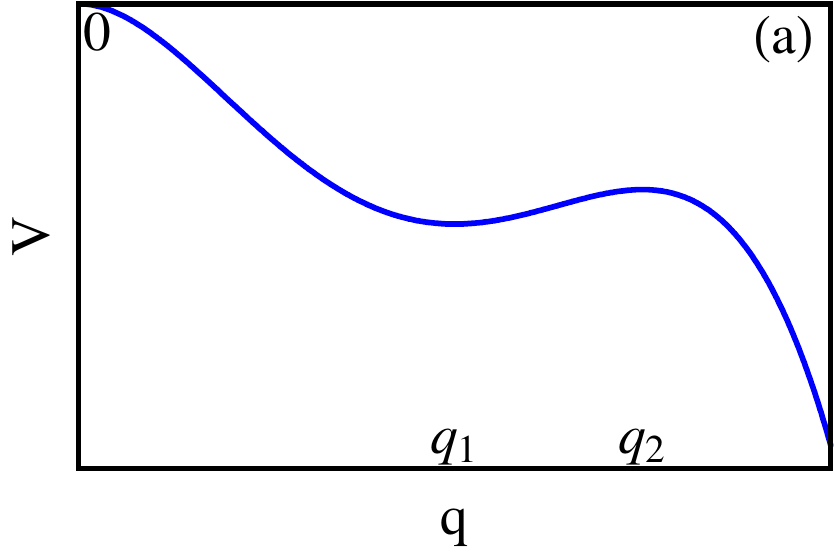}
\includegraphics[width=2.5 in,clip=]{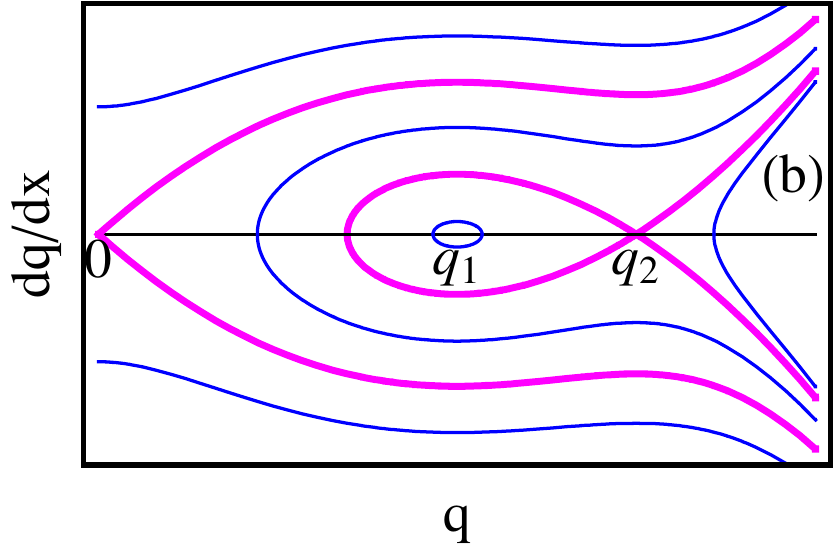}
\caption{Effective potential $V(q)$ (a) and phase portrait $(q,q^{\prime})$ (b) for steady-state solutions of Eq.~(\ref{rateeq3}) for a strong Allee effect, $V(0)>V(q_2)$.}
\label{Bstrong1}
\end{figure}

\begin{figure}[ht]
\includegraphics[width=2.5 in,clip=]{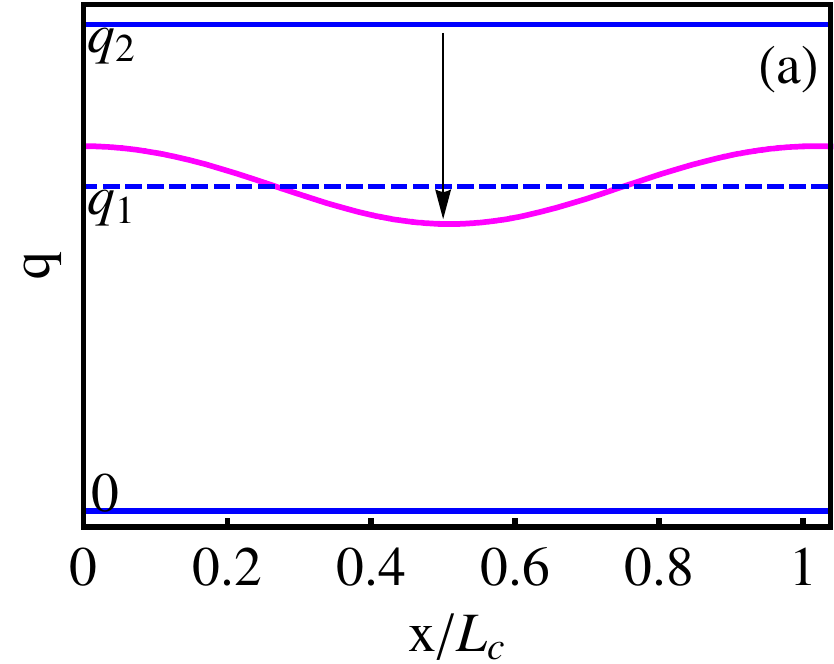}
\includegraphics[width=2.5 in,clip=]{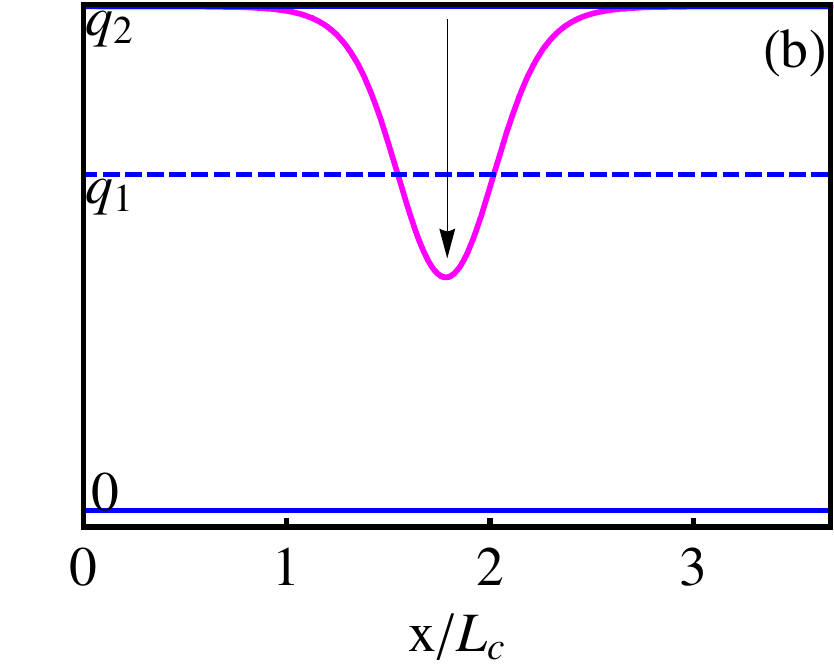}
\caption{Linearly stable states $q=q_2$ and $q=0$, linearly unstable state $q=q_1$ and critical nucleus for a strong Allee effect and periodic boundary conditions. The system size is weakly (a) and strongly (b) supercritical. The thick curves show the critical nucleus. The arrows indicate  transitions, driven by rare large fluctuations and leading to a quick population extinction. }
\label{Bstrong2}
\end{figure}

In the presence of a sufficiently strong Allee effect the second ``fixed point" of the heteroclinic trajectory is \textit{deterministic}. It describes the \textit{critical nucleus}  for extinction:  \cite{MS2011}. Let us assume periodic boundaries, see Figs. \ref{Bstrong1} and \ref{Bstrong2}. Here the only linearly stable nontrivial deterministic steady-state solution is the $x$-independent solution $q=q_2$, where $q_2>0$ is the attracting point of the deterministic equation $\dot{q}=\mu_0 f(q)$. A sufficiently large perturbation, however, triggers a \textit{deterministic} transition from $q=q_2$ to the trivial solution $q=0$
that is also linearly stable. The critical nucleus is an $x$-dependent deterministic solution $q_c(x)$ of Eq.~(\ref{steady}) which is linearly unstable under the dynamics of Eq.~(\ref{rateeq3}). The critical nucleus corresponds to a local \textit{maximum}
of free energy (\ref{freeenergy}) \cite{Mikhailov}. A small perturbation around the critical nucleus drives the system either to $q=0$ or to $q=q_2$. In a finite-size system the critical nucleus depends on the system size $L$ and corresponds to a phase trajectory inside the internal separatrix shown in Fig. \ref{Bstrong1}b \cite{MS2011}. The critical nucleus exists only for sufficiently large systems, $L>L_c$, where the critical system size $L_c$ can be obtained by linearizing Eq. (\ref{steady}) around $q=q_2$.  At $L\gg L_c$ the critical nucleus approaches the internal separatrix in Fig. \ref{Bstrong1}b. If  $f(q)$  is a cubic polynomial, the critical nucleus can be found analytically,  otherwise it can be found numerically.  Figure~\ref{Bstrong2} shows the critical nuclei for two different values of $L>L_c$.

For extinction to occur a large fluctuation of the size of a stochastic population residing around $q=q_2$ only needs to  create the critical nucleus. The further population dynamics toward extinction proceed essentially deterministically. What happens at $L\gg L_c$, see Fig.~\ref{Bstrong2}b, is intuitively clear. Once having passed the critical nucleus, the solution $q(x,t)$ of Eq.~(\ref{rateeq3}) develops, on a fast time scale $\sim \mu_0^{-1}$, two outgoing deterministic ``extinction fronts" that drive the whole population to extinction on a time scale $\sim L/(\mu_0 D)^{1/2}$. As a result, the MTE is determined by the mean creation time of the critical nucleus. This quantity does not include an exponential dependence on the system size $L$ (unless $L$ is exponentially large in $K$ \cite{MS2011}) and is therefore much shorter than the MTE in the absence of Allee effect.

For a \textit{very} strong Allee effect -- when the system is close to the saddle-node bifurcation, corresponding to the appearance of the stable fixed point $q=q_2$ of the deterministic equation $\dot{q}=\mu f(q)$ -- the WKB formulation of the population extinction problem turns out to be completely integrable, similarly to the complete integrability of the WKB problem for population explosion close to the saddle-node bifurcation, earlier studied in Ref. \cite{EK2004}. This integrability can be explained by establishing a direct mapping between this problem and the overdamped limit of the theory of homogeneous nucleation due to Langer \cite{Langer}. We refer the reader to papers \cite{EK2004,MS2011} for details, derivations and additional examples.

Nucleation via fluctuations also plays a critical role in the formation of macroscopic \textit{clusters} (for example, above the critical point in certain lattice gas models). This fundamental problem is of interest in several biological contexts such as quorum sensing and clustering of motile adhesive cancer cells \cite{KKS}. In the latter case, clustering via nucleation can explain growth of recurrent tumors. Ref. \cite{KKS} studied this problem numerically and via a WKB approximation to a phenomenological master equation.

\section{Large Velocity Fluctuations of Population Invasion Fronts}

Biological invasions is one of the key issues in contemporary
ecology \cite{Williamson}. The term describes a broad variety of phenomena
related to introduction and spread of non-native species
in ecosystems. Biological invasions often have a dramatic negative impact on native
eco-communities, and are considered as one of the main reasons
for biodiversity loss in the world. They may also jeopardize agriculture and lead
to economic losses. Mathematical modeling plays an important role in the understanding
of biological invasions \cite{Allen,Petrovskii}.  Here we will consider simple one-population
invasion fronts that, at the level of deterministic theory, are described by a traveling front
solution of a reaction-diffusion equation. When the demographic noise is taken into account,
the reaction front position fluctuates.   What
is the probability that the fluctuating reaction front moves slower or faster
than its deterministic counterpart?  The WKB approximation can be very useful in providing
insightful answers.
This theory assumes many particles in the transition region of the front.
As it turns out, details crucially depend on whether the reaction front
describes propagation into a state that is deterministically unstable or stable. We will only
consider here fronts propagating into an unstable state, where the effects of noise are remarkably strong.

A celebrated example of a reaction front propagating into an unstable state is provided by the Fisher-Kolmogorov-Petrovsky-Piscounov (FKPP)
equation \cite{Fisher},
\begin{equation}\label{FKPP}
    \partial_t q = q-q^2+\partial_{x}^2 q,
\end{equation}
where $q(x,t)$ is the particle density. This equation describes invasion of an unstable state $q(x\to \infty,t)=0$ by a stable state $q(x\to -\infty,t)=1$.
The FKPP equation is a fundamental model in population biology and mathematical genetics \cite{Fisher}, but it also appears in many other applications.
Equation~(\ref{FKPP}) has a family of  traveling-front solutions (TFSs) parameterized by their velocity $c$: $q(x,t)=Q_{0,c}(\xi)$, where $\xi=x-ct$.
$Q_{0,c}(\xi)$ obeys the  equation
\begin{equation}\label{MFeq}
    Q_{0,c}^{\prime\prime}+c Q_{0,c}^{\prime}+Q_{0,c}-Q_{0,c}^2=0.
\end{equation}
and boundary conditions $Q_{0,c}(-\infty)=1$ and $Q_{0,c}(\infty)=0$. The invasion front can have any velocity from the interval
$c_0\leq c<\infty$, where $c_0=2$. It is known, however,  that for sufficiently steep initial conditions the solution of the time-dependent equation~(\ref{FKPP}) approaches at long times a limiting TFS, $Q_{0,2}$, of Eq.~(\ref{MFeq}) with the velocity $c_*=2$ \cite{Fisher,Bramson,Saarloos}.
Importantly, the special velocity  $c_0=2$
is determined only by the dynamics of the leading edge of the front, where Eq.~(\ref{FKPP}) can be linearized around $q=0$. That is,  the full nonlinear front, described by Eq.~(\ref{FKPP}), is ``pulled" by its leading edge, hence the term ``pulled fronts" \cite{Saarloos}, of which the FKPP equation (\ref{FKPP}) is the best known example.

The intrinsic (demographic) noise, unaccounted for by Eq.~(\ref{FKPP}), causes deviations of the front position with time. These include a systematic front velocity shift and fluctuations that \textit{typically} are diffusion-like. If $N\gg 1$ is the characteristic number of particles in the front region, the front velocity shift scales as  $(\ln N)^{-2}$ \cite{Derridashift,Levine,Derridanum}, whereas the front diffusivity scales as $(\ln N)^{-3}$  \cite{Derridanum,Panja1,Derridatheory}. The logarithmic $N$-dependencies are markedly different from the more expected $1/N$ dependencies of the same quantities for the fluctuating fronts propagating into \textit{locally stable} states \cite{Kessler,Panja2,MSK2011,KM2013,Frey2}. This fact makes the pulled fronts much more interesting.

The starting point of the stochastic treatment of population invasion fronts can be the same as
in the spatial extinction problem, see Sec. \ref{spatialextinction}. One starts from the exact master equation
in the lattice formulation. Then, assuming a large $N$, one can apply a WKB ansatz directly to the time-dependent multivariate distribution of population sizes. Neglecting the second derivatives of the action and assuming a fast diffusion in space, one again arrives at the continuous Hamilton's
equations~(\ref{p110}) \cite{MSK2011,MS2011}. Ref. \cite{MS2011} studied (negative) velocity fluctuations of a
pulled front on the example of reversible reactions $A\rightleftarrows 2A$ and independent random walk.  In this case
the (properly rescaled) Eqs.~(\ref{p110}) become
\begin{eqnarray}
\partial_t q&=&q e^p - q^2 e^{-p}+ \partial_x^2q -2\partial_x\left(q\partial_x p\right),
\label{p101}\\
\partial_t p&=& 1-e^p -2q (e^{-p}-1)
-\partial_x^2 p -\left(\partial_x p\right)^2\,.
\label{p1100}
\end{eqnarray}
The specifics of the problem of fluctuations of the front position are reflected in the boundary conditions in space and in time \cite{MS2011,MVS2012}.
At $x\to -\infty$ there is a stationary distribution of the particle density, peaked at $q=1$.
Therefore, the boundary conditions at $x\to -\infty$ are $q(-\infty,t)=1$ and
$p(-\infty,t)=0$ which correspond to the fixed point $(q=1,p=0)$ of the \textit{on-site} Hamiltonian
$H_0(q,p)= q (e^p-1) +q^2 (e^{-p}-1)$. The boundary condition at $x\to \infty$ is $q(\infty,t)=0$. The momentum
$p$ can be unbounded at $x=\infty$, but it must be bounded at finite $x$ \cite{MS2011}. The boundary conditions
in time involve two kink-like profiles of $q$ at $t=0$ and $t=T\gg 1$ which are at distance $X\gg 1$ apart (where $X$ can
be positive, negative or zero). The front velocity
is defined as $c=X/T$.  A natural constraint in the context of invasion
is  that the initial particle density is zero to the right of some finite point in space.

\begin{figure}[ht]
\includegraphics[width=3 in,clip=]{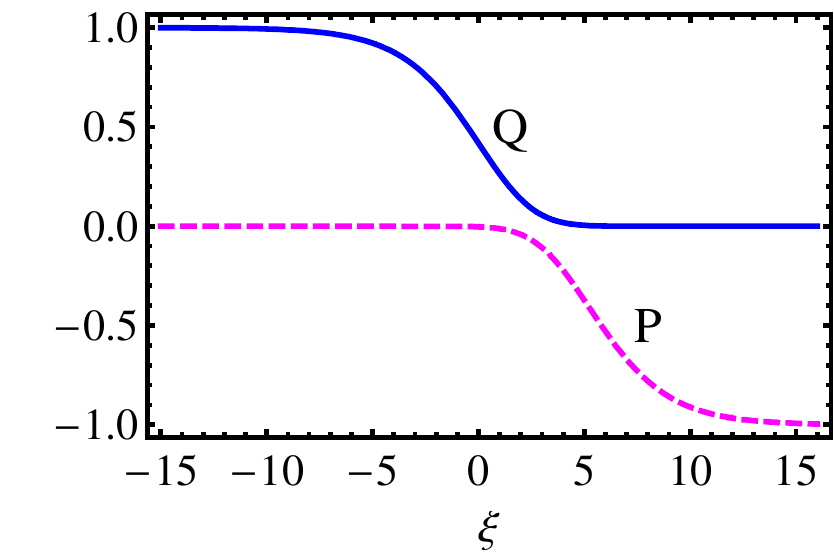}
\caption{Numerical $Q(\xi)$ and $P(\xi)$ profiles for $c=1.5$ obtained in Ref. \cite{MS2011}.}
\label{1.5}
\end{figure}

\begin{figure}[ht]
\includegraphics[width=2.4 in,clip=]{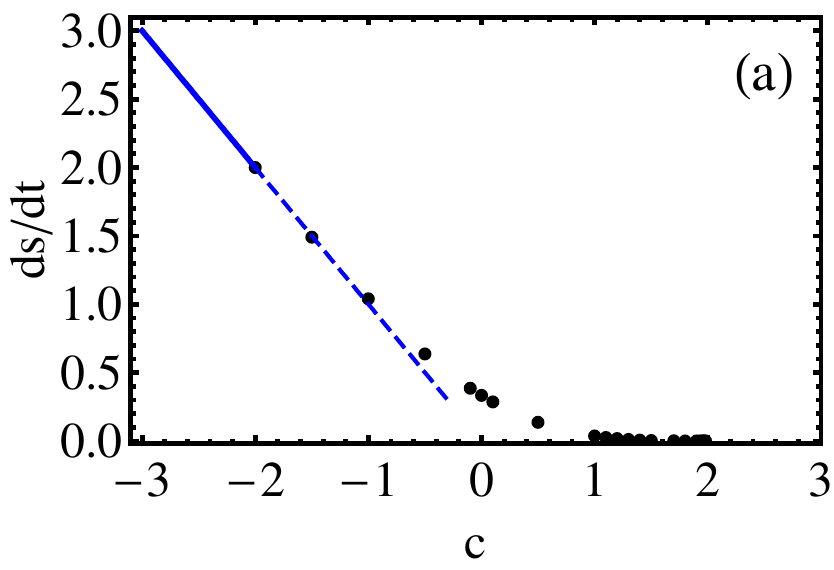}
\includegraphics[width=2.4 in,clip=]{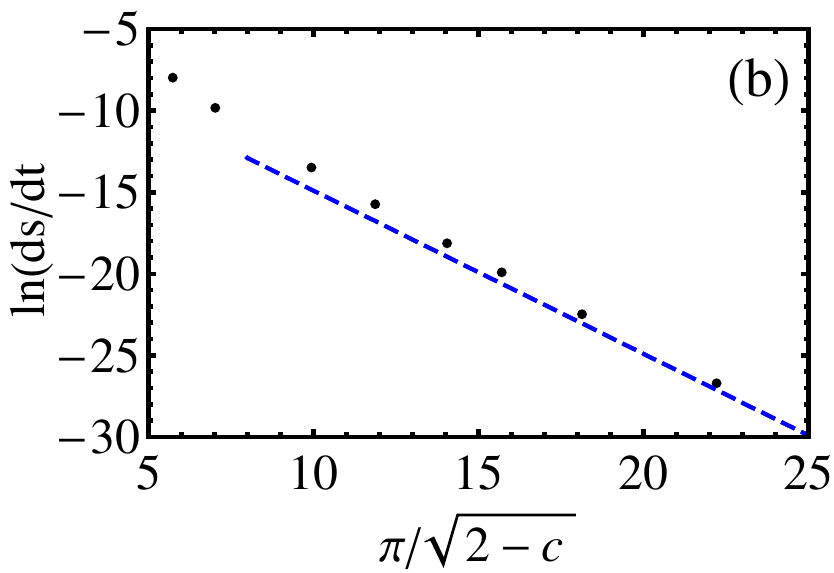}
\caption{(a) Symbols: the
large deviation function ${\cal F}(c)$, see Eq.~(\ref{accumrate}), found numerically in Ref. \cite{MS2011},
versus the front velocity $c$. Straight line: the asymptote
${\cal F}(c)=-c$ which becomes exact at $c\leq -2$ \cite{MS2011}.
(b) $\ln {\cal F}(c)$ versus $\pi/\sqrt{2-c}$ for $c\geq 1.5$. Symbols: numerical results. Dashed
line: asymptote (\ref{asymplead}).}
\label{numaction}
\end{figure}

Equations (\ref{p101}) and (\ref{p1100}) look simpler in the new variables $Q=q e^{-p}$ and $P=e^p-1$.  But a crucial simplification appears
when one realizes that, except in narrow boundary layers in time at $t=0$ and $t=T\gg 1$, the
optimal path of the system approaches a traveling-front solution (TFS),  $Q(x,t)=Q(x-ct)$ and $P(x,t)=P(x-ct)$, of Eqs.~(\ref{p101}) and (\ref{p1100}).
As a result, these PDEs become ODEs.   The probability density of observing the front velocity $c$ is, in the leading order,
\begin{equation}\label{scaling}
 - \ln {\cal P}(X,T) = N T \,{\cal F}(X/T)=  N T \,{\cal F}(c), \quad |c|<2.
\end{equation}
The optimal TFS, and the large deviation function (or rescaled action) $\mathcal{F}(c)$,
\begin{equation}\label{accumrate}
  {\cal F}(c) =\int_{-\infty}^{\infty} \!  d\xi  \left[Q(\xi)-Q^2(\xi)\right] P^2(\xi),
\end{equation}
can be easily found numerically \cite{MS2011}. Notice also, that the ODEs have an integral of motion that can be found explicitly
\cite{MS2011}. Figure \ref{1.5} shows the numerical solution for  $c=3/2$. As one can see, for this unusually slow front
the $P$-front is well ahead of the $Q$-front. Essentially, the $P$-front modifies the effective reaction rates so as to slow down the front in comparison to its deterministic counterpart at $P=0$.   Figure~\ref{numaction}a shows the numerically evaluated ${\cal F}(c)$, see Eq.~(\ref{accumrate}), at different $c$.

One interesting regime where an analytic solution -- by a matched asymptotic expansion -- can be obtained is $c_0-c \ll 1$ \cite{MS2011}.
In this case the main contribution to the
large deviation function comes from the leading edge of the front, $\xi\gg 1$, where $Q\ll 1$.
As $|P| \sim 1$ there, this theory
must be nonlinear.
On the other hand,  $|P|\ll 1$ in the left region, whereas $Q\ll 1$ in the right region. An analytical theory is
possible because there is a \textit{joint} region, where the strong inequalities $|P|\ll 1$ and $Q\ll 1$ hold
simultaneously. We refer the reader for details to Ref. \cite{MS2011}. Finally one obtains
\begin{equation}\label{asymplead}
    {\cal F}(c) \simeq 0.0074 \, \exp\left(-\frac{\pi}{\sqrt{2-c}}\right) .
\end{equation}
As a result,
\begin{equation}\label{P}
    \ln \mathcal{P}(X,T) \simeq  -0.0074\, N T \exp\left(-\frac{\pi}{\sqrt{2-c}}\right) .
\end{equation}
As one can see, the negative tail of the front speed distribution around $c_0$ is strongly non-Gaussian,
and rapidly falls as $c$ goes down. It also strongly depends on $N$. The latter property reflects the fact that, in order to considerably slow down the invasion front,
a \textit{multi-particle} optimal fluctuation is required.  Being a WKB asymptotic, Eq.~(\ref{P}) holds only when there are many particles at the leading edge of the front, where $\xi\simeq \xi_0\simeq \pi/\sqrt{2-c}$. This condition leads to the strong inequality
\begin{equation}\label{strongineq}
  c_{*}-c\gg 2\pi^2 \ln^{-3} N,
\end{equation}
where $c_{*}=2-\pi^2 \ln^{-2} N$ describes the \textit{systematic} correction to the front velocity, coming from the noise \cite{Derridashift}. By putting $c=c_*-\delta c$,
where $\delta c \ll \ln^{-2} N$, one can see that the linear dependence on $N$ in Eq.~(\ref{P}) cancels out,
giving way to a much slower logarithmic dependence, compatible with a phenomenological theory of typical fluctuations of a pulled front's position \cite{Derridatheory}.

It was realized in Ref. \cite{MS2011} that the result~(\ref{P})
holds, up to a $c$-independent   numerical coefficient, for all pulled reaction fronts,
where $A\to 2A$ is the only first-order birth process. It also holds, close to the transcritical bifurcation, for a whole class of reactions. The (properly rescaled) on-site Hamiltonian for this class of models, $H_0(Q,P)=QP(P-Q+1)$ \cite{Kamenev2}, was discussed in Secs. \ref{catastrophe} and \ref{extendemic}. Fluctuations of the front velocity for this class of models
were studied in Ref. \cite{MVS2012}. In particular, the authors solved numerically the complete time-dependent WKB equations -- two coupled PDEs similar to Eqs.~(\ref{p1100}). As they observed,  the TFS is indeed the true optimal history of the system, except in the narrow boundary layers at $t=0$ and $t=T\gg 1$,  and it yields the leading-order contribution to $\ln {\cal P}$ as described by the scaling relation (\ref{scaling}).

Large \textit{positive} deviations of the pulled front velocity have an entirely different nature. In this case the distribution $\mathcal{P}(X,T)$ is independent of $N$ in the leading order of theory \cite{DS2016,DMS2016}. The positive deviations are of a non-WKB character, as they are determined by only a few particles that run ahead of the front.

\section{Summary and Some Open Questions}
We have reviewed recent progress in applying the WKB approximation to the study of large deviations in stochastic population dynamics.  The WKB approximation yields a versatile and universal framework for analysis of large deviations. In many cases it enables one to determine the optimal path to the specified large deviation and to evaluate, analytically or numerically, the mean time to extinction/fixation/switching, and/or the corresponding probabilities. Different variants of the WKB approximation, developed by physicists, are steadily ``invading" communities of mathematical biologists, and this process is likely to continue.

There are still many unresolved questions. Here are some of them. It would be interesting to search for additional instances of time-scale separation in multi-population systems in order to have a better analytical insight.  Large deviations on heterogeneous networks remain largely unexplored beyond the mean-field-like ``annealed network" assumption. In spatially explicit large-deviation problems, time-scale separation seems to be the only hope for analytical progress, except when the optimal path is describable in terms of traveling wave solutions of the WKB equations, and the problem simplifies dramatically. Finally, it would be very interesting to investigate the effect of static disorder (``good" and ``bad" regions of space) on the large deviations in spatially-explicit populations.

\section*{Acknowledgments}

We are very grateful to colleagues and students who worked with us on different aspects of stochastic population dynamics. They are too many to be listed here. We are also grateful to Joachim Krug -- who suggested that we write this review -- for useful discussions. We were supported by the Israel Science Foundation (Grant No 408/08) and by the U.S.-Israel Binational Science Foundation
(Grant No. 2008075). M.A. was also supported by the Israel Science Foundation through Grant No. 300/14. B.M. acknowledges support from the University of Cologne through the Center of Excellence ``Quantum Matter
and Materials."

\end{document}